\documentclass[12pt]{article}
\usepackage[T1]{fontenc}

\usepackage[utf8]{inputenc}

\usepackage{geometry}
\usepackage{amssymb}
\usepackage{amsfonts}
\usepackage{graphicx}
\usepackage{amsmath}
\usepackage{endnotes,harvard}
\usepackage{rotating}
\usepackage{pdfpages}
\usepackage{setspace}
\usepackage{pdfsync,ulem}
\usepackage{dcolumn}
\usepackage{longtable}

\usepackage[space]{grffile}
\usepackage[lofdepth,lotdepth]{subfig}
\usepackage{url}

\def\spvec#1{\left(\vcenter{\halign{\hfil$##$\hfil\cr \spvecA#1;;}}\right)}
\def\spvecA#1;{\if;#1;\else #1\cr \expandafter \spvecA \fi}
\newcommand{\bs}{\boldsymbol}
\usepackage{algorithm}
\usepackage{algorithmic}

\begin{document}

\title{Demand Modeling, Forecasting, and Counterfactuals, Part I\thanks{We thank
Kamal Chavda, Dr. Carol Johnson, Peter Sloan, Kim Rice, and Jack Yessayan and the staff at Boston Public Schools
for their expertise and for facilitating
access to the data used in this study.  Matt Gentzkow, Ali Hortacsu, Jon Levin, and Ariel Pakes provided helpful
feedback.  Financial support provided by the National Science Foundation.}}
\author{Parag A. Pathak and Peng Shi\thanks{Pathak: Massachusetts Institute of
Technology and NBER, Department of Economics, email: \textsf{ppathak@mit.edu}
and Shi: Massachusetts Institute of Technology, Operations Research Center, email: \textsf{pengshi@mit.edu}.}}
\date{First draft: January 2014 \\This draft: January 2015}
\maketitle

\begin{abstract}
There are relatively few systematic comparisons of the ex ante counterfactual predictions
from structural models to what occurs ex post. This paper  uses a large-scale policy change
in Boston in 2014 to investigate the performance of discrete choice models of demand compared to simpler
alternatives.  In 2013, Boston Public Schools (BPS) proposed alternative zone configurations in their school choice plan,
each of which alters the set of schools participants are allowed to rank. \citeasnoun{pathak/shi:13} estimated discrete choice models
of demand using families' historical choices and these demand models were used to forecast the outcomes under alternative plans.
BPS, the school committee, and the public used these forecasts to compare
alternatives and eventually adopt a new plan for Spring 2014.
This paper updates the forecasts using the most recently available historical data on participants' submitted preferences
and also makes forecasts based on an alternative statistical model not based a random utility foundation.
We describe our analysis plan, the methodology, and the target forecast outcomes.  Our
ex ante forecasts eliminate any scope for post-analysis bias because they are made before
new preferences are submitted.  Part II
will use newly submitted preference data to evaluate these forecasts and assess the strengths and limitations
of discrete choice models of demand in our context.
\end{abstract}

\newpage

\section{Introduction}

The aim of developing models capable of quantitatively forecasting the effects of policy changes has been an objective
of economic science since at least \citeasnoun{hurwicz:50} and \citeasnoun{marschak:53}.
Recent years, in contrast, have seen the popularity of design-based research strategies
that focus on estimating particular parameters or causal effects.
The design-based approach does not typically lend itself to ex ante policy evaluations
involving changes outside of historical experience.  Both \citeasnoun{angrist/pischke:10} and \citeasnoun{heckman:10}
attribute the growth of the design-based
research program to skepticism on the value of structural modeling
for counterfactual analysis given its reliance on parametric and/or behavioral assumptions.

Though opinions vary greatly on the credibility of structural models,
there is one area of consensus: there are relatively few systematic comparisons of
the ex ante counterfactual predictions of structural models after policy changes have occurred
to predictions made from simpler
methods. For instance, \citeasnoun{angrist/pischke:10} lament:
\begin{quote}
``\textsf{Many new empirical industrial organization studies forecast counterfactual outcomes based on models and simulations, without a clear foundation in experience.  [...] At minimum, we'd expect such a judgment to be based on evidence showing that the simulation-based approach delivers reasonably accurate predictions.  As it stands, proponents of this work seem to favor it as a matter of principle.''}\footnote{There is a small literature in the context of merger analysis.
\citeasnoun{peters:06}
examines the predictive value of airline mergers and find that structural simulation methods yield poor predictions of post-merger ticket prices.
\citeasnoun{ashenfelter/hosken:08} argue that ``transparently'' identified design-based estimates
of the mergers differ markedly from those produced by the structural approach.
\citeasnoun{einav/levin:10}
acknowledge that there should be more retrospective analysis of past mergers, but question
whether information from particular mergers can be extrapolated for other ones.}
\end{quote}
The goal of this paper is to fill this void by evaluating the performance of predictions
from discrete-choice models of demand, which underlie much of the work in the
new empirical industrial organization,  using a large-scale policy change
taking place in Boston in 2014.

Each year, thousands of Boston's families submit rank order lists of public schools
in the city's school choice plan.  In 2013, Boston Public Schools (BPS), the Mayor, and members
of the school committee wanted to change this plan to encourage students to attend schools closer
to home.
BPS put forth a number of plans that redraw the boundaries of the city
and modify the set of schools applicants are allowed to rank by eliminating
some choices and adding other choices.  After these plans were described
to the public, there was widespread
interest in forecasting the choices families would make and in comparing the final assignments under these alternative
proposals.\footnote{For more details, see
the materials available at \textsf{http://bostonschoolchoice.org} and press accounts by \citeasnoun{goldstein:12} and \citeasnoun{handy:12}.}
\citeasnoun{pathak/shi:13} used historical choices to predict the choices that
participants would express under these alternative plans.  They reported
estimates of models of school demand based on historical
choices and used those estimates as a basis for extrapolating how
schools would be ranked under different choice menus.  The analysis
played a significant role in evaluating trade-offs among the
proposals and ultimately led to the selection of a new plan.
In 2014, families throughout Boston will rank schools
under new choice menus, with the first deadline to submit
preferences on January 31.

The methods employed to make counterfactual forecasts in Boston are based on discrete choice models of demand.
These methods are commonly used to forecast from structural models and were
initially designed with this aim in mind \cite{mcfadden:01}.  McFadden and co-authors studied the impact of BART,
a new fixed-rail rapid transit system in San Francisco Area.  They collected data on the travel 	
behavior of a sample of individuals in 1972, prior to the introduction of BART, and estimated
models that were used to predict the behavior of the same individuals in 1975 after BART began.
Multinomial logit models were estimated using pre-BART commuter data at the realized attributes
of the alternatives (auto alone, carpool, bus).  \citeasnoun{mcfadden:01} reports that
\begin{quote}
``\textsf{our overall forecasts for BART were
quite accurate, particularly in comparison to the official 1973 forecast, obtained from aggregate gravity
models, that BART would carry 15 percent of commute trips. We were lucky to be so accurate,
given the standard errors of our forecasts, but even discounting luck, our study provided strong evidence
that disaggregate RUM-based models could outperform conventional methods.}''
\end{quote}
Based on the BART experience, random utility models became widely adopted
in travel analysis and many other areas of economics involving consumer choice \cite{mcfadden:01}.

More than three decades have progressed since the BART analysis and there has
been considerable progress in demand modeling during this time.  For instance,
\citeasnoun{mcfadden:01} states that the methods used to account for substitution
between modes of transportation were inferior to the modeling methods used today.
The aim of this paper is to compare discrete-choice based forecasts of
changes in school demand in Boston's school
choice plan to alternative forecasts that come from a statistical model not founded on utility maximization.
We will study how well ex ante counterfactual predictions match
actual submitted choices of participants in the first year of the new system,
and also examine other outcomes related to the assignments students obtain.
This paper lays out our methodology and forecasts before new choices
are submitted.  A companion follow-up paper will compare these forecasts to newly
reported choices when they become available.

There are many reasons our exercise has the potential to provide unusually
compelling evidence on the forecasting performance of structural demand models.  First,  the
forecasts are based on flexible models of demand exploiting the revealed preferences
that families expressed historically.  The data not only includes student's
top choices, but their entire submitted rank order list of schools. Rank order list data can
potentially reveal richer information about substitution patterns between choices (see, e.g., \citeasnoun{berry/levinsohn/pakes:04}).
Moreover, our dataset includes a large number of observables, including information on student characteristics and exact
geographic location.  Furthermore, Boston's choice plan has been in existence
for more than two decades, meaning that there is a wealth of knowledge and shared experience about the system.
The current strategy-proof system, in place since 2005, eliminates the need for participants to be strategic about their
choices and the advice BPS provides participants in their school guide reflects this feature.\footnote{For
instance, the 2012 School Guide states: ``List your school choices in your true order of preference.  If you list a popular school
first, you won't hurt your chances of getting your second choice school if you don't get your first choice.}
Our exercise should be particularly informative on substitution patterns since the policy change
predominantly involves a change in choice menus and relatively small changes in the characteristics of choices.
As preference data become more widely available, there are also a growing number of papers that estimate models of school demand, and our results
will speak to their reliability as a policy planning tool for school districts (see, e.g., \citeasnoun{abdulkadiroglu/agarwal/pathak:13}, \citeasnoun{hastings/kane/staiger:05},
\citeasnoun{walters:12}).
Finally, since discrete choice models of demand are a building-block for many structural models that examine more complicated situations
involving dynamics and strategic interactions, the relatively simple counterfactual environment
should allow us to easily compare our predictions and understand reasons for their performance
abstracting away from these additional complications.

On the other hand, the premise of our exercise, and other forecasts based on discrete choice
models of demand is that preferences are stable, can be estimated adequately, and can be used to make predictions
in different environments.  In a field experiment, \citeasnoun{hastings/weinstein:08}
provide evidence that choice behavior in Charlotte's school choice plan can easily be swayed
by informational cues.  In other contexts, there is similar
evidence that information interventions that give people the same information that is already available in a simpler format
change choice behavior, and are evidence of ``comparison frictions'' \cite{kling:12}.    If these
features of choice dominate decision-making, then they may interfere with the reliability
of any forecasts based on historical revealed preferences.  Indeed, if the details of how counterfactuals are presented matter more than what we can learn from past data using demand models, it may motivate a reevaluation of the use of demand models for analyzing counterfactuals in our context.
 Our work also has the potential to
provide evidence on how particular post-BART developments in discrete choice modeling are
important for accurate counterfactual prediction.

There are only a small number of papers that make ex ante forecasts and compare them ex post to
what actually occurs.  The closest project is the study of BART and its ex post validation described above \cite{mcfadden/talvitie:77}.
\citeasnoun{carrell/sacerdote/west:14} take reduced form estimates of
peer effects and use these estimates to design an experiment grouping peers together to increase
student achievement.  Their results show that the reduced form estimates do not provide
an adequate guide to predict the effects of their peer grouping experiment.
Other work uses social experiments as a validation tool for structural modeling.
\citeasnoun{wise:85} estimates a model of housing demand and compares
the predicted impacts of a subsidy program to those from a randomized experiment.
\citeasnoun{todd/wolpin:06} estimate a model on control households from a randomized
social experiment without using post-program data and compare the model
predictions about program impacts to the experimental impact estimates.
They report that their model's predicted program impacts track the experimental results.

This project proceeds in two steps.   First, we report on demand models used to guide the policy change in Boston in 2013 in the redesign of their school choice plan, updating the estimates in \citeasnoun{pathak/shi:13}.
We build on those methods and report on ex-ante forecasts on the performance of demand models.  This report and our analysis plan is
being reported in January 2014, before the completion of preference submission in Boston.
Part II of this project will use data submitted on schools to evaluate the forecasts and measure the strengths and limitations of discrete choice
analysis in our context.

\section{School Choice in Boston}

Boston Public Schools is home to one of the nation's most iconic school choice plans, which initially evolved out of a court-ordered busing plan in the 1970s.
Until 2014, the city was divided into the North, West, and East Zone for elementary school admissions.  There
are about 25 elementary schools in each zone.
Students residing in a zone are allowed to rank any school choice in the zone as well as any school
within a 1 mile walk zone and a handful of city-wide schools on their application
form.  At each school, students are prioritized as follows:  continuing students have
the highest priority, followed by students who had a sibling at the school, followed
by other students.  Within each group, for half of the program seats, students residing in
the walk-zone obtain priority (but this priority does not apply to the other half of the school seats).
A single lottery number draw serves as a tie-breaker.\footnote{\citeasnoun{dkps:12} present
additional details on Boston's implementation of this algorithm.}

Since 2005, after students submit their choices, they are processed through the student-proposing deferred acceptance algorithm,
which works as follows:
\begin{itemize}
\item Round 1: Each student applies to his first choice school. School $s$
ranks applicants by their priority, rejecting the
lowest-ranking students in excess of its capacity, with the rest provisionally
admitted (students not rejected at this step may be rejected in later steps).

\item Round $\ell>1$: Students rejected in Round $\ell$-1 apply to their next
most preferred school (if any). School $s$ considers these students
\textit{and} provisionally admitted students from the previous round, ranks
them by their priority, rejecting the lowest-ranking
students in excess of capacity, producing a new provisional admit list (again,
students not rejected at this step may be rejected in later steps).
\end{itemize}
The algorithm terminates when either every student is matched to a school or
every unmatched student has been rejected by every school he has ranked.

There have been many attempts to reform Boston's school choice plan including
community-wide task forces in 2003 and 2009.  Aside from changing the assignment algorithm
in 2005, however, there have been no significant changes to the three zone plan since 1999
(see \citeasnoun{aprs:05} for more details).     In 2012, due to concerns about transportation costs and the
overall merits of busing children far from home, outgoing Mayor Menino spent the last
year of his administration advocating for a ``radically different
school assignment process---one that puts priority on children attending schools closer to their
homes'' \cite{menino:12}.

In Fall 2012, BPS proposed five different plans
that all restricted participant choice by reducing the number of schools students could rank.\footnote{The initial plans suggested dividing the city into 6, 9, 11, or 23 zones, or assignment based purely on neighborhood.}  The idea behind each of these plans was to
reduce competition from non-neighborhood applicants at each school.
When these plans were publicly unveiled in September 2012, they were met with widespread criticism
(see, e.g., \citeasnoun{seelye:12}).   These plans and other proposals from the community became the center of a year-long, city-wide discussion on school choice.

The plan that was eventually chosen was devised by \citeasnoun{shi:13} and became
known as the Home Based plan.  Initially, BPS categorizes each school into Tiers,
which are computed using a combination of standardized test score growth and levels
on the Massachusetts Comprehensive Assessment System (MCAS) tests for the past two years.
For the 2014 admissions cycle, Tiers were finalized as of January 2013.
Under the new plan, every family is allowed to choose from any
school within a mile (as the crow flies), along with the two closest Tier 1,  the
four closest Tier 1 or 2, the six closest Tier 1, 2 or 3 schools, and the three
closest ``option schools'' chosen by BPS using internal simulations.  The menu of choices
also includes the closest early learning center (ELC) and closest school with an advanced
work class (AWC) program.\footnote{There are two idiosyncratic exceptions to the choice menu
composition due to Boston's unique geography.  First, students residing in parts
of Roxbury, Mission Hill and Dorchester are allowed  to rank the Jackson Mann school.   Second, students in East Boston are eligible
to apply to any school in East Boston.  East Boston students have priority over
non-East Boston students at East Boston schools.  Non-East Boston students have priority
over East Boston students for Non-East Boston Schools.  Finally, there are
certain provisions in the plan for students who are limited English proficient or special needs.
 Limited English proficiency students
of ELD level 1, 2, 3 are allowed to apply to any compatible ELL program within
their ELL zone, which is a specially constructed six-zone overlay
of Boston. Substantially separated special education students do not apply in Round 1,
the focus of our investigation.}  Families can easily access their choice menu via an online
portal, which shows a map of all schools in the choice menu and a summary of their attributes.
As before, choices are processed through the student-proposing deferred
acceptance algorithm, though the new plan also eliminates walk zone priority.

Our analysis focuses on Round 1, where the vast majority of students obtain their
initial placement.  Students who do not obtain an assignment after the algorithm is run
are allowed to participate in an administrative assignment round where BPS places
these students to remaining schools based on proximity. For the purposes of our study,
we do not model the placement of administratively assigned students and only consider the assignments
produced by the assignment mechanism.

\section{Data}

The main data sources for this project are BPS Round 1 choice data and enrollment data for 2010-2013. For each year, the Round 1 choice data was collected in January or February of that calendar year, for application to the school year that began in September of the same calendar year. The enrollment data is a snapshot taken in December of the same year, 11 months after the choice data and three months after the school year began.

The choice data contains for each student who participated in round 1 his/her student identification number; grade; English proficiency status and first language; special education or disability status; geocode (a geographic partitioning of the city used by BPS); school program to which the student has guaranteed priority (designation for continuing students); student identification numbers of the student's siblings currently enrolled in BPS; lottery number; first 10 choices and priorities to each; school program to which the student was assigned and the priority under which he/she was assigned.
 Using the assigned school and program codes, we infer the capacity available for Round 1 assignment for each school program. We assume that this reflects true Round 1 capacities.

 The enrollment data, which covers the vast majority of the students in the Round 1 choice data contains student identification numbers; enrolled school and program; grade; geocode; address; gender; race; languages spoken at home; dates of entrance and, if applicable, withdrawal from BPS;
 food service code (whether the student's socio-economic situation qualifies for free or reduced lunch). (Since BPS began offering free lunch to all students after September 2013, the food service code will not be available in the future.)

 In addition, we have access to a data set of school characteristics for each of the four years. The school dataset includes for each year and each school the school code, address, school type, \% of students of each race, \% of students who are English Language Learners (ELL), \% of students who have Special Education (SPED) requirements, and \% of students who scored Advanced or Proficient in grades 3, 4 and 5 for MCAS math and English in the previous year.

Since the assignment reform is mainly for elementary school assignment, we focus on the entry grades kindergarten 0 to 2 (K0, K1, K2).  K2 is the main entry grade to elementary schools in Boston. Table~\ref{tab:totalSupplyDemand} shows the total number of Round 1 applicants in each of the kindergarten grades, as well as the total
Round 1 capacity. As can be seen, only a small fraction of seats are available in grade K0, while more than half the seats are available in K1.

 \begin{table}[!htbp]
\centering
\caption{Aggregate supply and demand in grades K0-2, in years 2010-2013.}
\label{tab:totalSupplyDemand}
\begin{tabular}{l c c c| c c c}
 & \multicolumn{3}{c}{Applicants} & \multicolumn{3}{c}{Inferred Capacity} \\
Year & K0 & K1 & K2 & K0 & K1 & K2\\ \hline
2010 & 803 & 2134 & 3473 & 148 & 1676 & 3139\\
2011 & 704 & 2202 & 3556 & 170 & 1689 & 3328\\
2012 & 1001 & 2660 & 3985 & 181 & 1921 & 3689\\
2013 & 913 & 2599 & 4038 & 155 & 1890 & 3979\\
\end{tabular}
\end{table}

\begin{figure}[!htbp]
 \centering
 \caption{Number of new and continuing applicants to BPS (K0-2)}
 \label{fig:totalTrend}
 \includegraphics[width=0.8\textwidth]{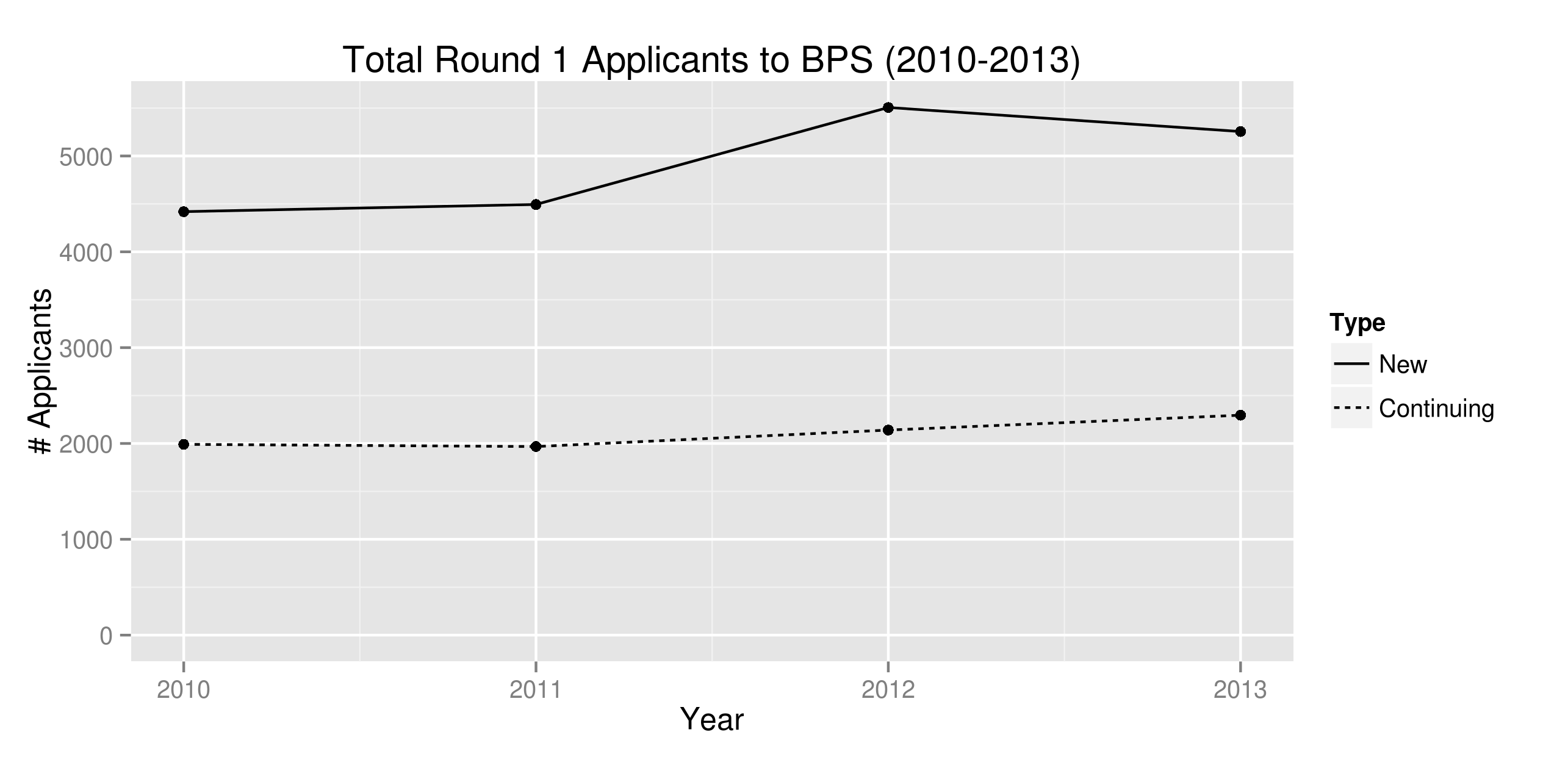}
\end{figure}

\noindent Students who are assigned to K0 or K1 in the previous grade enter the assignment system the next year as continuing students. These have priority to their current seat over new students. We define every non-continuing student as ``new.''  Figure~\ref{fig:totalTrend} plots the total number of new and continuing applicants to BPS for four years.

 To make use of geocode data, we take the latitude and longitude centroids of each geocode (provided by BPS) and compute a mapping from geocodes to ``neighborhoods,'' which gives a geographic partition of Boston into 14 regions.\footnote{Originally BPS has 16 neighborhoods, but we combined three small contiguous neighborhoods Central Boston, Back Bay, and Fenway/Kenmore into one neighborhood that we call ``Downtown.'' This is because these neighborhoods approximately make up the Boston downtown, and they each have very few number of applicants. Combining them still yields one of the smallest neighborhoods by number of applicants.} Table~\ref{tab:applicantDemo} shows the neighborhood breakdown of the Round 1 applicants and demographic profiles for each neighborhood, including an estimate of  household income using the median household income from the 2010 Census of the census block group where the centroid of the student's geocode lies\footnote{We used the 2012 ESRI demographics data set, which is available 
at 
\url{http://www.esri.com/data/esri_data/demographic-overview/demographic.}}, the percentage of applicants who are English Language Learners (ELL), and the racial composition of the applicant pool from this neighborhood. This table
 aggregates all four years of Round 1 choice data for all 3 kindergarten grades.

\begin{table}[!htbp]
 \centering
 \caption{Applicants' demographics across neighborhoods.}
 \label{tab:applicantDemo}
 \footnotesize
 \begin{tabular}{l c c c c c c c c}
Neighborhood & \% Total & Income Est. (K) & ELL & Black & Hispanic & White & Asian & Other\\ \hline
Allston-Brighton & 5\% & 53.9 & 51\% & 8\% & 36\% & 28\% & 22\% & 5\%\\
Charlestown & 2\% & 60.7 & 24\% & 12\% & 27\% & 49\% & 10\% & 2\%\\
Downtown & 3\% & 58.6 & 41\% & 10\% & 18\% & 36\% & 30\% & 5\%\\
East Boston & 14\% & 33.8 & 76\% & 2\% & 77\% & 16\% & 3\% & 2\%\\
Hyde Park & 6\% & 53.4 & 28\% & 40\% & 42\% & 13\% & 2\% & 3\%\\
Jamaica Plain & 6\% & 47.6 & 31\% & 13\% & 44\% & 30\% & 5\% & 8\%\\
Mattapan & 7\% & 36.0 & 30\% & 56\% & 40\% & 1\% & 1\% & 2\%\\
North Dorchester & 6\% & 40.5 & 49\% & 28\% & 40\% & 11\% & 17\% & 4\%\\
Roslindale & 8\% & 54.9 & 28\% & 17\% & 42\% & 34\% & 3\% & 5\%\\
Roxbury & 14\% & 29.6 & 36\% & 45\% & 49\% & 2\% & 1\% & 2\%\\
South Boston & 3\% & 37.8 & 36\% & 16\% & 38\% & 36\% & 6\% & 3\%\\
South Dorchester & 14\% & 44.2 & 35\% & 33\% & 35\% & 13\% & 16\% & 3\%\\
South End & 4\% & 47.0 & 38\% & 29\% & 40\% & 15\% & 11\% & 4\%\\
West Roxbury & 7\% & 65.6 & 21\% & 16\% & 27\% & 47\% & 6\% & 4\%\\ \hline
All Neighborhoods & 100\% & 44.3 & 40\% & 26\% & 44\% & 19\% & 8\% & 3\%\\
\end{tabular}
 \end{table}

Our analysis also uses distance estimates between each student's home and each school. To account for geographic barriers and road availabilities, we use walking distances provided by Google Maps API. For students for whom we cannot find a valid address by matching to the enrollment data\footnote{This occurs when the student is either not found in the enrollment data or the spelling of the address does not yield a valid result on Google maps, or the result ended up being more than 0.5 mile from the centroid of the applicant's geocode, indicating a data error.}, we used the centroid of the student's geocode as a proxy for home location.

\section{Methodology}
\label{sec:methodology}

The aim of our analysis is to explore alternative approaches of forecasting outcomes and to evaluate the accuracy of each approach.
We target outcomes that are important to BPS operations and decision-making.  We make these predictions for 2014 Round 1, where
the deadline to submit preferences is January 31, 2014. We hoped to commit to numerical forecasts before the outcome data is collected, but because of computational challenges, 
we were only able to calibrate two out of three of our demand models by January 2014. At that time, we published the partial forecasts in an earlier version 
of this report, in which we also committed to the specification for the third model, which we were not able to calibrate yet\cite{earlier-version}. In this current version, we include 
estimates and forecasts from all three models, and we do not depart from the specification for the third model previously committed to. 
All these efforts were to make sure that we are free from multiple-hypothesis testing or post-analysis bias. 

We forecast three assignment outcomes. The first is the number of unassigned students per neighborhood.  This outcome is important for BPS
because they have publicly committed to assign each K2 student to a set within his/her Home Based choice menu.
 Two other assignment outcomes are average access to quality, as defined by students' chances of getting into a Tier 1 or 2 school, and average distance to assigned school for each neighborhood. These were the two most important metrics by which the city committee during the 2012-2013 school assignment reform made their decisions, and the numbers they examined were based on forecasts arising from demand modeling. This analysis examines the accuracy of such an approach.

We also forecast market shares, as more direct measures of the choice patterns themselves, apart from interactions with the assignment system. We examine school market shares for each neighborhood, for top 1, top 2, and top 3 choices. This represents the demand and substitution patterns of families' choices across different neighborhoods. Because most of the available data is for K1 and  K2, and because these grades are more important for BPS strategic and operational policies, we focus on these two grades in the analysis.

Given the actual choice data and lottery numbers, and given a table of school program capacities, the above moments can be computed deterministically. Program capacities are control variables that BPS often varies over the assignment cycle.  Our simulation engine, which can be seen as a function mapping program capacities to outcome forecasts.  To commit to specific forecasts, it is necessary to specify program capacities. For simplicity, we use Round 1 inferred capacities from the previous year.

Forecasting the above moments involves forecasting the application pool in 2014, how applicants choose schools, and simulating the BPS assignment algorithm to yield final outcomes. We describe these steps in detail in the following subsections.

\subsection{Demand Models}
The focus of our study is alternative approaches to predict families' demand. While a full demand model should include families' decisions to apply to BPS, we do not have sufficient data about each family's outside options to precisely estimate such a model, so in this project we choose to focus on choices among BPS alternatives, and forecast the application pool using an ad-hoc approach described in section~\ref{sec:populationForecast}.

We consider three types of demand models, which are ordered in increasing complexity.
\begin{itemize}
 \item {\em Naive Model:} Assume that all families choose based on a simple rule that agrees with intuition and naturally arises from how the new assignment system has been presented.
 \item {\em Multinomial Logit Model:} This model is one of the simplest and most widely used approaches in demand modeling, especially for industrial organization applications. The 2012-2013 Boston school assignment reform heavily leaned on an analysis based on such a model, described in \citeasnoun{pathak/shi:13}.
 \item {\em Mixed-Logit Model:} This model is a popular alternative to the multinomial logit model due to its greater flexibility in capturing complex substitution patterns that violate the Independence of Irrelevant Alternatives (IIA) property of pure logit models.  One theoretical justification of such a model is that any Random Utility Maximization (RUM) consistent demand model can be approximated to arbitrary accuracy by a mixed-logit model \cite{mcfadden/train:00}.
\end{itemize}
\noindent We now describe the details and specifications of each model.
\subsubsection{Naive Model}

There are many possibilities for specifying statistical
models of demand, which are not based on the random utility framework.  We chose one particular model so that our investigation
of more sophisticated models can be compared to an alternative.   For instance, \citeasnoun{nevo/whinston:10} point out that when evaluating the performance of simulation based merger analysis, it is
important to compare these methods to other possibilities.   Even though it necessarily requires some ad hoc choices, our naive benchmark represents such an alternative.

From interactions with parents and BPS staff, we learned that many people expect families to simply choose schools with the best Tier schools first, breaking ties with distance. For ELL students, BPS staff stated that if given sufficient information, families would place a premium on ELL programs since they
offer targeted programming, especially language-specific ELL programs in their home language.
Other patterns are suggested by the choice data. For example, the vast majority of continuing students (91\%) select the next grade level of their current program first,
 an expected pattern since families may not like having to move schools. Furthermore, for students who have a sibling currently attending BPS, 66\% of them rank a school first that a sibling goes to, an expected pattern if families value having siblings attend the same school for transportation or other reasons.

If families' choices are not based on coherent and stable preferences, but are strongly influenced by BPS' publicity or framing efforts, the following simple model may adequately approximate families' choices. Each family ranks the school programs in their personalized menu based on the following hierarchy:
\begin{center}
\begin{tabular}{c l}
Hierarchy & Criteria \\ \hline
1 (most important) & present school program \\
2 & another program in current school \\
3 & school where sibling attends \\
4 & (for ELL students) ELL program \\
5 & (for ELL students) ELL program in home language \\
6 & better Tier school \\
7 & closer walking distance \\
\end{tabular}
\end{center}
\noindent Students only consider the hierarchy that pertain to them. For example, new applicants do not consider hierarchies 1 or 2, and non-ELL students do not consider hierarchies 4 and 5.

Outside of a random utility framework, a model of this type is a natural choice: for example, such a hierarchical model was used by the independent consulting group WXY, when commissioned by the BPS to analyze various counterfactuals. 

\subsubsection{Multinomial Logit}
\label{sec:logit}

This model assumes that rankings of each student $i$ are induced by underlying utilities for each school program $j$, and that the utilities can be approximated by the following model:

\[u_{ij} = \boldsymbol{\beta} \cdot \mathbf{F}_{ij} + \epsilon_{ij},\]
\noindent where $\mathbf{F}_{ij}$ is a vector of observable characteristics pertaining to student $i$ and choice option $j$, $\epsilon_{ij}$'s are iid random variables following a standard Gumbel distribution, and $\boldsymbol{\beta}$ is a vector of parameters to be fitted. As usual, we standardize the error term without loss
of generality since the model is invariant under multiplication or addition by constants.  By the same reason we normalize one of the components of $\boldsymbol{\beta}$ to zero.

A key implication of this model is that choices follow the Independence of Irrelevant Alternatives (IIA) property: the relative market shares of two programs does not depend on whether a third option is available. This means that substitution between programs follows the same proportional pattern across all choices. Although this property may be unrealistic for school choice, as two choices made by the same family may be correlated due to common, unobservable characteristics, it is plausible that the model may nevertheless provide a reasonable forecast of the moments that matter for decision making.

We fit this model by Maximum Likelihood Estimation (MLE) and obtain the covariance matrix of estimated coefficients by taking the inverse of the Hessian of the log likelihood function at the maximum. Table \ref{tab:logitCoefficents} shows the estimated coefficients for various specifications, using each of the 2012 and 2013 Round 1 choice data for grades K1 and K2. There are three specifications \textit{Simple} (which does not use students' demographics), \textit{Full} (which fully interacts students' race and income estimates with several key school characteristics), and \textit{Reduced} (which removes the insignificant terms in Full and combines terms for efficiency). The features used are the following:

\begin{itemize}
 \item distance: walk distance from home to school.
 \item continuing: indicator for whether the student has guaranteed status for the school program.
 \item sibling: indicator for whether student has sibling at the school.
 \item ell match: indicator for the student being English Language Learner (ELL) and the program being specialized for ELL.
 \item ell language match: indicator for the student being ELL and the program having a language-specific ELL program in the student's first language.
 \item walk zone: indicator for whether student lives in the school's walk-zone, which is approximately a 1-mile circle around the school.\footnote{The one mile circle is only approximate because the walk zones in our data are defined by drawing a one-mile circle from the school and including all geocodes that intersect the circle. A family from a geocode on the circle's boundary may be a little further than one-mile from the school, but still in the walk zone.}
 \item black/hispanic: indicator for whether the student is black or hispanic.
 \item mcas: the proportion of students at the school who scored Advanced or Proficient in the previous year's MCAS standardized test for math, averaging the proportions for grades 3, 4 and 5. (The MCAS test begins at grade 3. Grade 5 is the highest grade in many elementary schools. We only choose math because it is highly correlated with English.\footnote{This correlation is about .84 in years 2012 and 2013.})
 \item \% white/asian: the proportion of students at the school who are white or asian.
\end{itemize}

In each model, we include a fixed effect for each school, which captures any common propensities to choose a school, due to perceived school quality, facilities, and other unobserved characteristics. In the \textit{Full} specification, we interact the student's income estimate, along with indicators for race (black, asian, hispanic, other, or unknown), with the school's mcas, \% white/asian and distance. (This represents $6 \times 3 = 18$ terms).

\begin{table}[!htbp] \centering
  \caption{Estimated coefficients for logit models. Each model is estimated using maximum likelihood, using the choice data for grades K1 and K2. The standard errors are estimated using the inverse of the Hessian of the log likelihood function at the maximum.}
  \label{tab:logitCoefficents}
    \scriptsize

\begin{tabular}{@{\extracolsep{-5pt}}lD{.}{.}{-3} D{.}{.}{-3} D{.}{.}{-3} | D{.}{.}{-3} D{.}{.}{-3} D{.}{.}{-3} }

\\[-1.8ex]\hline
\hline \\[-1.8ex]
\\[-1.8ex] & \multicolumn{3}{c}{2012 Data} &  \multicolumn{3}{c}{2013 Data}  \\
\\[-1.8ex] & \multicolumn{1}{c}{\textit{Simple}} & \multicolumn{1}{c}{\textit{Full}} & \multicolumn{1}{c|}{\textit{Reduced}} & \multicolumn{1}{c}{\textit{Simple}} & \multicolumn{1}{c}{\textit{Full}}  &\multicolumn{1}{c}{\textit{Reduced}}\\
 & \multicolumn{1}{c}{(1)} & \multicolumn{1}{c}{(2)} & \multicolumn{1}{c|}{(3)} & \multicolumn{1}{c}{(4)} & \multicolumn{1}{c}{(5)} & \multicolumn{1}{c}{(6)}\\
\hline \\[-1.8ex]
distance & -0.395^{***} & -0.438^{***} & -0.365^{***} & -0.459^{***} & -0.499^{***} & -0.403^{***}\\
  & (0.005) & (0.018) & (0.014) & (0.006) & (0.019) & (0.015)\\
  &  &  &  &  &  & \\
continuing & 4.070^{***} & 4.029^{***} & 4.027^{***} & 4.401^{***} & 4.347^{***} & 4.354^{***}\\
  & (0.053) & (0.052) & (0.052) & (0.054) & (0.054) & (0.054)\\
  &  &  &  &  &  & \\
sibling & 2.143^{***} & 2.101^{***} & 2.104^{***} & 2.141^{***} & 2.107^{***} & 2.102^{***}\\
  & (0.037) & (0.037) & (0.037) & (0.038) & (0.038) & (0.038)\\
  &  &  &  &  &  & \\
ell match & 1.548^{***} & 1.550^{***} & 1.548^{***} & 1.210^{***} & 1.202^{***} & 1.211^{***}\\
  & (0.035) & (0.035) & (0.035) & (0.040) & (0.040) & (0.040)\\
  &  &  &  &  &  & \\
ell language match & 0.719^{***} & 0.610^{***} & 0.606^{***} & 0.791^{***} & 0.671^{***} & 0.672^{***}\\
  & (0.043) & (0.044) & (0.043) & (0.049) & (0.049) & (0.049)\\
  &  &  &  &  &  & \\
walk zone & 0.570^{***} & 0.497^{***} & 0.500^{***} & 0.474^{***} & 0.396^{***} & 0.399^{***}\\
  & (0.019) & (0.019) & (0.019) & (0.020) & (0.020) & (0.020)\\
  &  &  &  &  &  & \\
distance * black/hispanic &  &  & 0.115^{***} &  &  & 0.114^{***}\\
  &  &  & (0.010) &  &  & (0.011)\\
  &  &  &  &  &  & \\
distance * income est. &  & -0.233^{***} & -0.262^{***} &  & -0.252^{***} & -0.296^{***}\\
  &  & (0.022) & (0.021) &  & (0.024) & (0.023)\\
  &  &  &  &  &  & \\
mcas * black &  & -0.599^{***} & -0.874^{***} &  & -0.904^{***} & -1.062^{***}\\
  &  & (0.166) & (0.105) &  & (0.175) & (0.111)\\
  &  &  &  &  &  & \\
mcas * income est. &  & 0.506^{**} & 0.424^{*} &  & 0.959^{***} & 0.906^{***}\\
  &  & (0.232) & (0.221) &  & (0.267) & (0.252)\\
  &  &  &  &  &  & \\
\% white/asian * black/hispanic &  &  & -2.581^{***} &  &  & -2.666^{***}\\
  &  &  & (0.097) &  &  & (0.094)\\
  &  &  &  &  &  & \\
\% white/asian * income est. &  & 1.908^{***} & 1.982^{***} &  & 1.447^{***} & 1.778^{***}\\
  &  & (0.218) & (0.211) &  & (0.228) & (0.219)\\
  &  &  &  &  &  & \\
school fixed effects & \multicolumn{1}{c}{Yes} & \multicolumn{1}{c}{Yes} & \multicolumn{1}{c|}{Yes} & \multicolumn{1}{c}{Yes} & \multicolumn{1}{c}{Yes} & \multicolumn{1}{c}{Yes} \\
  & & & & & & \\
full interaction &  & \multicolumn{1}{c}{Yes} &  &  & \multicolumn{1}{c}{Yes} &  \\
  & & & & & & \\
\hline \\[-1.8ex]
Log likelihood& \multicolumn{1}{c}{-70,969}& \multicolumn{1}{c}{-70,013}& \multicolumn{1}{c|}{-70,090}& \multicolumn{1}{c}{-65,944}& \multicolumn{1}{c}{-64,763}& \multicolumn{1}{c}{-64,829}\\
\# of Parameters& \multicolumn{1}{c}{81}& \multicolumn{1}{c}{99}& \multicolumn{1}{c|}{87}& \multicolumn{1}{c}{83}& \multicolumn{1}{c}{101}& \multicolumn{1}{c}{89}\\
\# Students & \multicolumn{1}{c}{6,644} & \multicolumn{1}{c}{6,644} & \multicolumn{1}{c|}{6,644} & \multicolumn{1}{c}{6,627} & \multicolumn{1}{c}{6,627} & \multicolumn{1}{c}{6,627} \\
\# Choices & \multicolumn{1}{c}{27,905} & \multicolumn{1}{c}{27,905} & \multicolumn{1}{c|}{27,905} & \multicolumn{1}{c}{26,991} & \multicolumn{1}{c}{26,991} & \multicolumn{1}{c}{26991} \\
\hline \\[-1.8ex]
\textit{Note:}  & \multicolumn{6}{l}{$^{*}$p$<$0.1; $^{**}$p$<$0.05; $^{***}$p$<$0.01} \\

\end{tabular}
\end{table}

All of the specifications in Table \ref{tab:logitCoefficents} yield a significant and negative coefficient on distance. The magnitudes of the coefficients can be interpreted as follows: for the \textit{Simple} specification fitted using 2012 data, the distance coefficient is $-0.395$, which means that for two school programs that are otherwise identical but one is one mile further from home, the student is more likely to choose the closer one $\frac{e^{0.395}}{1+e^{0.395}} \approx 60\%$ of times.
All estimates yield highly significant and positive coefficients for continuing, sibling, ell match, and ell language match. To gain further intuition, one can examine the ratios of these estimates with the distance coefficient. For example, all else being equal, students in the \textit{Simple} specification in 2012 are on average willing to travel $\frac{4.070}{0.395} \approx 10.3$ extra miles to go to a continuing program, $5.4$ extra miles to go to school with a sibling, $3.9$ extra miles to go to an ELL program (if the student is ELL), and $1.8$ extra miles to go to an ELL program specialized to his/her home language.

Being in the walk zone is relevant because only students who live outside the walk zone are provided busing. Moreover, it is correlated with extreme proximity. Because of these potentially conflicting influences, the positive coefficient for walk zone is difficult to interpret. Another complication is that before 2014, students in the walk zone get walk zone priority to go to the school. Although this should not theoretically affect choice rankings because the mechanism is strategyproof, families may not fully appreciate this property and rank walk-zone schools higher because they think they have better chances to get into them.   Alternative, the significance of this variable may indicate
that student's perceive a significant fixed costs to having to attend a school that requires a bus ride.

In creating the \textit{Reduced} specification, we first fit the \textit{Full} model. However, we found that the estimates for black and hispanic students are statistically indistinguishable, except for interaction with mcas, in which the coefficient for black students are significantly negative while the coefficient for hispanic students is insignificant. Moreover, the results for the other races are unstable between the years or insignificant, most likely due to lack of data because as seen in Table~\ref{tab:applicantDemo} few students are asian (8\%) or other (3\%). Thus, in the \textit{Reduced} specification, we group black and hispanics together, except for the interaction with mcas, and we remove the other race dummies, implicitly grouping them with whites. We included all the terms involving income estimates since they tend to be statistically significant. The coefficients suggest that black/hispanic students are willing to travel further than other students, and tend to choose schools that have lower
\% white/asian.
This may be due to demographic preferences, or preferences for unobserved characteristics that are correlated with demographics, such as school culture or environment. Black students seem to disproportionately choose schools with lower math scores even with our controls for distance and neighborhood income.

Because the \textit{Reduced} specification captures all of the significant and stable interaction terms in the \textit{Full} specification, but has more precise estimates, so we opt for this specification and simply refer to it by ``Logit'' in the rest of this paper.

\subsubsection{Mixed Logit (MLogit)}
\label{sec:mixed-logit}

This model adds random coefficients to multinomial logit. Specifically, we use the following formulation,
\[u_{ij} = \boldsymbol{\beta}\cdot\mathbf{F}_{ij} + \boldsymbol{\gamma_i}\cdot \mathbf{G}_{ij} + \epsilon_{ij},\]
where $\mathbf{F}_{ij}$, $\epsilon_{ij}$ and $\boldsymbol{\beta}$ are observed features, iid taste shocks, and fixed coefficients as before. However, we allow the addition of a subset of features $\mathbf{G}_{ij}$ to interact with random coefficients $\boldsymbol{\gamma_i}$, which we assume to be zero mean jointly Gaussian distributed random variables, with possible covariance restrictions. The assumption of zero mean is without loss of generality since the means are captured by the fixed coefficients. The assumption of Gaussian distribution is for convenience.

In terms of features we include, we use the fixed coefficients as in the \textit{Reduced} specification of the logit model, but add random coefficients to the following features, which we organize into ``blocks,'' assuming independence across blocks, but allowing arbitrary covariance within each block.

\begin{center}
\begin{tabular}{c l}
 Block & Features \\ \hline
 A & ell match \\
 B & walk zone \\
 C & distance, mcas, \% white/asian \\
\end{tabular}
\end{center}
\noindent This formulation allows students to have heterogeneous preferences of going to an ELL program (if applicable), of choosing schools in the one-mile walk zone\footnote{The walk zones are only approximately one mile disc, because they were originally defined using geocodes.}, and of trading off distance, academics, and school demographics. We also include school fixed effects in order to capture the many unobserved school characteristics, such as safety, reputation, facilities, environment, and teaching quality. 

Because the model no longer has closed form log-likelihood functions, and the log-likelihood functions are no longer guaranteed to be globally concave, we fit the model by Markov Chain Monte Carlo (MCMC), which is a commonly employed method for fitting such models in practice \cite{train:03}. One technical difficulty in our situation is that we have many school fixed effects. As far as we are aware, the state of the art MCMC techniques for including fixed effects in mixed-logit models is described in~\citeasnoun{train:03}, and it involves adding a layer of Gibbs sampling and simulating the conditional distribution of the fixed effects using the Random Walk Metropolis-Hasting algorithm. However, in our case, there are 75 schools, so this step requires simulating a 75-dimensional distribution, which is
prohibitively slow using Random Walk Metropolis (RWM).\footnote{See \citeasnoun{katafygiotis/zuev:2008} for geometric insight to why RWM breaks down in high dimensions when the dimensions are correlated.} Hence, we speed up computation by using Hamiltonian Monte Carlo (HMC)\footnote{See \citeasnoun{neal:2011}}, which incorporates the gradient of the log likelihood function, so can more quickly update the 75-dimensional estimate for fixed effects. We fit the above model by using 1,000,000 iterations of MCMC sampling, throwing out the first half as burn-in. To check for the convergence of the estimates, we repeated this 6 times with independent draws, sometimes with random starting values, and found the results to be near identical. Details of how we fit the mixed logit model are in Appendix~\ref{app:mixedlogit-mcmc}.

The estimates are in Table~\ref{tab:mlogitCoefficents}. Note that beside the fixed coefficients in the \textit{Reduced} specification of the simple logit model, we also estimate the standard deviations of the random coefficients, denoted by $\sigma(\text{ell match})$, $\sigma(\text{walk zone})$, $\sigma(\text{distance})$, $\sigma(\text{mcas})$, and $\sigma(\text{\% white/asian})$. These are the square roots of the respective variances. We also estimate the correlation coefficients $\rho(\text{distance},\text{mcas})$, $\rho(\text{distance},\text{\% white/asian})$, and $\rho(\text{mcas},\text{\% white/asian})$, which are computed by dividing the respective covariance terms by the product of the standard deviations. In the rest of the paper, we refer to this model as ``MixedLogit'' or ``MLogit'' for short.

\begin{table}[!htbp] \centering
  \caption{Coefficients for Mixed Logit (MLogit) models, compared to simple Logit.}
  \label{tab:mlogitCoefficents}
    \scriptsize

\begin{tabular}{@{\extracolsep{-5pt}}lD{.}{.}{-3} D{.}{.}{-3} | D{.}{.}{-3} D{.}{.}{-3}  }

\\[-1.8ex]\hline
\hline \\[-1.8ex]
\\[-1.8ex] & \multicolumn{2}{c}{2012 Data} &  \multicolumn{2}{c}{2013 Data}  \\
\\[-1.8ex] & \multicolumn{1}{c}{\textit{Logit}}  & \multicolumn{1}{c|}{\textit{MLogit}} & \multicolumn{1}{c}{\textit{Logit}} & \multicolumn{1}{c}{\textit{MLogit}}  \\
 & \multicolumn{1}{c}{(3)} & \multicolumn{1}{c|}{(7)} & \multicolumn{1}{c}{(6)} & \multicolumn{1}{c}{(8)}\\
\hline \\[-1.8ex]
distance & -0.365^{***} & -0.638^{***} & -0.403^{***} & -0.674^{***}\\ 
  & (0.014) & (0.037) & (0.015) & (0.039)\\ 
  &  &  &  & \\ 
continuing & 4.027^{***} & 4.777^{***} & 4.354^{***} & 4.966^{***}\\ 
  & (0.052) & (0.069) & (0.054) & (0.068)\\ 
  &  &  &  & \\ 
sibling & 2.104^{***} & 2.478^{***} & 2.102^{***} & 2.451^{***}\\ 
  & (0.037) & (0.045) & (0.038) & (0.045)\\ 
  &  &  &  & \\ 
ell match & 1.548^{***} & 1.892^{***} & 1.211^{***} & 1.311^{***}\\ 
  & (0.035) & (0.058) & (0.040) & (0.059)\\ 
  &  &  &  & \\ 
ell language match & 0.606^{***} & 0.610^{***} & 0.672^{***} & 0.967^{***}\\ 
  & (0.043) & (0.052) & (0.049) & (0.060)\\ 
  &  &  &  & \\ 
walk zone & 0.500^{***} & 0.339^{***} & 0.399^{***} & 0.185^{***}\\ 
  & (0.019) & (0.028) & (0.020) & (0.028)\\ 
  &  &  &  & \\ 
distance * black/hispanic & 0.115^{***} & 0.188^{***} & 0.114^{***} & 0.183^{***}\\ 
  & (0.010) & (0.024) & (0.011) & (0.024)\\ 
  &  &  &  & \\ 
distance * income est. & -0.262^{***} & -0.295^{***} & -0.296^{***} & -0.343^{***}\\ 
  & (0.021) & (0.049) & (0.023) & (0.052)\\ 
  &  &  &  & \\ 
mcas * black & -0.874^{***} & -1.100^{***} & -1.062^{***} & -1.371^{***}\\ 
  & (0.105) & (0.153) & (0.111) & (0.144)\\ 
  &  &  &  & \\ 
mcas * income est. & 0.424^{*} & 1.065^{***} & 0.906^{***} & 0.925^{***}\\ 
  & (0.221) & (0.299) & (0.252) & (0.313)\\ 
  &  &  &  & \\ 
\% white/asian * black/hispanic & -2.581^{***} & -3.732^{***} & -2.666^{***} & -3.861^{***}\\ 
  & (0.097) & (0.162) & (0.094) & (0.148)\\ 
  &  &  &  & \\ 
\% white/asian * income est. & 1.982^{***} & 2.633^{***} & 1.778^{***} & 2.217^{***}\\ 
  & (0.211) & (0.322) & (0.219) & (0.311)\\ 
  &  &  &  & \\ 
$\sigma(\text{ell match})$ &  & 1.638^{***} &  & 1.358^{***}\\ 
  &  & (0.058) &  & (0.063)\\ 
  &  &  &  & \\ 
$\sigma(\text{walk zone})$ &  & 0.981^{***} &  & 0.878^{***}\\ 
  &  & (0.030) &  & (0.030)\\ 
  &  &  &  & \\ 
$\sigma(\text{distance})$ &  & 0.392^{***} &  & 0.409^{***}\\ 
  &  & (0.011) &  & (0.011)\\ 
  &  &  &  & \\ 
$\sigma(\text{mcas})$ &  & 2.275^{***} &  & 2.121^{***}\\ 
  &  & (0.086) &  & (0.101)\\ 
  &  &  &  & \\ 
$\sigma(\text{\% white/asian})$ &  & 2.672^{***} &  & 2.512^{***}\\ 
  &  & (0.093) &  & (0.106)\\ 
  &  &  &  & \\ 
$\rho(\text{distance} ,\text{mcas})$ &  & -0.232^{***} &  & -0.285^{***}\\ 
  &  & (0.041) &  & (0.043)\\ 
  &  &  &  & \\ 
$\rho(\text{distance} ,\text{\% white/asian})$ &  & -0.089^{**} &  & -0.055\\ 
  &  & (0.039) &  & (0.040)\\ 
  &  &  &  & \\ 
$\rho(\text{mcas} ,\text{\% white/asian})$ &  & 0.035 &  & -0.110^{*}\\ 
  &  & (0.056) &  & (0.061)\\ 
\hline \\[-1.8ex]
\textit{Note:}  & \multicolumn{4}{l}{$^{*}$p$<$0.1; $^{**}$p$<$0.05; $^{***}$p$<$0.01} \\
\end{tabular}
\end{table}

\subsection{Forecasting the Applicant Pool}
\label{sec:populationForecast}

An important driver of the number of unassigned students and the average access to Tier 1 or 2 schools is the number of applicants from each neighborhood. Regardless of how applicants choose schools, a large influx of new applicants from a neighborhood would drive up the number of unassigned from that neighborhood and drive down the average access to top Tier schools from that neighborhood. If we had data on all potential applicants and their non-BPS options, we might include this aspect as part of the
structural model. In the absence of such data, we still need to reflect this uncertainty and to capture any first-order trends in the neighborhood participation patterns.

\begin{figure}[!htbp]
 \centering
 \caption{Trend in total number of applicants to BPS.}
 \label{fig:totalNewTrends}
 \includegraphics[width=0.8\textwidth]{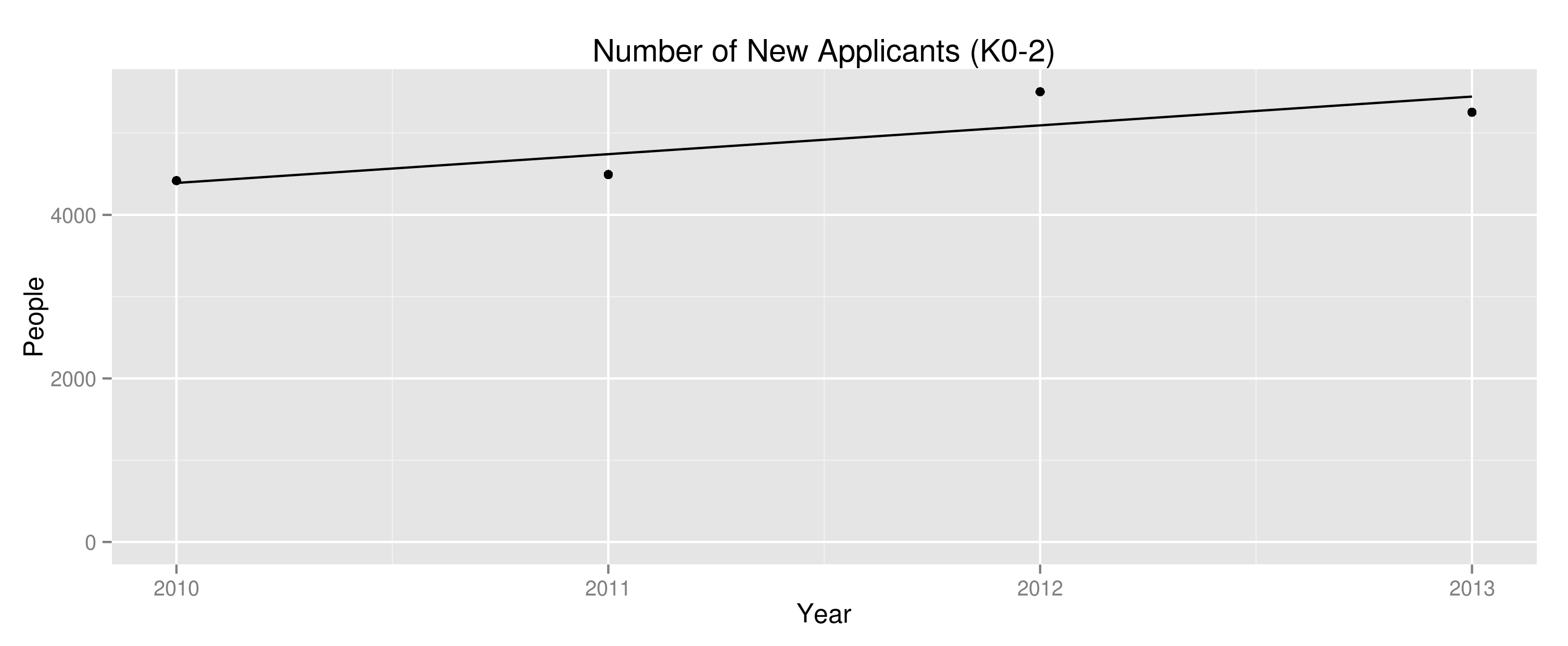}
\end{figure}

In forecasting the applicant pool, we consider new and continuing students separately. This is because continuing students are already in the enrollment data of the previous year, while for new students we need to use previous year's applicants' demographics as proxies. Figure~\ref{fig:totalNewTrends} plots the number of new applicants to BPS in Round 1 for grades K0-2 for years 2010-2013, as well as the regression line with respect to the year. As seen, the number of applicants is on average increasing each year, at a rate of 6\% a year on average, although it is not steady. For example, in 2012 there was an above-expected number of applicants.\footnote{The influx of applicants in 2012 Round 1 raised operational pressures for BPS, as it had to add about 10 new classrooms to accommodate.} We model the next year's total number of new applicants by a normal distributed random variable, having mean and standard deviation being the predicted mean and standard error of the regression line.

\begin{figure}
 \centering
 \caption{Proportion of new applicants distributed into each grade, neighborhood combination.}
 \label{fig:newRatio2014}
  \includegraphics[width=0.8\textwidth]{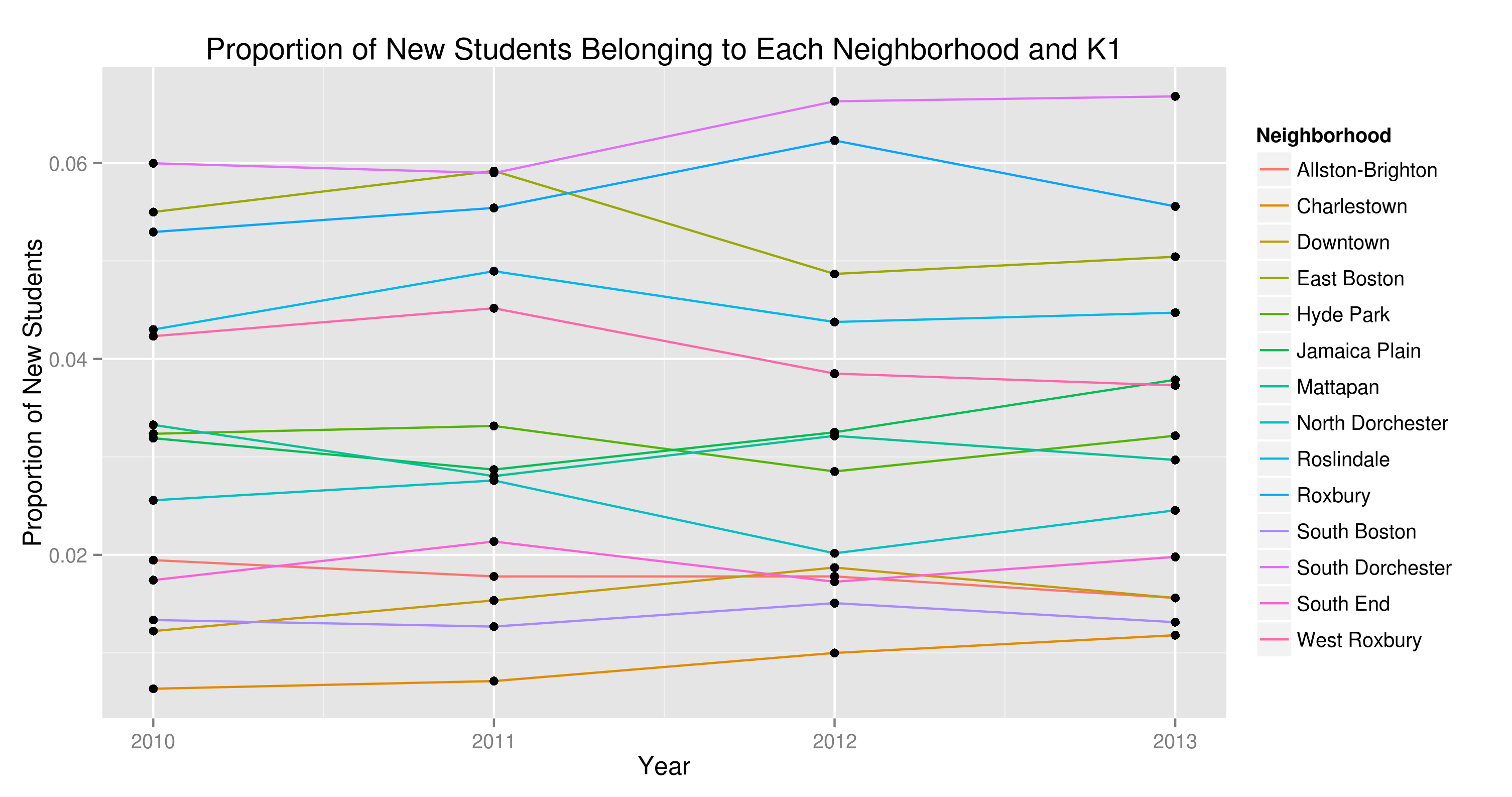} \\

 \includegraphics[width=0.8\textwidth]{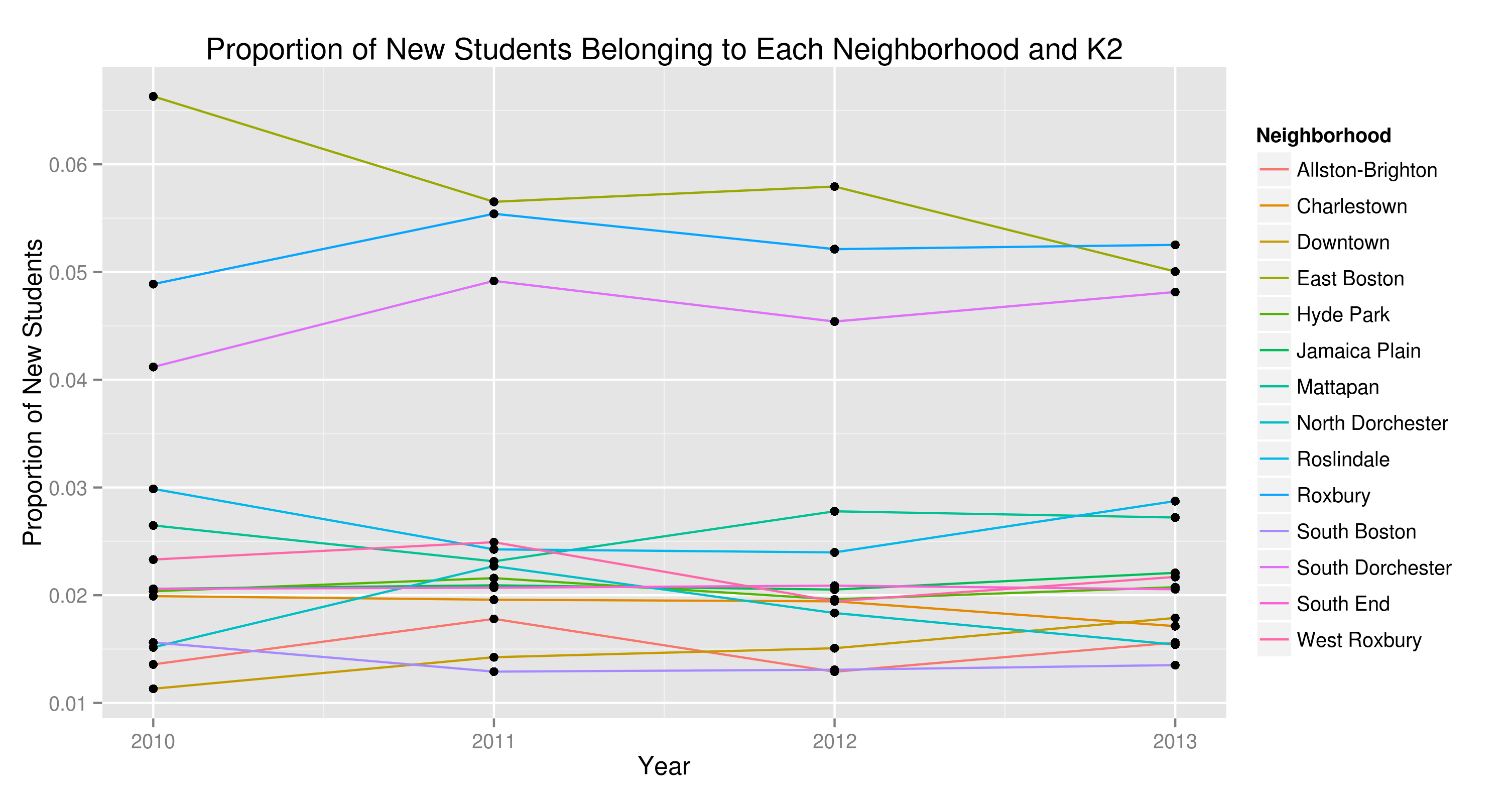}
\end{figure}

Figure~\ref{fig:newRatio2014} shows what proportion of this total is distributed into each grade and neighborhood combination. Since we study two grades and there are 14 neighborhoods, there are 28 time series in these plots. Most of the time series do not exhibit obvious trends. We model each of the 28 proportions next year as a normally distributed random variable. To estimate the mean and standard deviation, we run a regression with respect to year for each of these 28 time series, and discard all regressions for which the slope has less than 95\% significance level. For the neighborhood and grade combinations in which we discard the regression, we forecast next year's proportion using the previous 4 years' sample mean and sample standard deviation. For the neighborhood and grade combinations in which the regression slope has 95\% significance, we use the predicted mean and standard error of the regression.\footnote{Note that the standard error has one fewer degree of freedom than the sample
standard deviation.} The regression lines we kept are for K1 Charlestown and K2 Downtown, for which we detected a steady upward trend in the number of applicants.

\begin{figure}[!htbp]
 \centering
 \caption{The proportion of new applicants of each grade.}
 \label{fig:stableProportions}
 \includegraphics[width=0.8\textwidth]{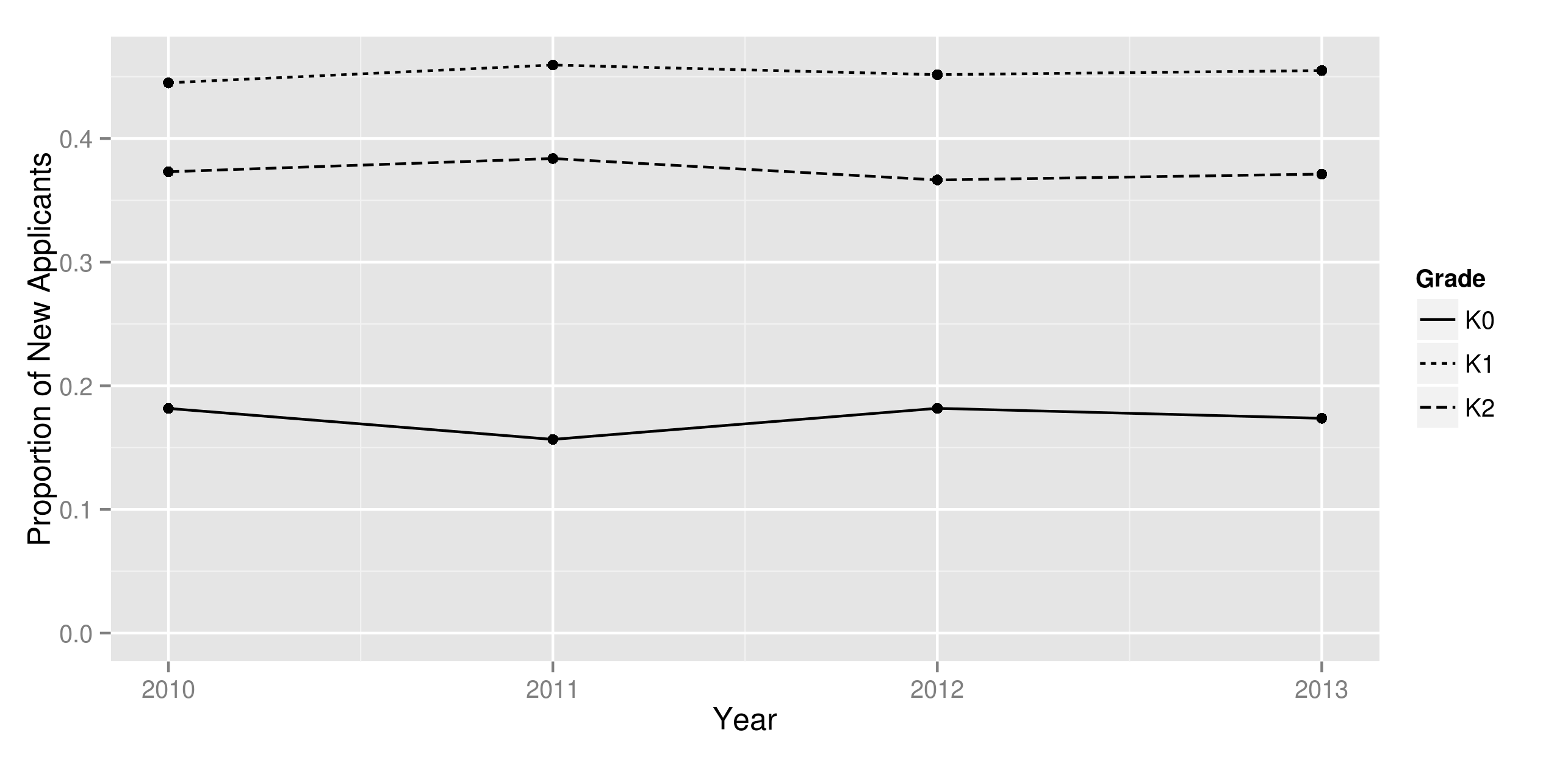}
\end{figure}

We therefore model the total number of new applicants from each neighborhood in the next year as a product of two independent normals, one representing a BPS wide shock and one a neighborhood shock. The common shock captures the uncertain effect that BPS publicity or policy initiatives have on the propensity for families to apply to BPS round 1. The neighborhood shock captures local population surges or unobserved reasons that affect participation. By using one common shock for all grades, we implicitly assume that different grades exhibit the same reactions to BPS policies, and are trending in the same directions.\footnote{The means and standard deviations of these estimates are tabulated in Appendix~\ref{app:populationForecast}. After multiplying the two normals, we truncate at zero if the product is negative and round to the nearest integer.} To check this, we plot in Figure~\ref{fig:stableProportions} the proportion of new applicants of each grade through the four years. As seen, the relatively 
horizontal lines suggest that modeling the aggregate participation of both K1 and K2 using the same random variable may be a reasonable approximation.

\begin{figure}[!htbp]
 \centering
 \caption{For each grade and neighborhood, the proportion of potentially continuing students from the previous year who apply as continuing students in the current year.}
 \label{fig:contRatio2014}
 \includegraphics[width=0.8\textwidth]{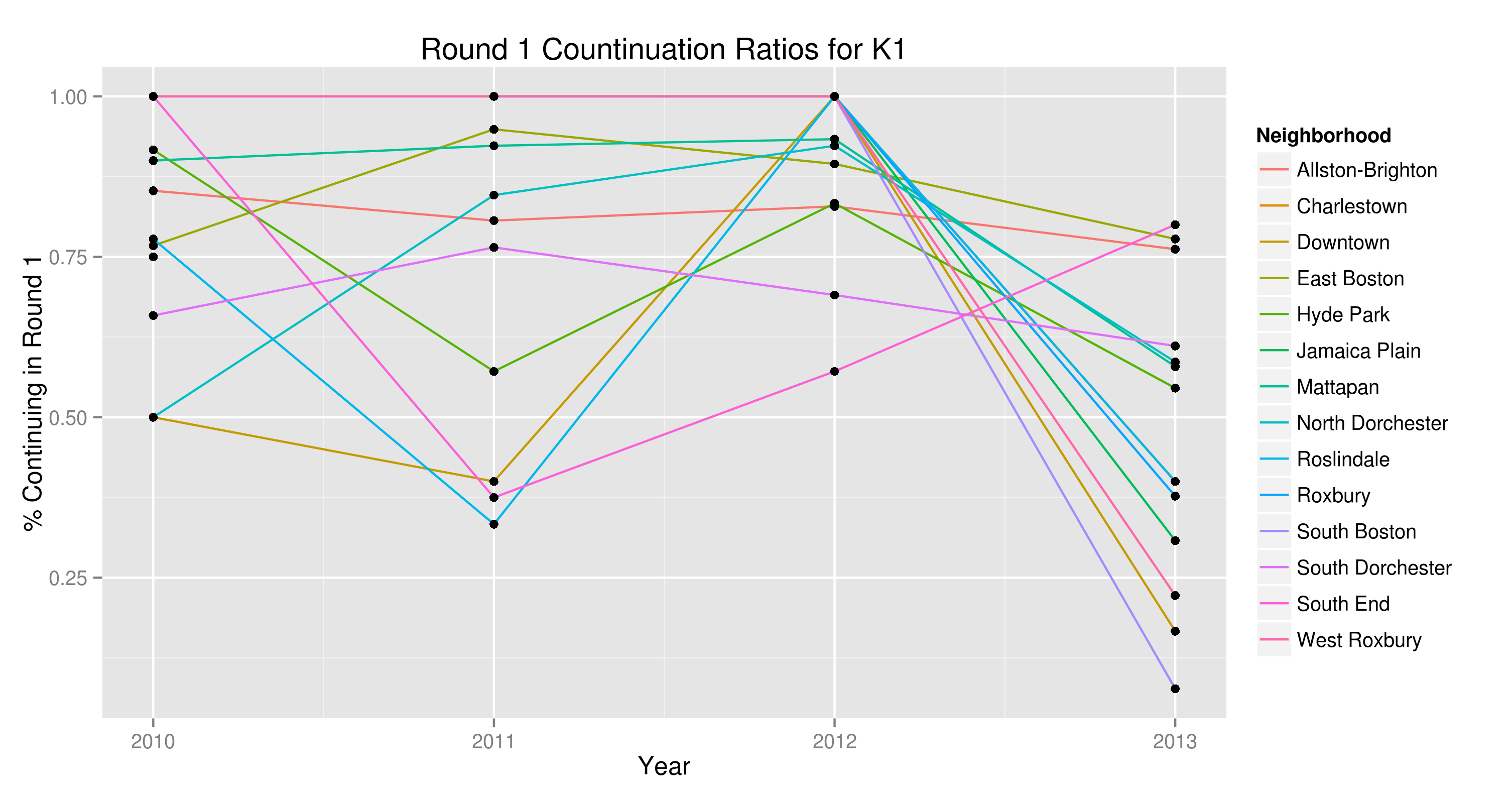} \\

 \includegraphics[width=0.8\textwidth]{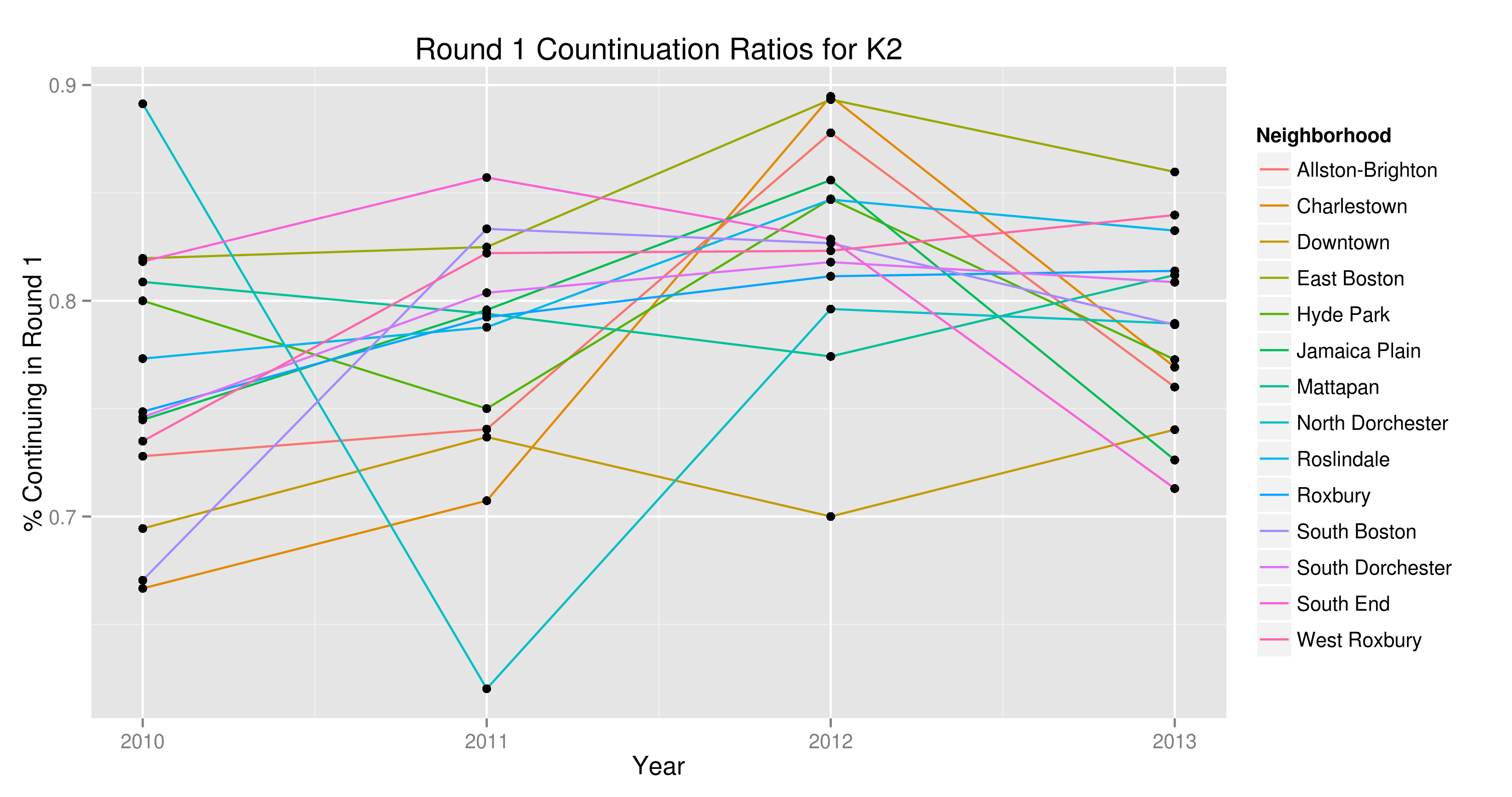}
\end{figure}

For continuing students, we define the Round 1 continuing ratio for a grade and neighborhood as the proportion of relevant students from the previous year's enrollment data who decide to continue in this year's Round 1. Figure~\ref{fig:contRatio2014} plots these for grades K1 and K2. As seen, due to the lower number of continuing students in K1 (recall that comparatively few students enroll in K0), the estimates for K1 are highly variable, while the K2 continuing ratios are around 70\% to 90\%.

We use the same approach to detect linear trends. However, in this data we failed to find any significant trends in continuing ratios, so we model them as normally distributed random variables, with mean and standard deviation according to the sample mean and sample standard deviation for years 2010-2013. The estimates are in Appendix~\ref{app:populationForecast}.

To simulate the pool applicant pool for next year, we independently draw the total number of new applicants, the proportion of this to allocate to each grade and neighborhood\footnote{Since the total includes K0 but we do not estimate proportions for K0, we do not require these draws to add up to 1.}, and the continuation ratio for each grade and neighborhood. Fixing these realizations for one sample of choice data, we draw new applicants by sampling with replacement the previous year's new applicants, but treating them as if they applied in the new year, and we sample the given number of new applicants from each neighborhood and grade. For continuing students, we go through all potentially continuing students and decide whether to include them independently, with probability according to the generated continuation ratio for that grade and neighborhood. We use this method to generate the applicant pool for all our simulations.

\subsection{Simulation}
\label{sec:simulation}
We use 4 layers of random draws for our simulations.
\begin{enumerate}
 \item {\em Population Draw}: Draw a pool of applicants according to the steps in Section~\ref{sec:populationForecast}. This represents uncertainty in participation rates.
 \item {\em Coefficient Draw}: For the logit and mixed-logit demand models, draw the coefficients as jointly normal random variables, using the estimated means and covariance matrix. This represents uncertainty in the demand model.
 \item {\em Preference Draw}: Having fixed a specific demand model and parameters, simulate for each student a complete ranking over his/her personalized menu of options, according to the randomness inherent in the demand model. Truncate this to the first ten choices.
 \item {\em Lottery Draw}: Generate iid lottery numbers for each student. The lottery numbers are distributed uniformly between zero and one, with lower lottery numbers being better.
\end{enumerate}

After doing these steps once, we have one set of simulated choice data, just as what we might have received from BPS. From this we can deterministically compute all of our outcome metrics by imitating the BPS assignment algorithm.\footnote{For validity of this project, we do not need to replicate the BPS internal system exactly.  In fact, we purposely deviate in our back testing simulations for 2013, by ignoring the walk-zone priority. This is because the year of interest, 2014, does not include such a priority, and our goal in back testing is to be able to quantify how well we expect various models to do for 2014. Having the walk-zone priority would skew the access to quality estimates for the Naive model, since under high competition for Tier 1 and 2 schools, the walk-zone priority would give students living near such schools a huge advantage. As a result, we will not be able to distinguish how much of the results for Naive is due to choice patterns or due to the obsolete priority rules.}

The reason we truncate to first 10 choices is as follows: currently our choice data from BPS truncates to first 10 choices, although students can rank arbitrarily many.  The previous report, \citeasnoun{pathak/shi:13}, provides evidence that assuming everyone ranks 10 choices yields reasonably accurate forecasts. Moreover, our earlier report on which the city committee based the 2012-2013 reform assumed that families ranked 10 options, and we keep the same assumptions to validate or invalidate the methodology of the earlier report.

An alternative approach is to model outside options and assume that unranked programs are inferior to the outside option. However, we observe in the data that often students end up enrolling to options they did not rank but could have ranked, suggesting that this assumption is invalid. In our interactions with parents and BPS staff, it
seems that many families are ranking few options not because they have better outside options, but because they feel confident they would get into the ones they picked and did not bother, or because they do not understand that ranking more options do not harm their chances to top choices. Future work is needed to better model this situation.

\subsection{Evaluation}
Having computed the assignment using the simulated choice data and lottery numbers, we compute the outcome measures as follows. In all of the analysis, we compute measures for grades K1 and K2 separately.
\begin{itemize}
 \item {\em Unassigned}: Tally the number of unassigned students from each neighborhood after each round.
 \item {\em Access to Quality}: For each student, define his/her access to quality as the highest (worst) lottery number he/she can have and still be able to be assigned to a Tier 1 or 2 school. (Recall that lottery numbers are uniformly distributed between zero and one, so this can be interpreted as a probability.) We estimate this by finding for each Tier 1 or 2 school program the highest (worst) lottery number he/she needs to obtain an offer. More precisely, if the program is not filled to capacity, the student can get in even with the worst lottery number; if it is filled to capacity, we look at the worst lottery number the student needs to be able to displace one student to obtain an offer.\footnote{For computing this metric, we ignore the possibility of that student displacing  someone else at his/her next choice and starting a chain reaction that cycles back to the first student, since this is unlikely to occur in markets with a large number of participants \cite{kojima/pathak:09}.} Then we take the 
maximum over all of his/her Tier 1 or 2 options and term this his/her ``access to quality.''
 Finally, we average the access to quality of all students of a neighborhood to compute the neighborhood average access.
 \item {\em Distance}: For each neighborhood, take all the assigned students from this neighborhood and average their walk distances to assigned school.
 \item {\em Top $k$ Market Share}: Take all the top $k$ choices of students from a neighborhood. For each school, find the proportion of these choices that are to this school.\footnote{If a student ranks multiple programs of the same school as part of his/her top $k$ choice, he/she contributes multiple ``votes'' for that school, since we treat each top $k$ choice as one ``vote.''}
\end{itemize}

We compute these metrics for each generated choice data and average across many simulations to find the mean prediction.
For a given choice data, we compute the above metrics and define the prediction error for each neighborhood as follows: for unassigned, access to quality, and distance, since we have a scalar value for each neighborhood, we simply take the absolute value of the difference between predicted mean and actual realization. For top $k$ market share, since for each neighborhood we have a vector of school market shares that add up to one, we define prediction error as the total variation distance between these probability vectors. Given vectors of market shares $\mathbf{p}$ and $\mathbf{q}$, define the total variation distance between them as
\[\text{total variation distance} = \frac{1}{2} \sum_{j} |p_j - q_j|.\]
This metric is standard for probability distributions, and can be interpreted as the least total movement needed to redistribute market shares by moving from one school to another, to turn the predicted shares into the actual shares.

Finally we judge the overall prediction accuracy by taking the Root Mean Squared Error (RMSE) over the neighborhoods. Specifically, for each metric, we take the prediction error for each neighborhood, square these numbers, take the mean of the squared errors, and take the square room of the mean. We choose this over Mean Absolute Deviation (MAD) because we want to penalize being far off for any specific neighborhood over being slightly off for many neighborhoods. This is because for capacity planning and for policy evaluation, the penalties are large for not foreseeing a large shortfall of seats in a neighborhood or not foreseeing vast inequities between neighborhoods.

\section{Results}
\subsection{Back Testing for 2013}
\label{sec:back-testing}
 We first test our methodology by using it to predict 2013 outcomes from 2010-2012 data, and quantifying the prediction error since we have the actual choice data in 2013. This gives us an idea of how well we might expect to predict future outcomes in 2014 Round 1. One caveat to keep in mind is that while the assignment system from 2010 through 2013 are the same, there are significant changes in 2014. Therefore, the results here may not reflect the results for 2014.

\begin{figure}[h!]
\centering
\caption{Back testing with assignment outcomes for 2013 K2. Tail distribution plots. \label{tab:prediction2013K2}}

\subfloat[][Unassigned (Naive)]{
 \includegraphics[width=0.33\textwidth]{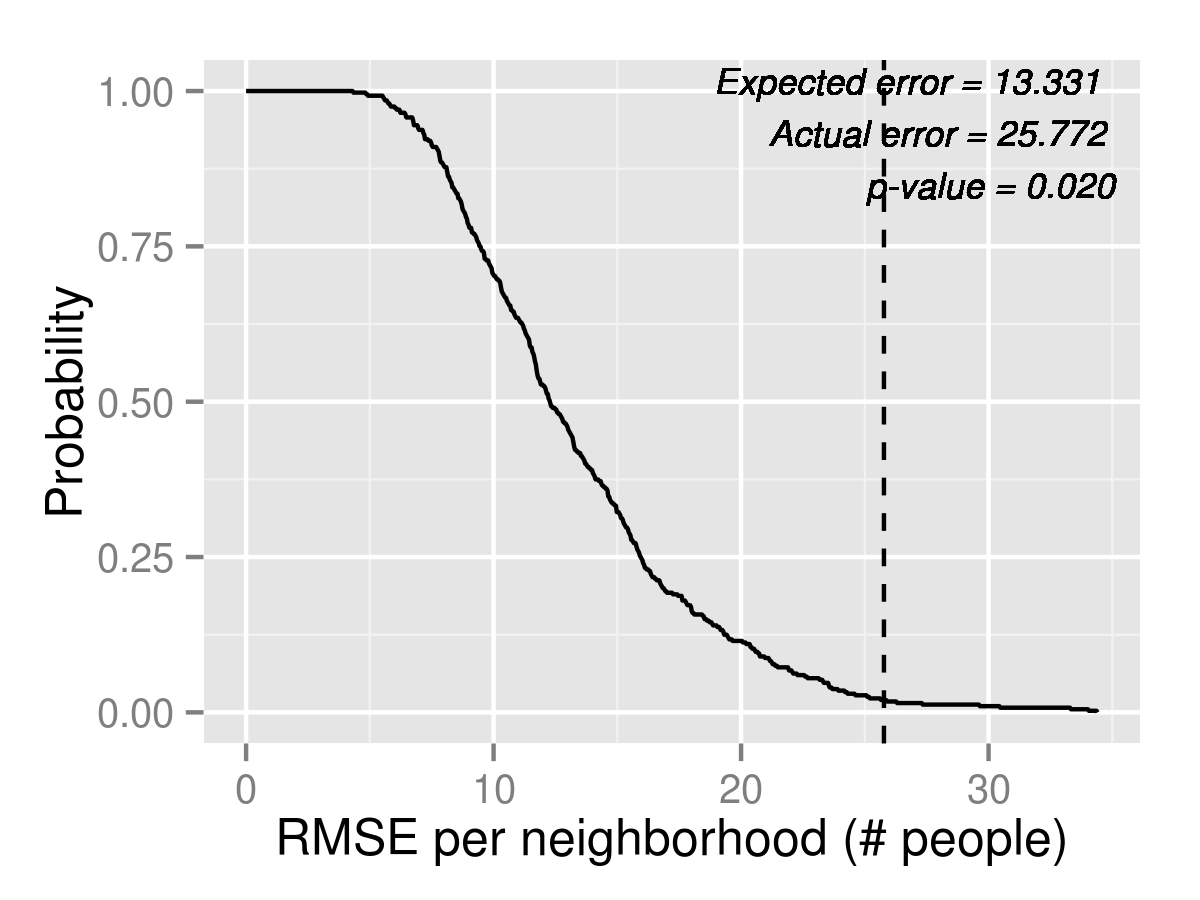}
} 
\subfloat[][Unassigned (Logit)]{
	\includegraphics[width=0.33\textwidth]{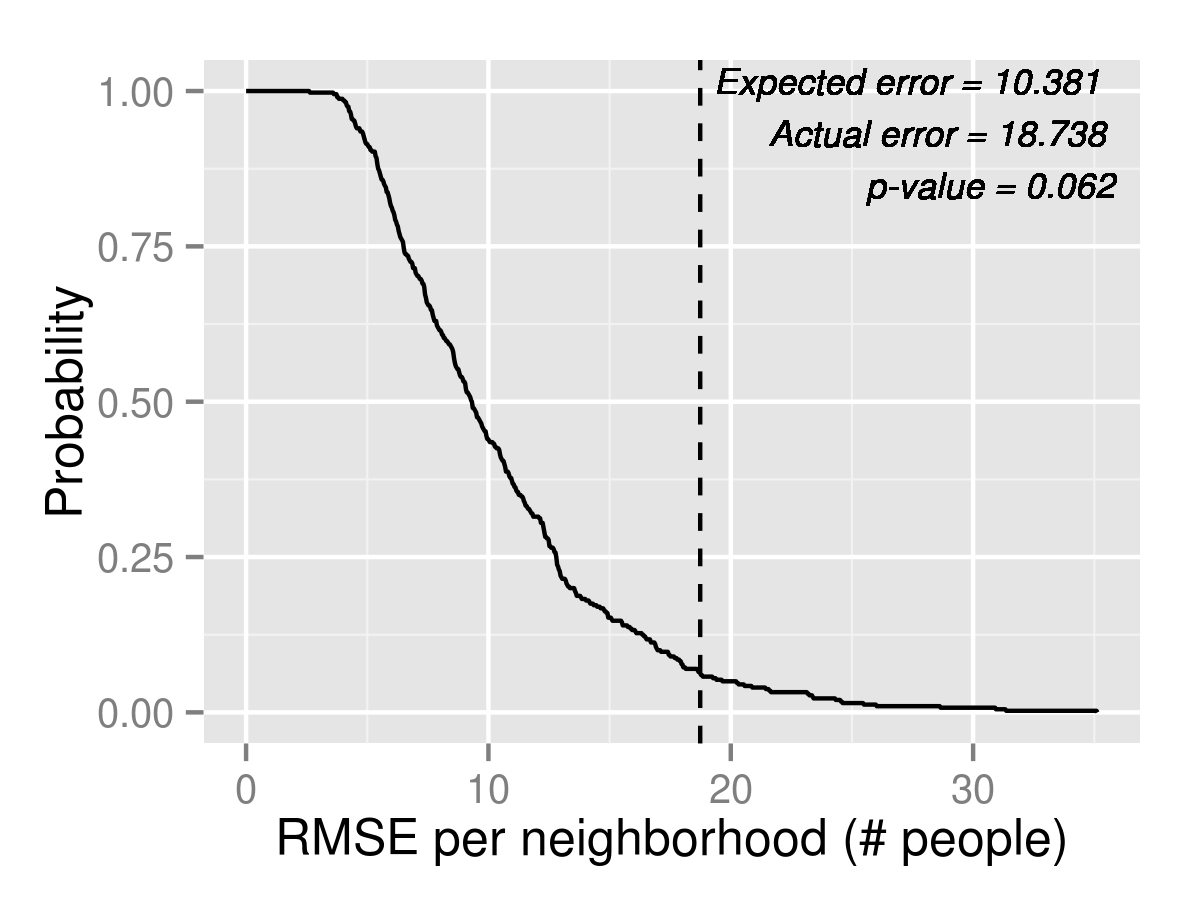}
}
\subfloat[][Unassigned (MLogit)]{
	\includegraphics[width=0.33\textwidth]{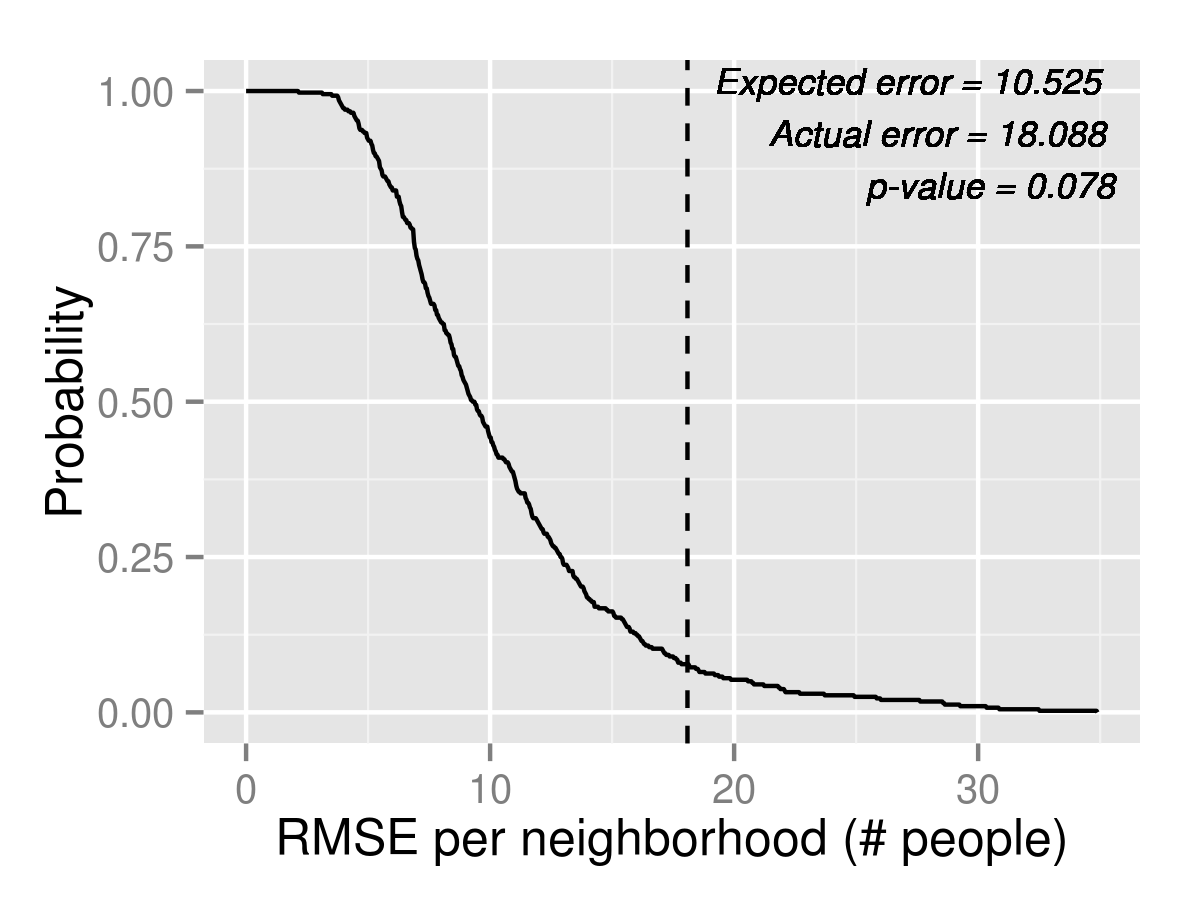}
}
\\
\subfloat[][Access to Quality (Naive)]{
 \includegraphics[width=0.33\textwidth]{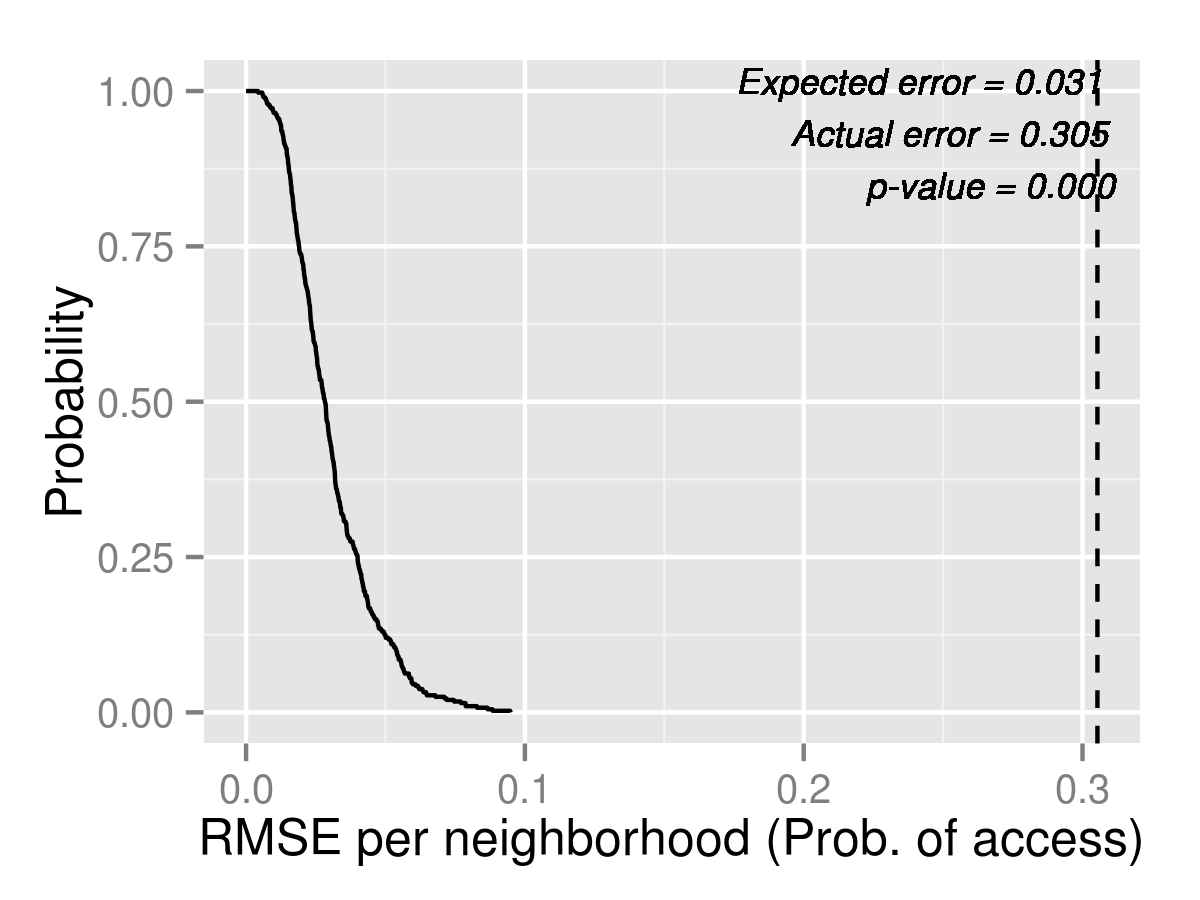}
} 
\subfloat[][Access to Quality (Logit)]{
	\includegraphics[width=0.33\textwidth]{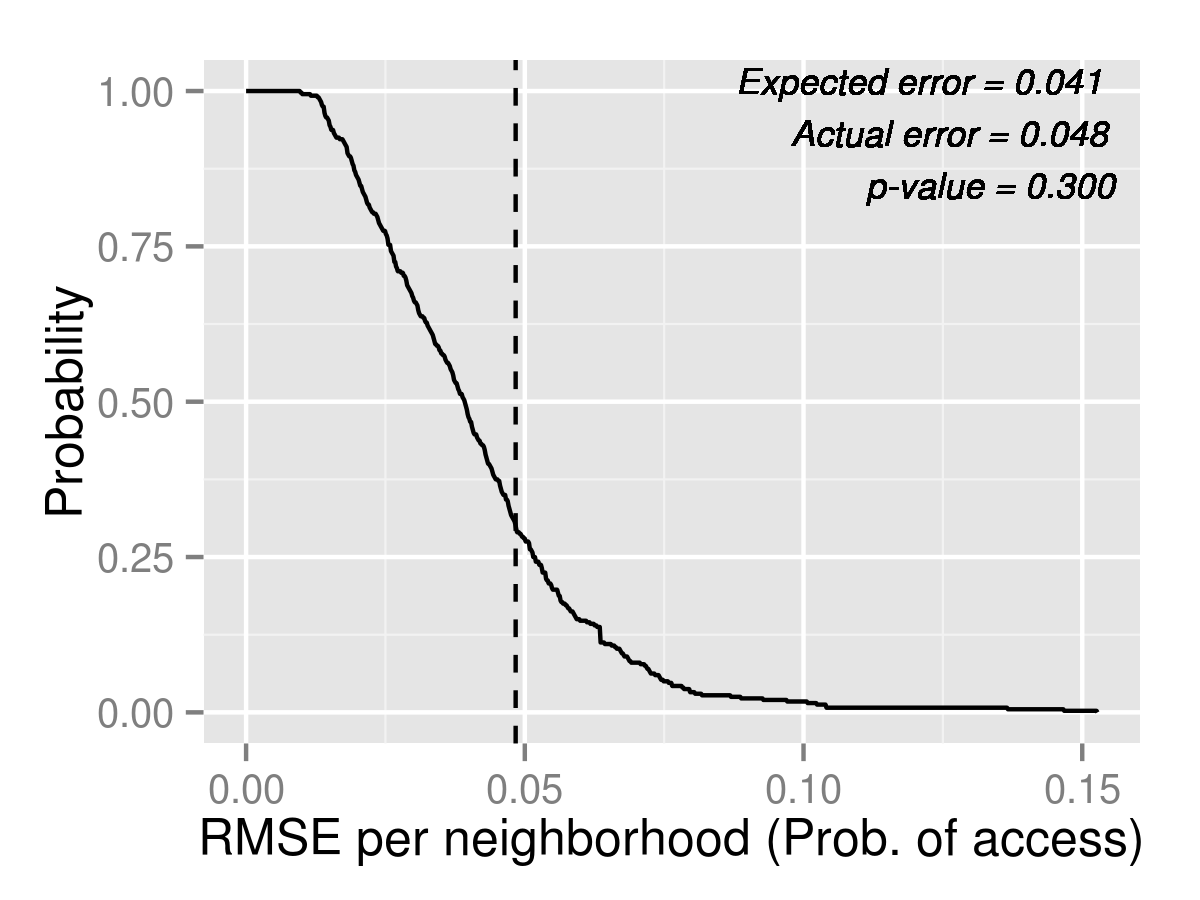}
}
\subfloat[][Access to Quality (MLogit)]{
	\includegraphics[width=0.33\textwidth]{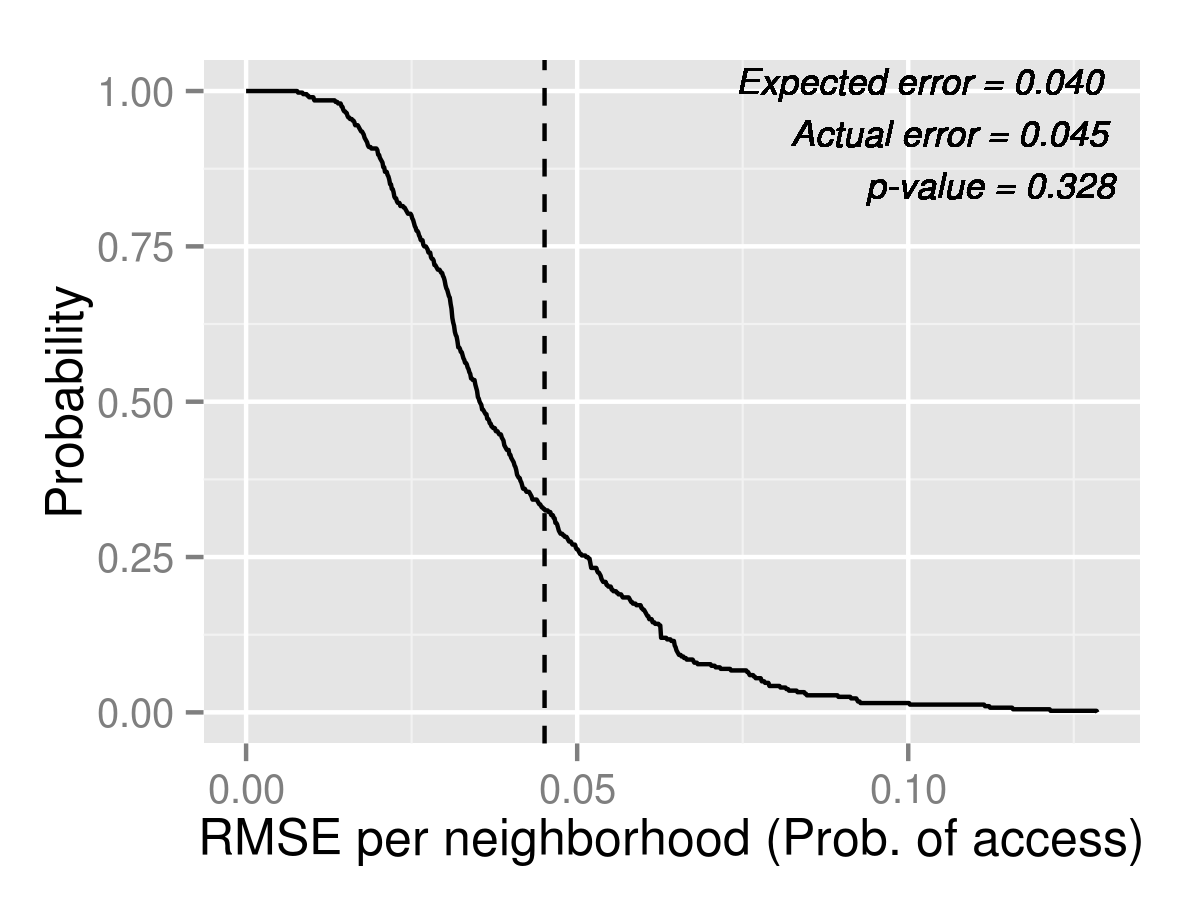}
}
\\

\subfloat[][Distance (Naive)]{
 \includegraphics[width=0.33\textwidth]{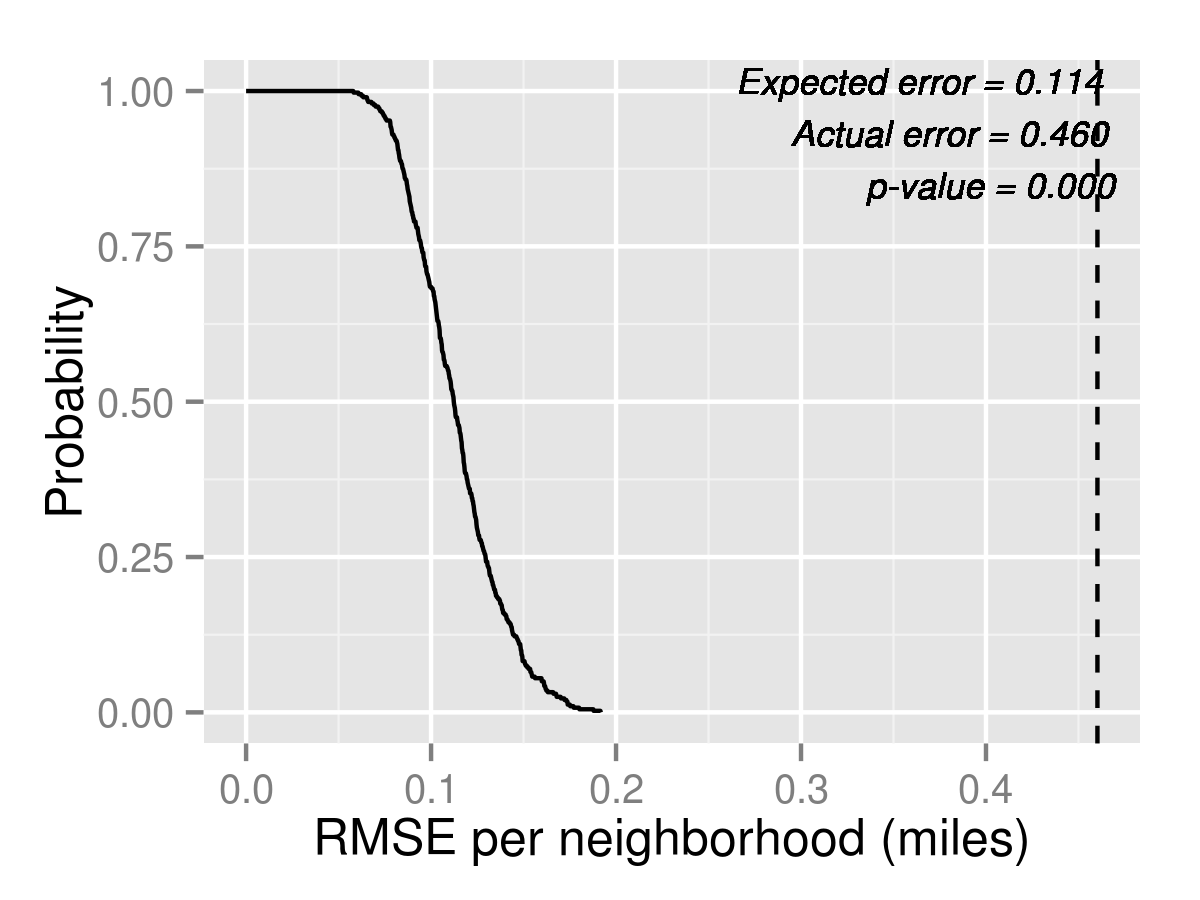}
}
\subfloat[][Distance (Logit)]{
	\includegraphics[width=0.33\textwidth]{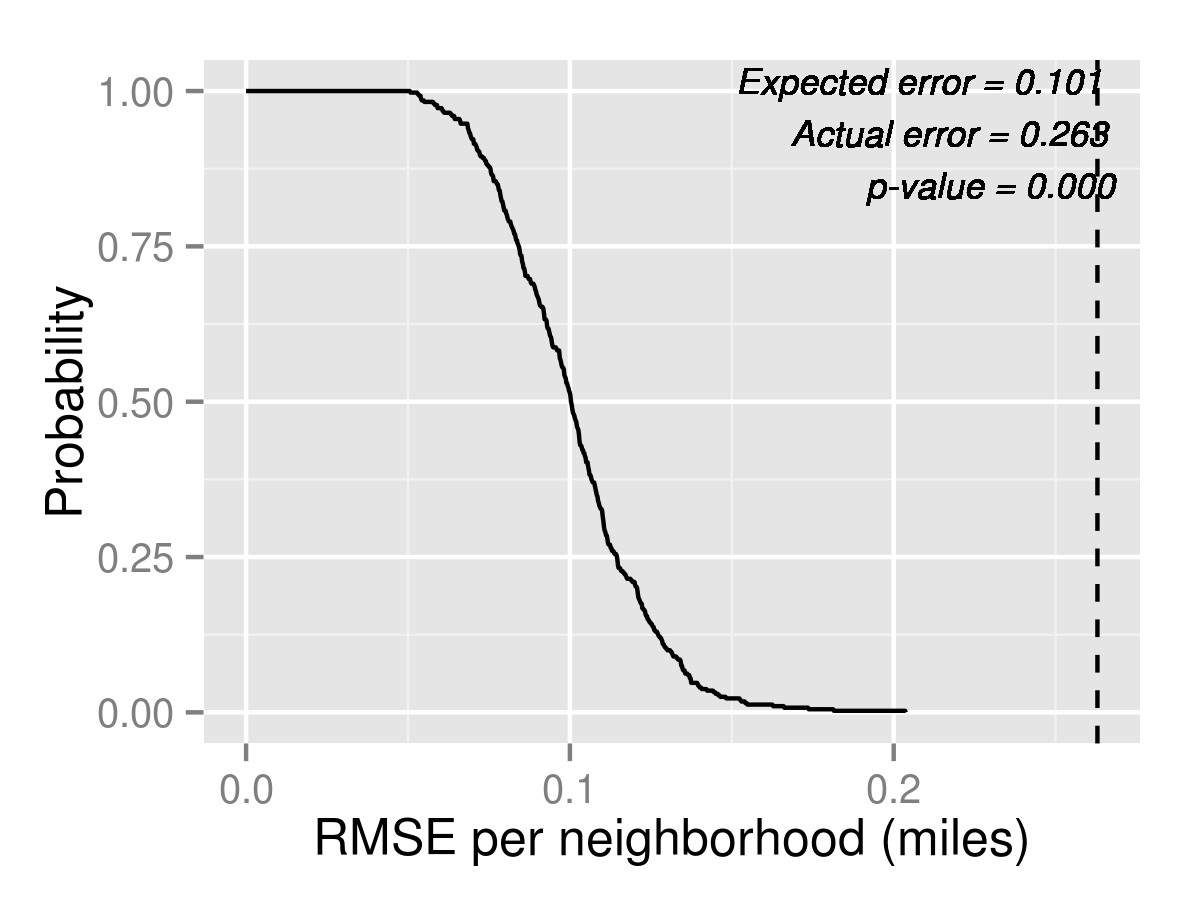}
}
\subfloat[][Distance (MLogit)]{
	\includegraphics[width=0.33\textwidth]{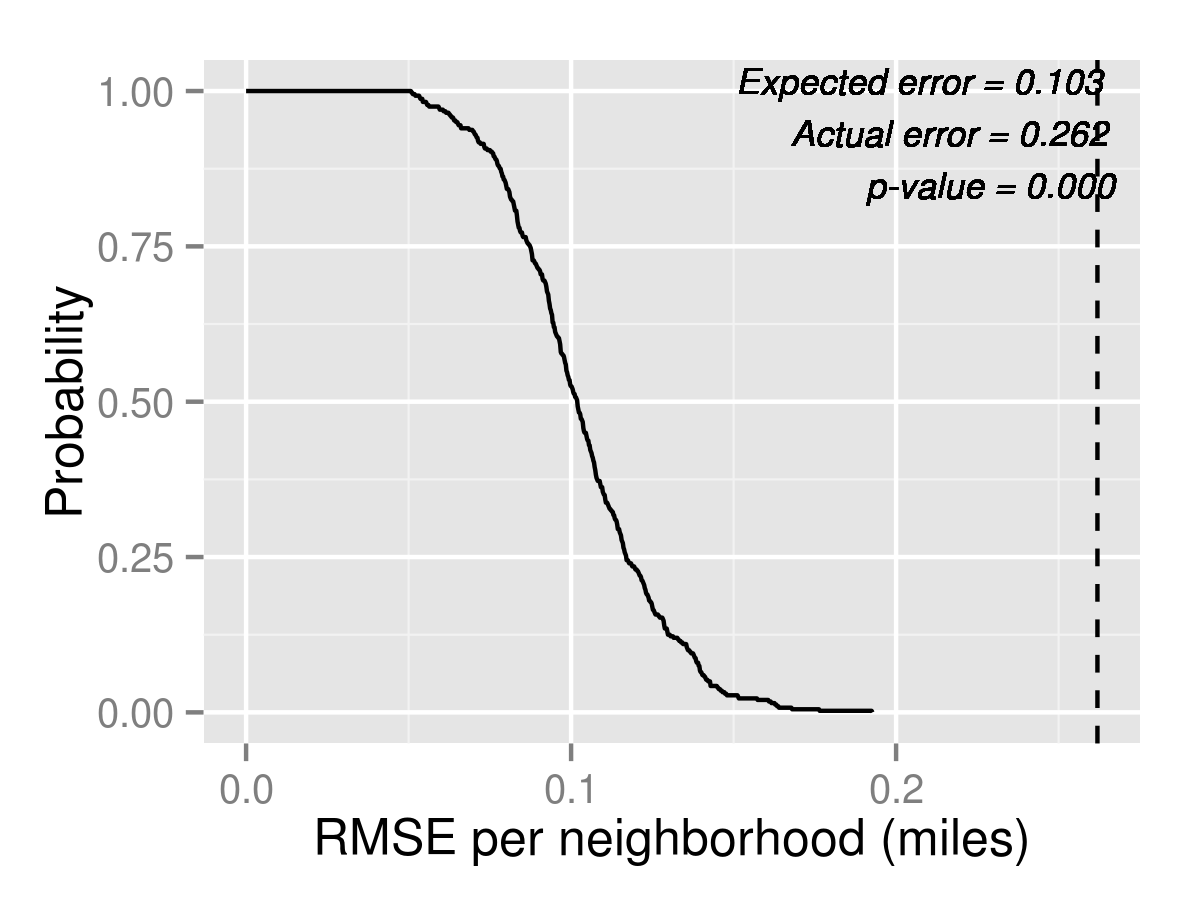}
}\\

\end{figure}

 As a uniform measure of how well the models predict each outcome, we compute the tail distribution of Root Mean Squared Error (RMSE) for each model and each outcome. This measure gives a sense of how much we expect to be off by on average. It's worth noting that even if a model is completely correct, there are multiple levels of randomness in the simulations themselves, so there will always be some forecast error. By evaluating the tail distribution at the actual RMSE, we estimate a p-value of the likelihood that this magnitude of error occurs when the model is correct. We plot the p-values for the first three outcome measures for 2013 K2 in Table~\ref{tab:prediction2013K2}. For brevity, we only show results for K2 in the section, but the corresponding plots for K1 are in Appendix~\ref{app:resultsK1}. All plots in our back testing exercise are computed using 400 independent simulations.
 Since we can no longer use 2013 data, we create population forecasts following the methodology in section~\ref{sec:populationForecast}, but using only 3 years of data.
The estimates are in Appendix~\ref{app:populationForecast}.

Figure~\ref{tab:prediction2013K2} shows that in 2013 the Naive model for unassigned has an expected RMSE of 13.3 students per neighborhood, while the actual data yields a RMSE of 25.8 students per neighborhood. (In other words, the model is expected to be off by about 13.3 students, while it's actually off by 25.8) If the Naive model were correct, such a large deviation occurs with 2\% probability, which indicates that while the Naive model may not be the most satisfactory, it is not totally implausible in terms of its predictions for number of unassigned per neighborhood. The Logit model on the other hand produces an actual RMSE of 18.7 students, which is much smaller than Naive. The p-value is larger at 6.2\%. This implies that the Logit model's predictions for unassigned cannot be rejected at 95\% confidence level. MixedLogit produces a smaller actual RMSE of 18.1 students and the p-value is 7.8\%. Hence, in terms of forecasting the numbers and locations of unassigned students, MixedLogit performs the 
best, followed closely by Logit, and Naive performs the worst.

For Access to Quality, the Naive model yields almost 10 times the expected error, with p-value of about zero, while the Logit model is reasonably on target, with p-value of 30\%. MixedLogit yields a smaller error and a larger p-value of 32.8\%. However, for distance, none of the models seem to explain the data, with all models having near zero p-values. The Naive model is off by 0.46 miles per neighborhood, and both Logit and MixedLogit are off by 0.26 miles per neighborhood.

Next we examine the neighborhood by neighborhood predictions themselves. These are shown in Tables~\ref{tab:unassigned2013K2}, \ref{tab:aoq2013K2}, and \ref{tab:distance2013K2}. For each model, we show the 95\% confidence interval as estimated in the 400 simulations. The Naive model over-predicts the number of unassigned everywhere, possibly due to some schools that should have been chosen much not being chosen often enough by the Naive rule. (Recall that we assume students rank at most 10 options, so low Tier schools far away would not be ranked at all under the Naive rule.) Logit performs better since school popularity levels are captured in the fixed effects. For the majority of neighborhoods, the actual realization is within 95\% confidence interval, suggesting that Logit could have been a reasonable model for capacity planning purposes, at least in 2013. MixedLogit yields very similar results to simple Logit, and the confidence intervals for all neighborhoods overlap significantly with Logit.

\begin{table}[!htbp]
 \centering
 \caption{Back testing unassigned predictions for 2013 K2.}
  \label{tab:unassigned2013K2}

 \footnotesize
 \begin{tabular}{l c c c c c c c}
Neighborhood & Naive & (95 \% C.I.) & Logit & (95 \% C.I.) & MLogit & (95 \% C.I.) & Actual\\ \hline  
Allston-Brighton & 32.80 & (17.00,51.02) & 8.83 & (0.00,24.00) & 7.66 & (0.00,23.02) & 11\\ 
Charlestown & 49.49 & (33.00,66.00) & 25.58 & (11.00,44.00) & 28.25 & (12.00,45.00) & 18\\ 
Downtown & 28.58 & (16.00,43.00) & 13.12 & (3.00,25.00) & 14.04 & (4.00,27.00) & 23\\ 
East Boston & 118.22 & (67.00,183.03) & 88.76 & (36.95,152.03) & 91.56 & (36.00,150.05) & 81\\ 
Hyde Park & 41.57 & (24.00,62.00) & 5.88 & (0.00,19.00) & 5.68 & (0.00,18.02) & 13\\ 
Jamaica Plain & 42.52 & (25.00,61.00) & 12.55 & (2.00,29.00) & 12.50 & (2.00,27.00) & 32\\ 
Mattapan & 33.88 & (14.00,58.00) & 5.46 & (0.00,20.02) & 5.86 & (0.00,23.00) & 21\\ 
North Dorchester & 36.54 & (18.00,60.05) & 6.41 & (0.00,24.02) & 7.48 & (0.00,23.02) & 16\\ 
Roslindale & 45.98 & (21.98,73.05) & 18.21 & (2.00,42.02) & 19.93 & (4.00,44.00) & 51\\ 
Roxbury & 109.14 & (73.97,154.03) & 18.78 & (2.00,54.08) & 20.56 & (3.00,56.02) & 47\\ 
South Boston & 20.35 & (6.00,39.00) & 3.90 & (0.00,16.00) & 4.12 & (0.00,17.00) & 7\\ 
South Dorchester & 72.47 & (37.00,113.03) & 12.98 & (0.00,43.00) & 14.65 & (0.00,49.02) & 52\\ 
South End & 45.78 & (32.00,62.00) & 18.82 & (5.00,35.00) & 19.75 & (6.97,36.02) & 26\\ 
West Roxbury & 53.11 & (27.98,84.03) & 26.14 & (7.00,51.00) & 27.28 & (8.00,54.00) & 48\\ 
\end{tabular}
 \end{table}

 \begin{table}[!htbp]
 \centering
 \caption{Back testing access to quality predictions for 2013 K2.}
  \label{tab:aoq2013K2}

 \footnotesize
 \begin{tabular}{l c c c c c c c}
Neighborhood & Naive & (95 \% C.I.) & Logit & (95 \% C.I.) & MLogit & (95 \% C.I.) & Actual\\ \hline  
Allston-Brighton & 1 & (1.00,1.00) & 0.94 & (0.82,1.00) & 0.95 & (0.85,1.00) & 1\\ 
Charlestown & 1 & (1.00,1.00) & 0.92 & (0.79,1.00) & 0.94 & (0.83,1.00) & 1\\ 
Downtown & 1 & (1.00,1.00) & 0.94 & (0.85,1.00) & 0.96 & (0.88,1.00) & 1\\ 
East Boston & 1 & (1.00,1.00) & 0.97 & (0.89,1.00) & 0.98 & (0.90,1.00) & 1\\ 
Hyde Park & 0.56 & (0.49,0.64) & 0.97 & (0.88,1.00) & 0.97 & (0.86,1.00) & 0.99\\ 
Jamaica Plain & 0.79 & (0.72,0.87) & 0.91 & (0.81,1.00) & 0.91 & (0.81,1.00) & 0.96\\ 
Mattapan & 0.52 & (0.44,0.60) & 0.97 & (0.88,1.00) & 0.97 & (0.86,1.00) & 1.00\\ 
North Dorchester & 0.58 & (0.50,0.66) & 0.97 & (0.87,1.00) & 0.97 & (0.86,1.00) & 1\\ 
Roslindale & 0.73 & (0.64,0.83) & 0.91 & (0.80,1.00) & 0.90 & (0.79,1.00) & 0.96\\ 
Roxbury & 0.74 & (0.66,0.81) & 0.91 & (0.82,1.00) & 0.91 & (0.81,1.00) & 0.96\\ 
South Boston & 0.44 & (0.36,0.53) & 0.98 & (0.87,1.00) & 0.97 & (0.85,1.00) & 1\\ 
South Dorchester & 0.53 & (0.46,0.61) & 0.98 & (0.89,1.00) & 0.98 & (0.87,1.00) & 1\\ 
South End & 1 & (1.00,1.00) & 0.93 & (0.82,1.00) & 0.95 & (0.85,1.00) & 1\\ 
West Roxbury & 0.73 & (0.63,0.83) & 0.90 & (0.78,1.00) & 0.89 & (0.77,1.00) & 0.96\\ 
\end{tabular}
 \end{table}

For access to quality, since the assignment system in 2013 gave each family at least about 25 choices in each zone, there were likely some Tier 1 or 2 school with some left over capacity, so the marginal student could have changed choices to selecting all Tier 1 or 2 schools and selecting them first, and gotten in a Tier 1 or 2 school with near certain probability. This results in the very high actual access to quality measures shown in Table~\ref{tab:aoq2013K2}. However, Naive forecasts a much tougher level of competition for Tier 1 or 2 schools, because it assumes every non-continuing students without siblings or ELL considerations go for such schools first. Although the MCAS data on which the Tiers were based were released months before, the Tiers themselves were finalized only around the time of the 2013 Round 1 choice. Hence, they were not salient at the time the choices were made. Again, Logit would have been a reasonable model for access to quality, with MixedLogit producing almost identical results.

\begin{table}[!htbp]
 \centering
 \caption{Back testing distance predictions for 2013 K2.}
  \label{tab:distance2013K2}

 \footnotesize
 \begin{tabular}{l c c c c c c c}
Neighborhood & Naive & (95 \% C.I.) & Logit & (95 \% C.I.) & MLogit & (95 \% C.I.) & Actual\\ \hline  
Allston-Brighton & 2.09 & (1.75,2.45) & 1.47 & (1.26,1.68) & 1.52 & (1.31,1.74) & 1.45\\ 
Charlestown & 1.52 & (1.18,1.84) & 1.77 & (1.43,2.07) & 1.71 & (1.43,2.00) & 0.98\\ 
Downtown & 1.59 & (1.30,1.91) & 1.51 & (1.30,1.77) & 1.52 & (1.27,1.78) & 1.57\\ 
East Boston & 2.88 & (2.63,3.12) & 2.30 & (2.00,2.61) & 2.37 & (2.03,2.70) & 1.82\\ 
Hyde Park & 2.93 & (2.67,3.23) & 2.22 & (2.01,2.48) & 2.22 & (2.02,2.43) & 2.19\\ 
Jamaica Plain & 1.66 & (1.53,1.80) & 1.57 & (1.45,1.71) & 1.57 & (1.43,1.72) & 1.43\\ 
Mattapan & 2.28 & (2.16,2.42) & 2.22 & (2.08,2.35) & 2.16 & (2.03,2.30) & 2.23\\ 
North Dorchester & 1.68 & (1.44,1.93) & 1.48 & (1.32,1.67) & 1.51 & (1.30,1.72) & 1.39\\ 
Roslindale & 1.84 & (1.73,1.96) & 1.80 & (1.68,1.91) & 1.77 & (1.66,1.90) & 1.65\\ 
Roxbury & 1.89 & (1.78,2.00) & 1.58 & (1.49,1.69) & 1.58 & (1.49,1.69) & 1.50\\ 
South Boston & 1.59 & (1.35,1.83) & 1.33 & (1.15,1.52) & 1.36 & (1.14,1.56) & 1.26\\ 
South Dorchester & 1.75 & (1.65,1.83) & 1.78 & (1.69,1.88) & 1.74 & (1.64,1.84) & 1.76\\ 
South End & 1.79 & (1.56,2.00) & 1.62 & (1.42,1.83) & 1.61 & (1.37,1.84) & 1.39\\ 
West Roxbury & 2.12 & (1.95,2.28) & 2.12 & (1.93,2.33) & 2.08 & (1.91,2.26) & 2.18\\ 
\end{tabular}
 \end{table}

For distance, Naive again over-predicts because of its preference on going for better Tier schools despite possibly further distances, which does not reflect the trade-off implicitly shown by most families in 2013. Logit produced more reasonable results, with the actual realization being within the 95\% confidence interval in the majority of neighborhoods. However, for the neighborhoods at the peripheries of the city, Charlestown and East Boston, which are separated by bridges from the rest of the city, Logit over-estimates distances, suggesting that it predicts families are more willing to cross the bridges than they actually are. Again, MixedLogit yields very similar results as Logit.

\begin{figure}[h!]
\centering
\caption{Back testing market share predictions for 2013 K2. Tail distribution plots. \label{fig:tail2013K2}}

\subfloat[][Top 1 (Naive)]{
 \includegraphics[width=0.33\textwidth]{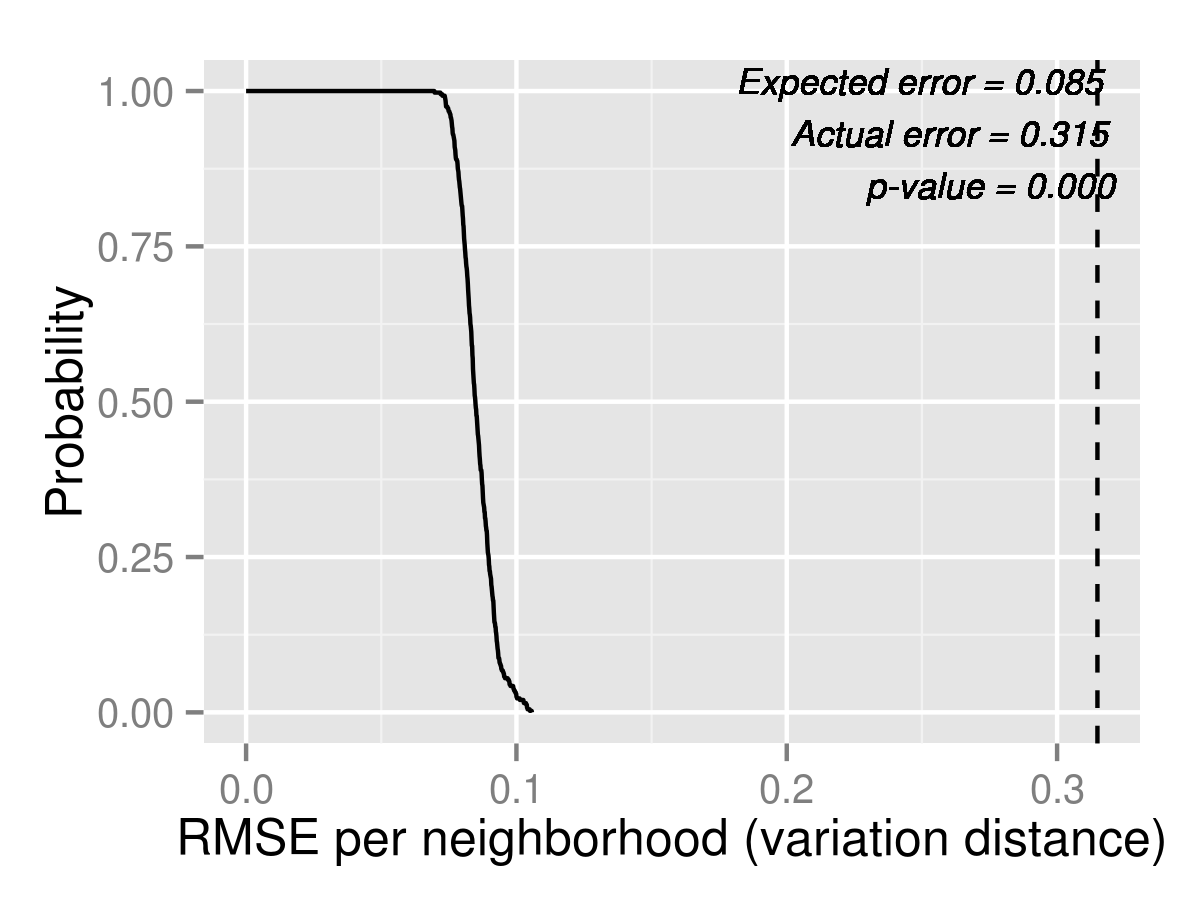}
}
\subfloat[][Top 1 (Logit)]{
	\includegraphics[width=0.33\textwidth]{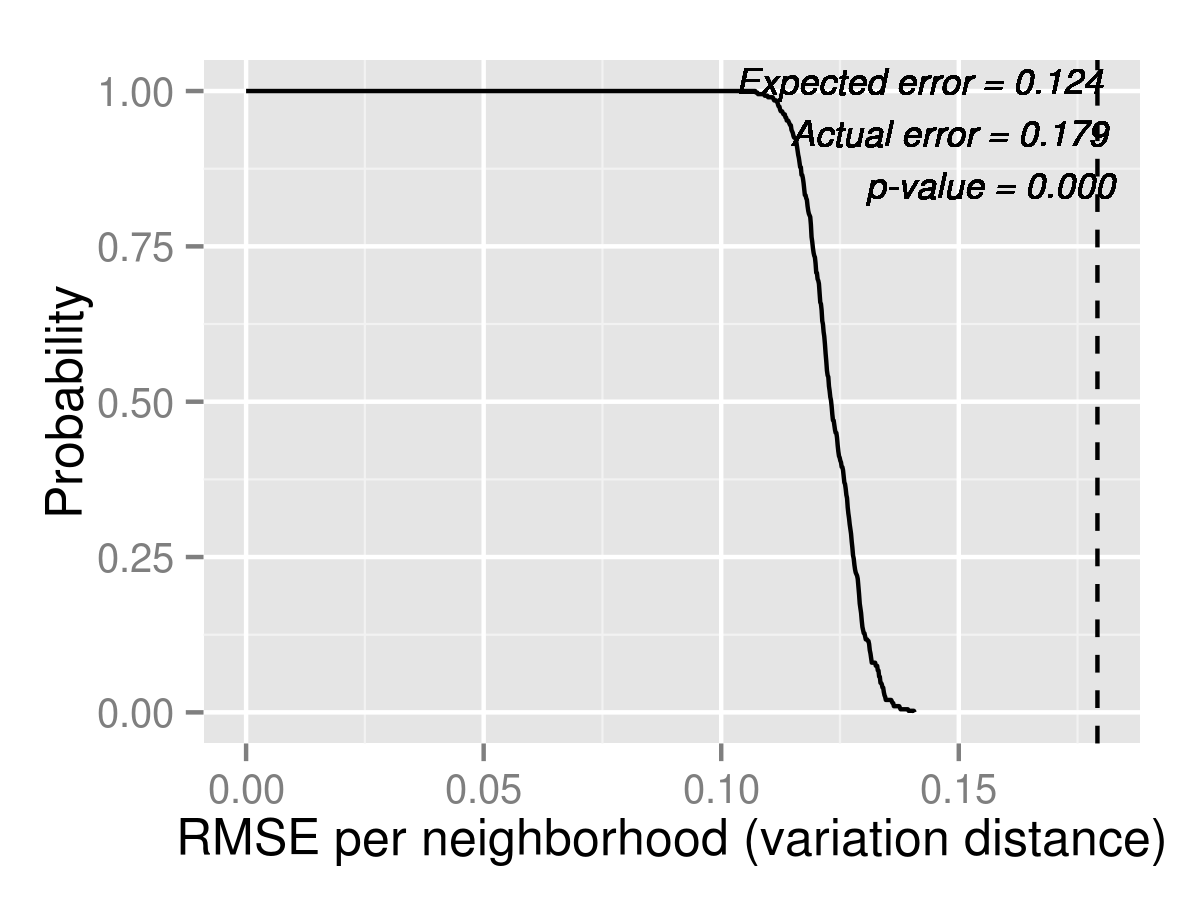}
}
\subfloat[][Top 1 (MLogit)]{
	\includegraphics[width=0.33\textwidth]{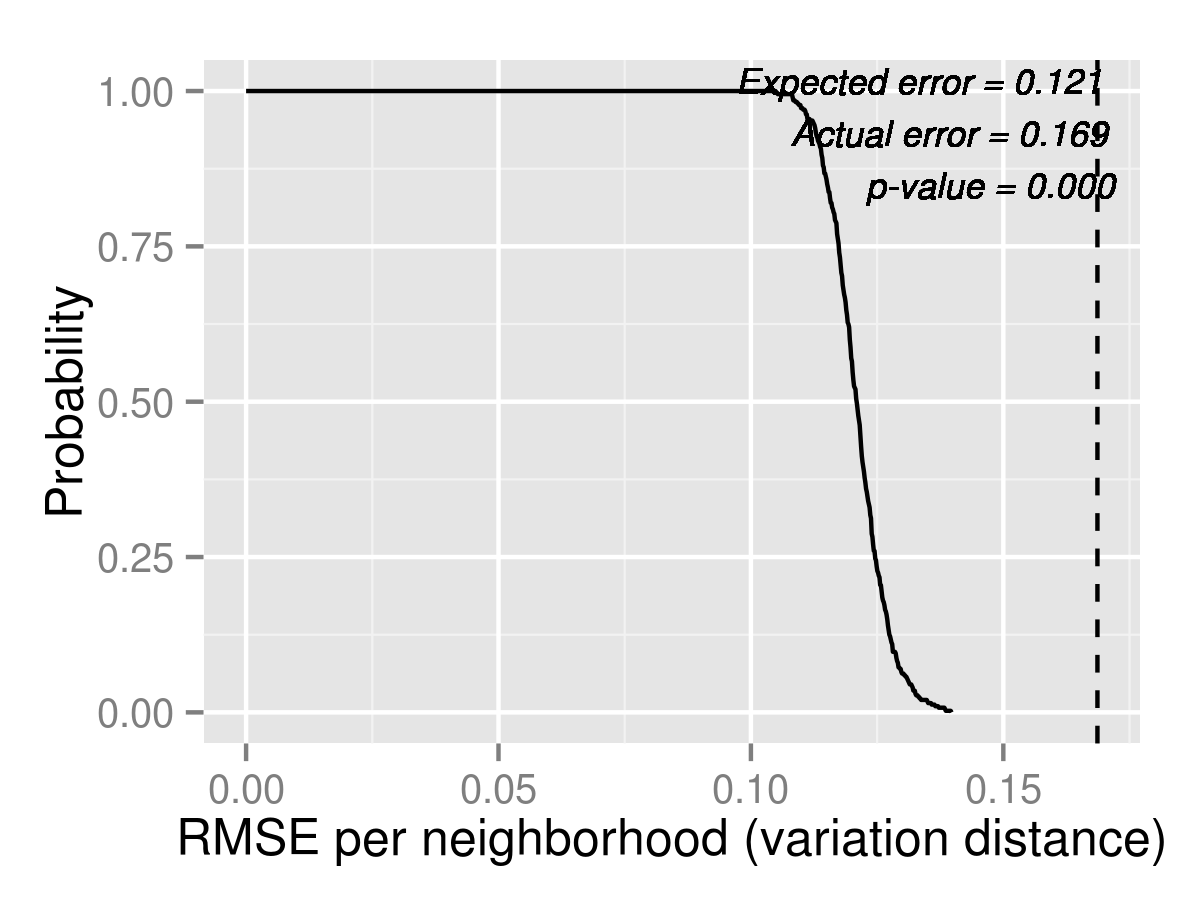}
}\\

\subfloat[][Top 2 (Naive)]{
 \includegraphics[width=0.33\textwidth]{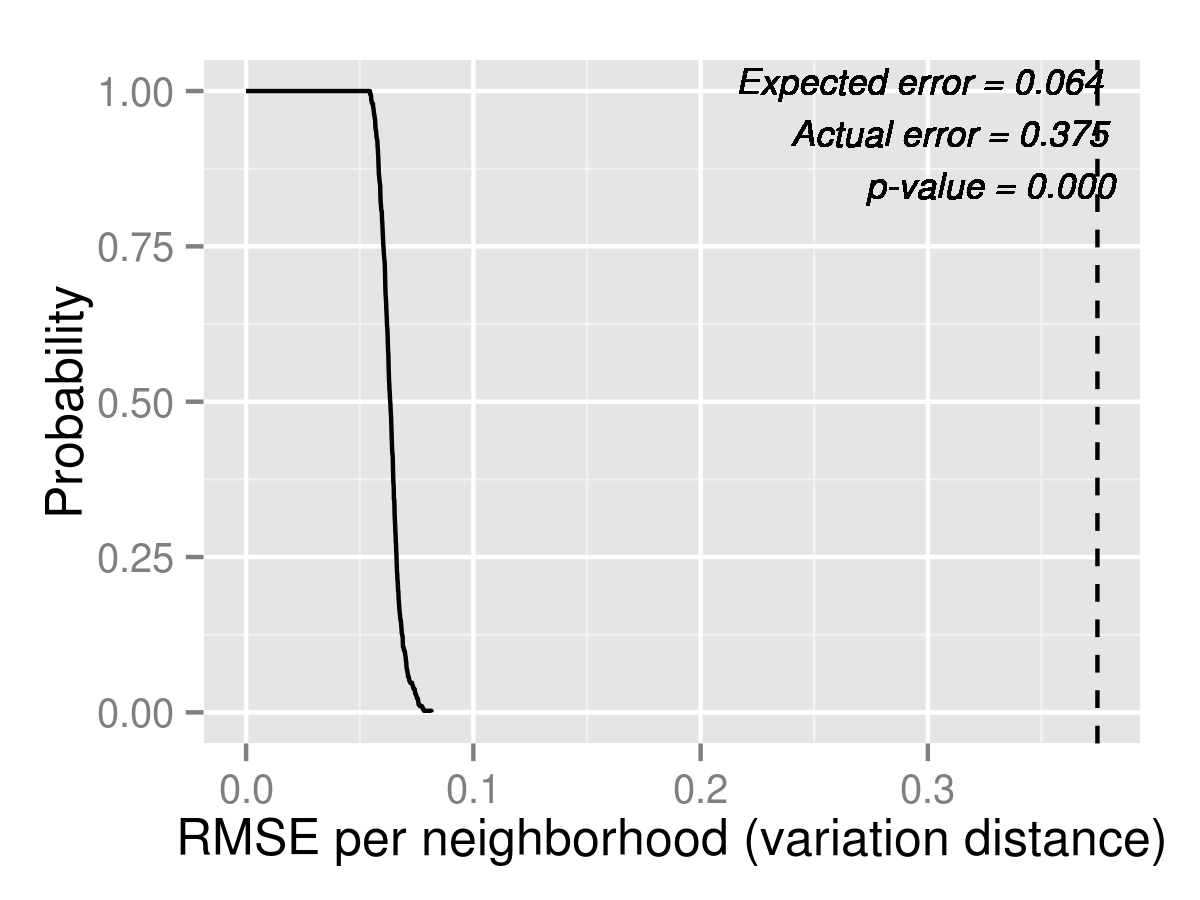}
}
\subfloat[][Top 2 (Logit)]{
	\includegraphics[width=0.33\textwidth]{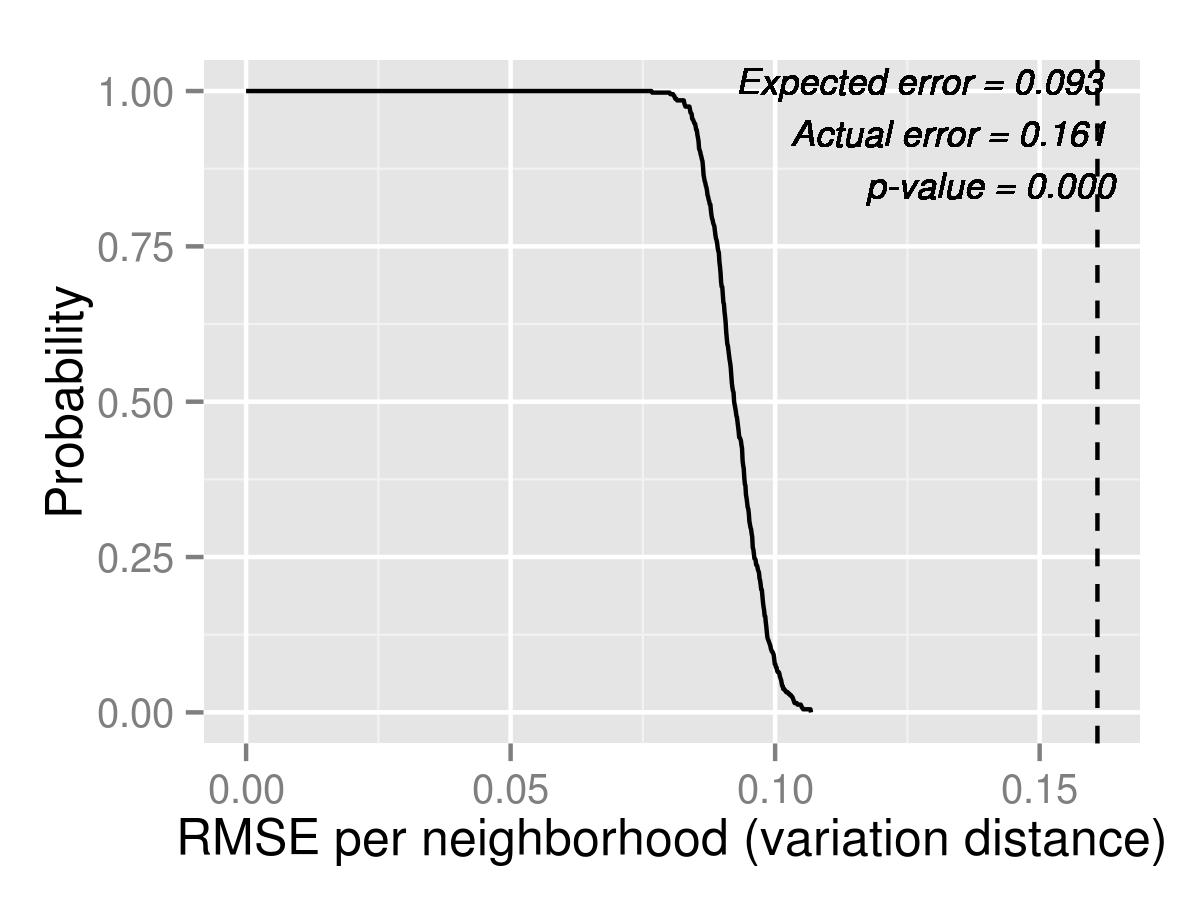}
}
\subfloat[][Top 2 (MLogit)]{
	\includegraphics[width=0.33\textwidth]{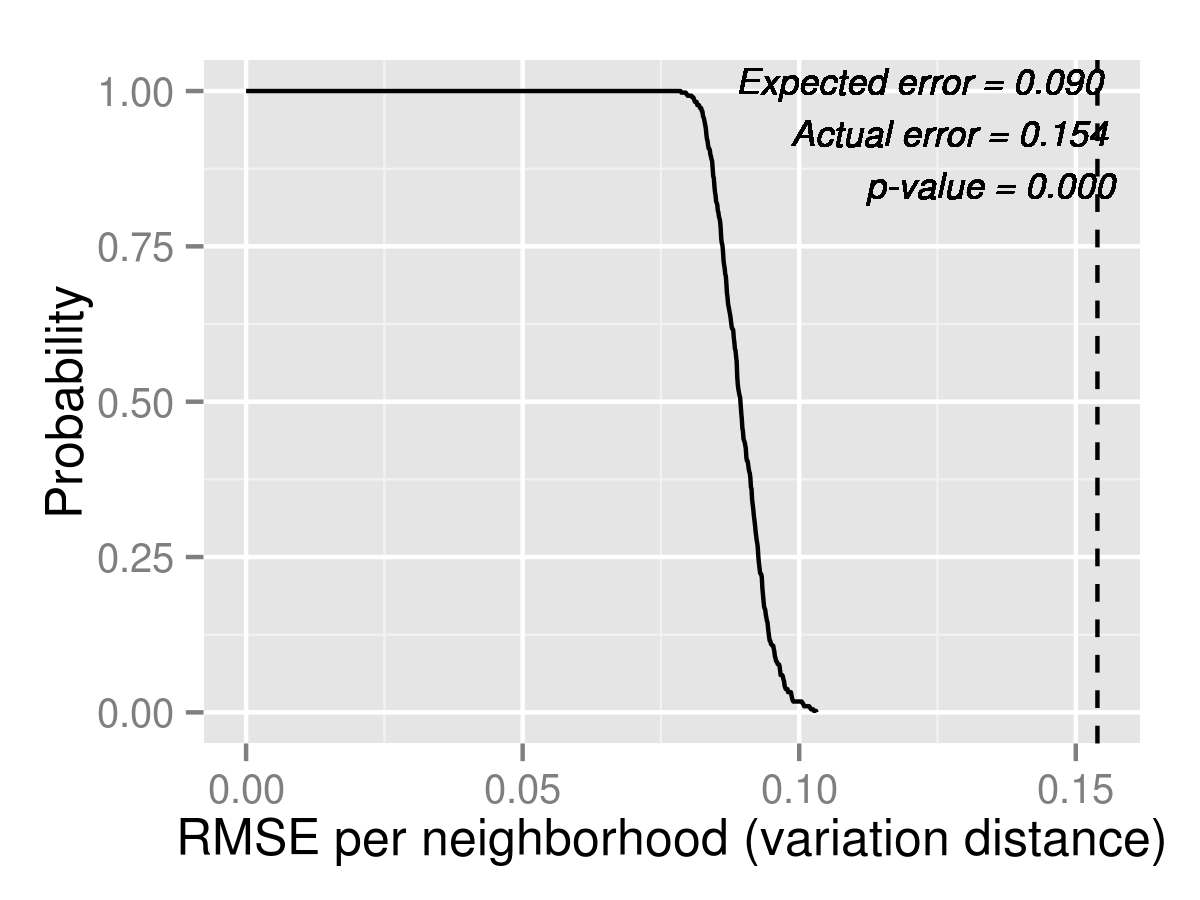}
}\\

\subfloat[][Top 3 (Naive)]{
 \includegraphics[width=0.33\textwidth]{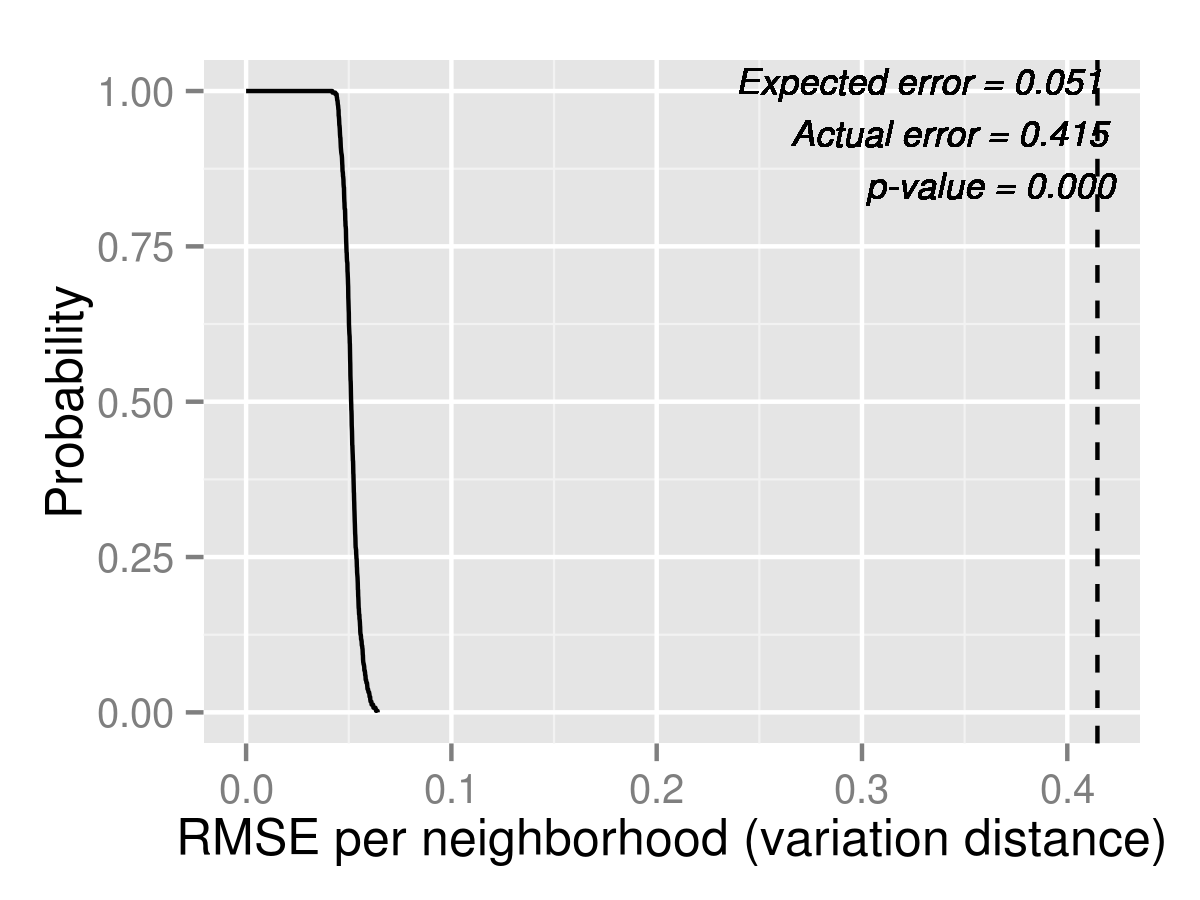}
}
\subfloat[][Top 3 (Logit)]{
	\includegraphics[width=0.33\textwidth]{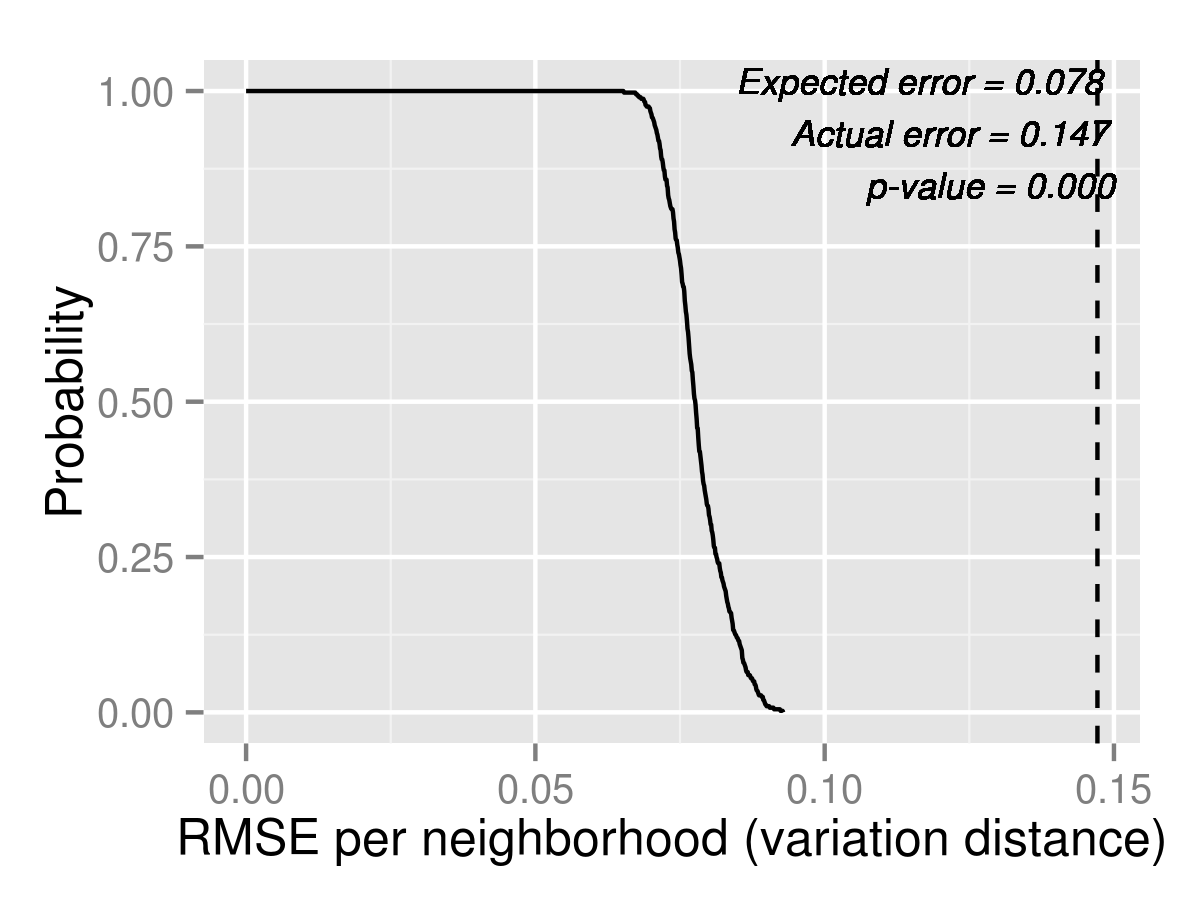}
}
\subfloat[][Top 3 (MLogit)]{
	\includegraphics[width=0.33\textwidth]{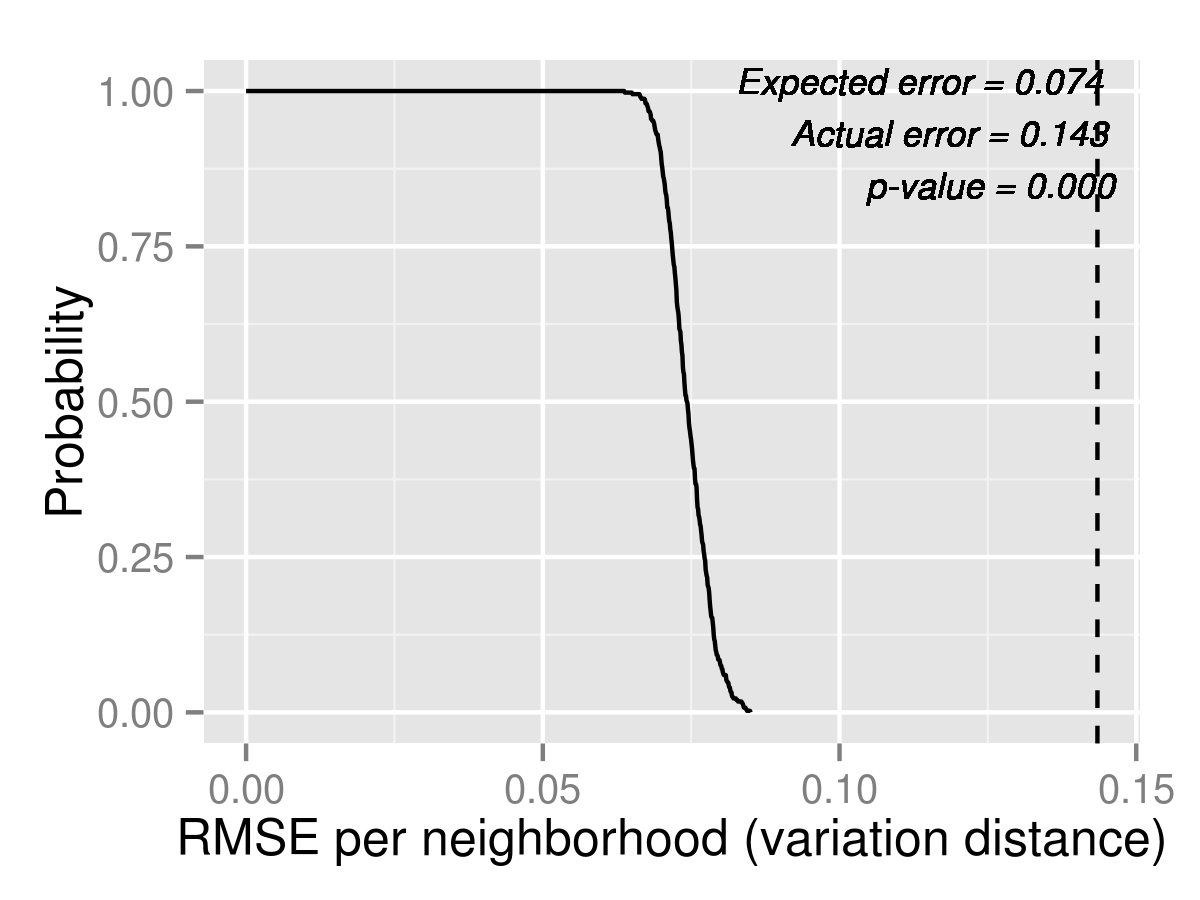}
}\\

\end{figure}

We repeat the same exercise using market shares, for top 1, 2 and 3 choices. The tail distribution plots are in Figure~\ref{fig:tail2013K2}. However, for this metric neither models seem to explain the data, with p-values being near zero for all cases. Nevertheless, the error in terms of total variation distance is about half as large with Logit compared to with Naive, suggesting that it is a much better model for market shares in 2013. MixedLogit improves over Logit in the actual error in all cases, but the improvements are small. The tables showing market share details from each neighborhood to each school are in Appendix D. Since these tables are much longer, we put them in a separate document titled ``Supplementary Data'' to this report, available as an ancillary file.

\subsection{Forecasts for 2014}
\label{sec:future}
\begin{figure}[h!]
\centering
\caption{Forecasts for assignment outcomes in 2014 K2. Tail distribution plots. \label{fig:outcomes2014K2}}

\subfloat[][Unassigned (Naive)]{
 \includegraphics[width=0.33\textwidth]{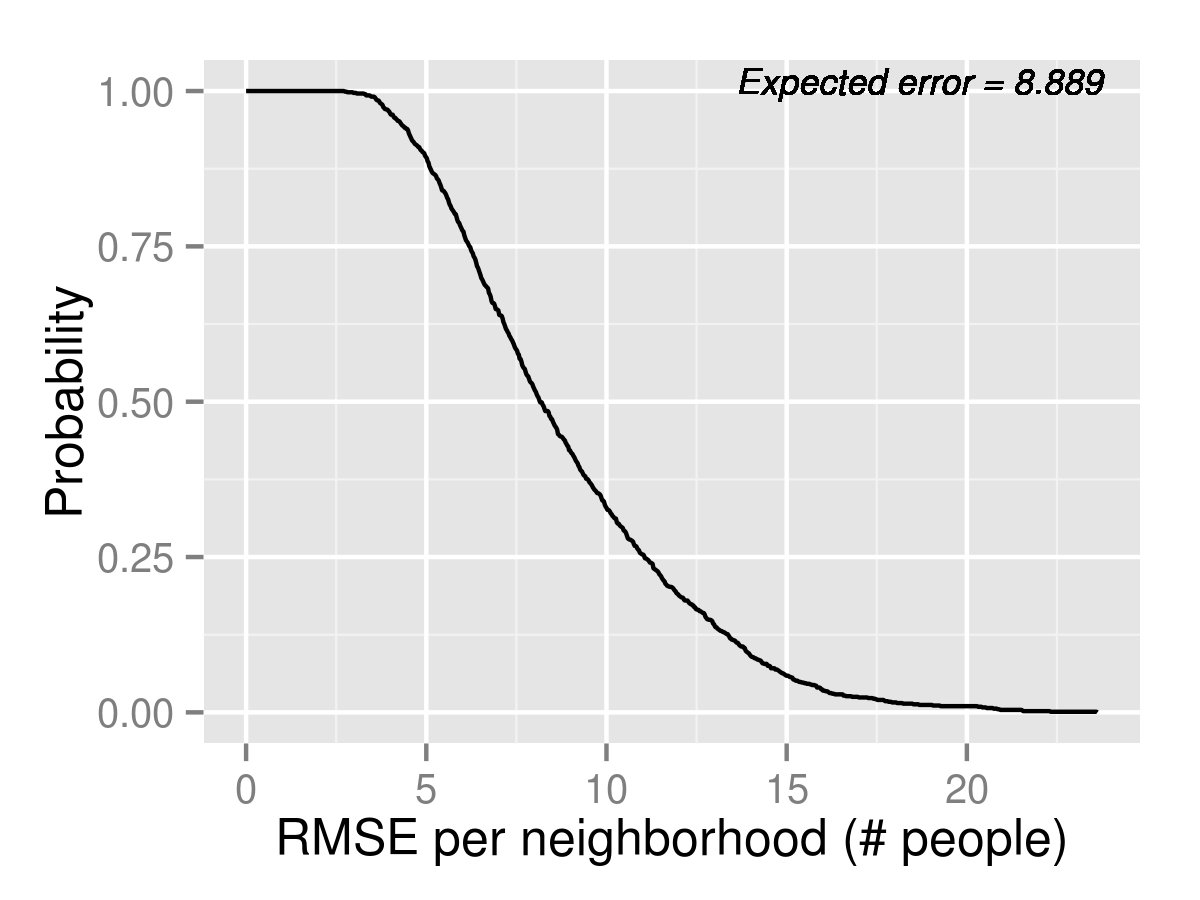}
}
\subfloat[][Unassigned (Logit)]{
	\includegraphics[width=0.33\textwidth]{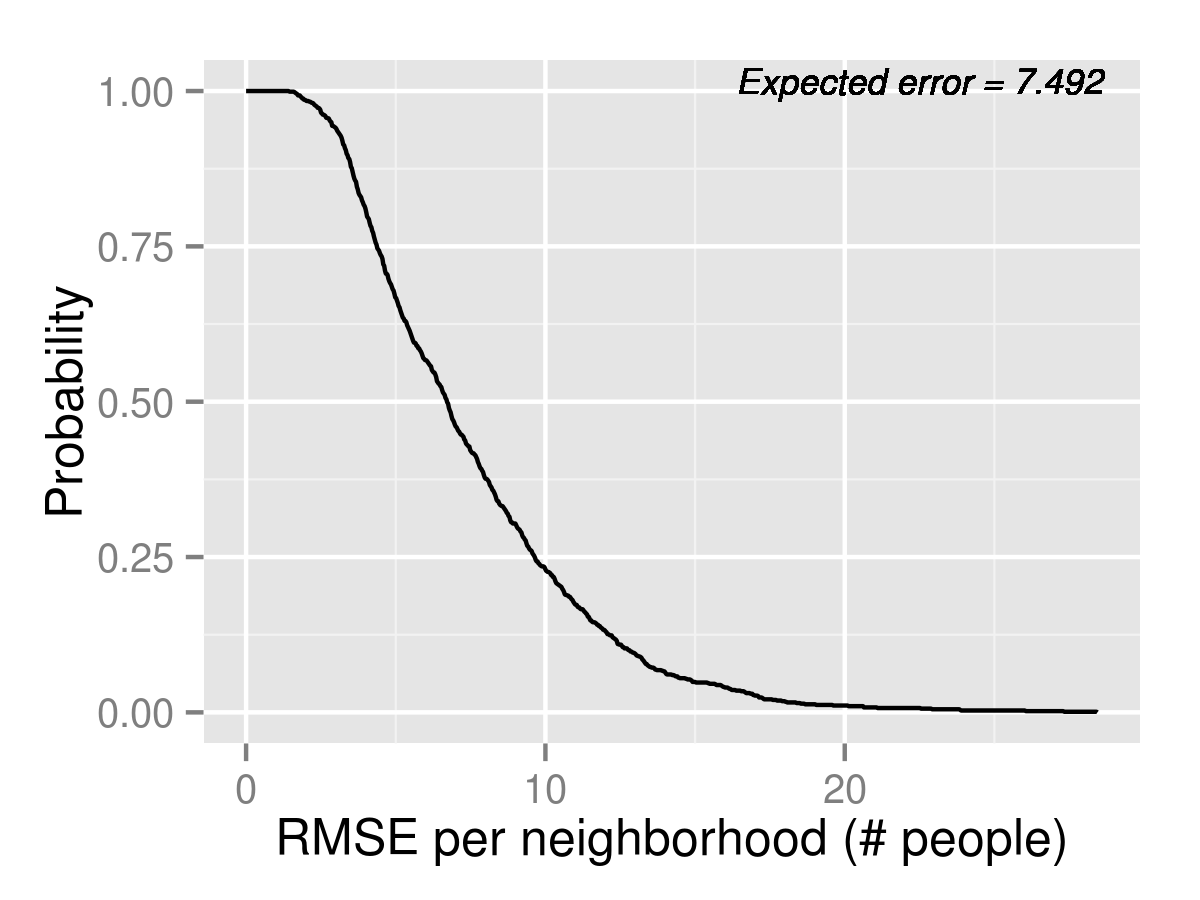}
}
\subfloat[][Unassigned (MLogit)]{
	\includegraphics[width=0.33\textwidth]{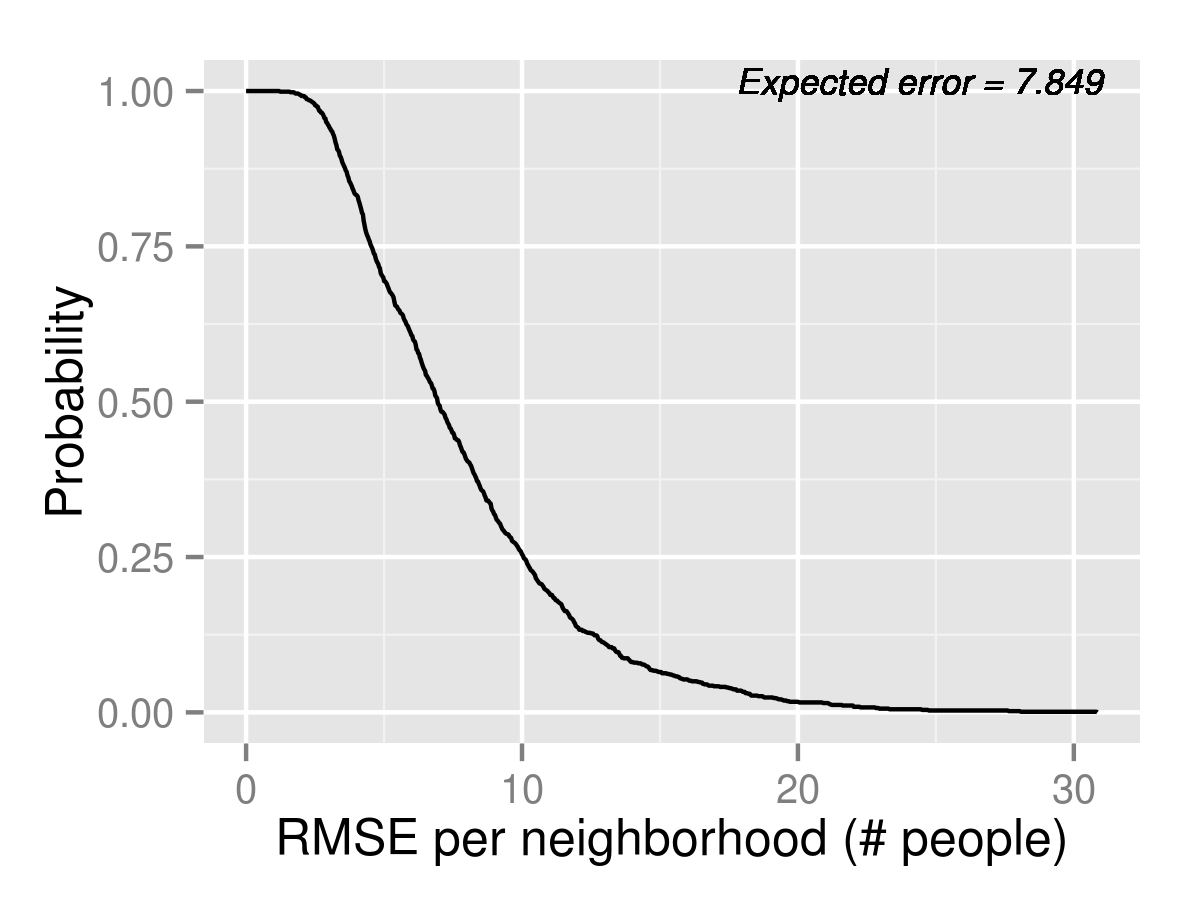}
}\\

\subfloat[][Access to Quality (Naive)]{
 \includegraphics[width=0.33\textwidth]{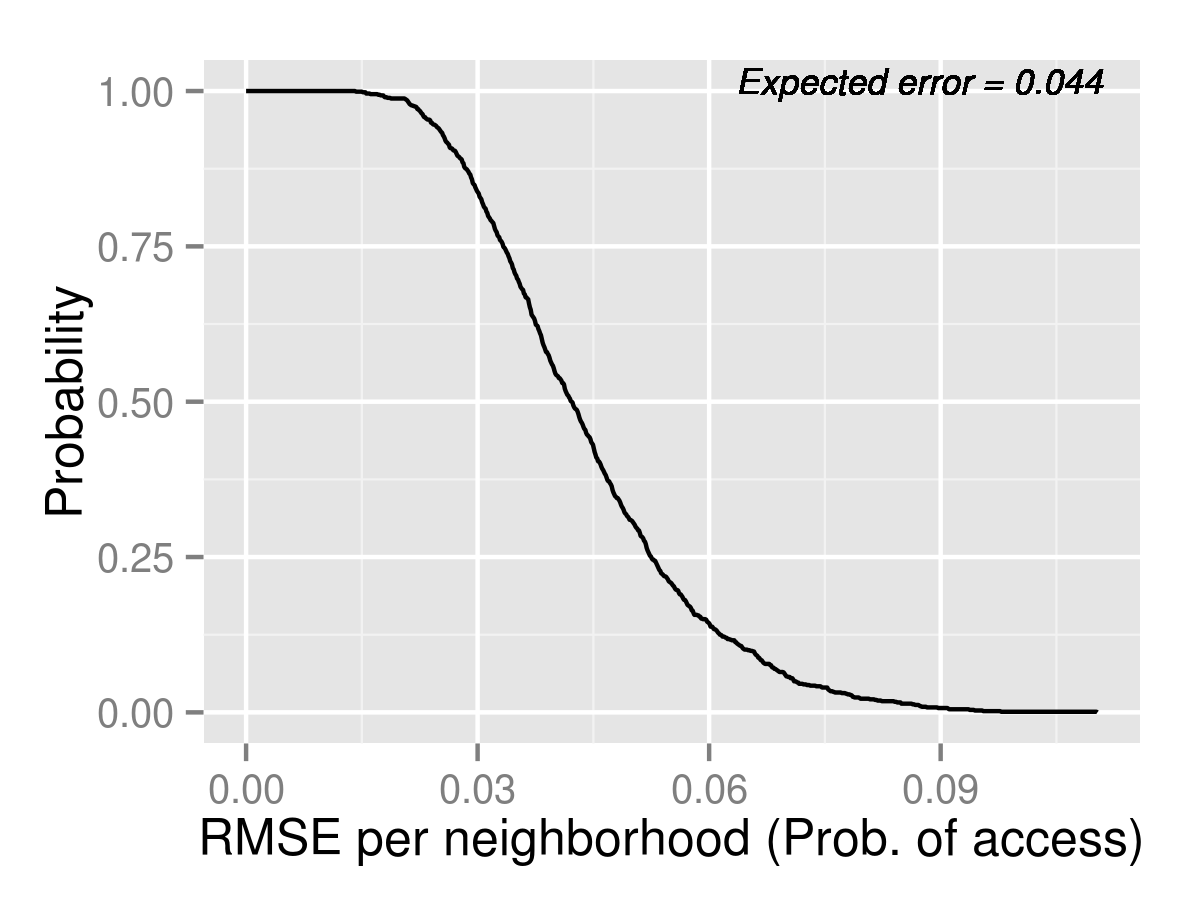}
}
\subfloat[][Access to Quality (Logit)]{
	\includegraphics[width=0.33\textwidth]{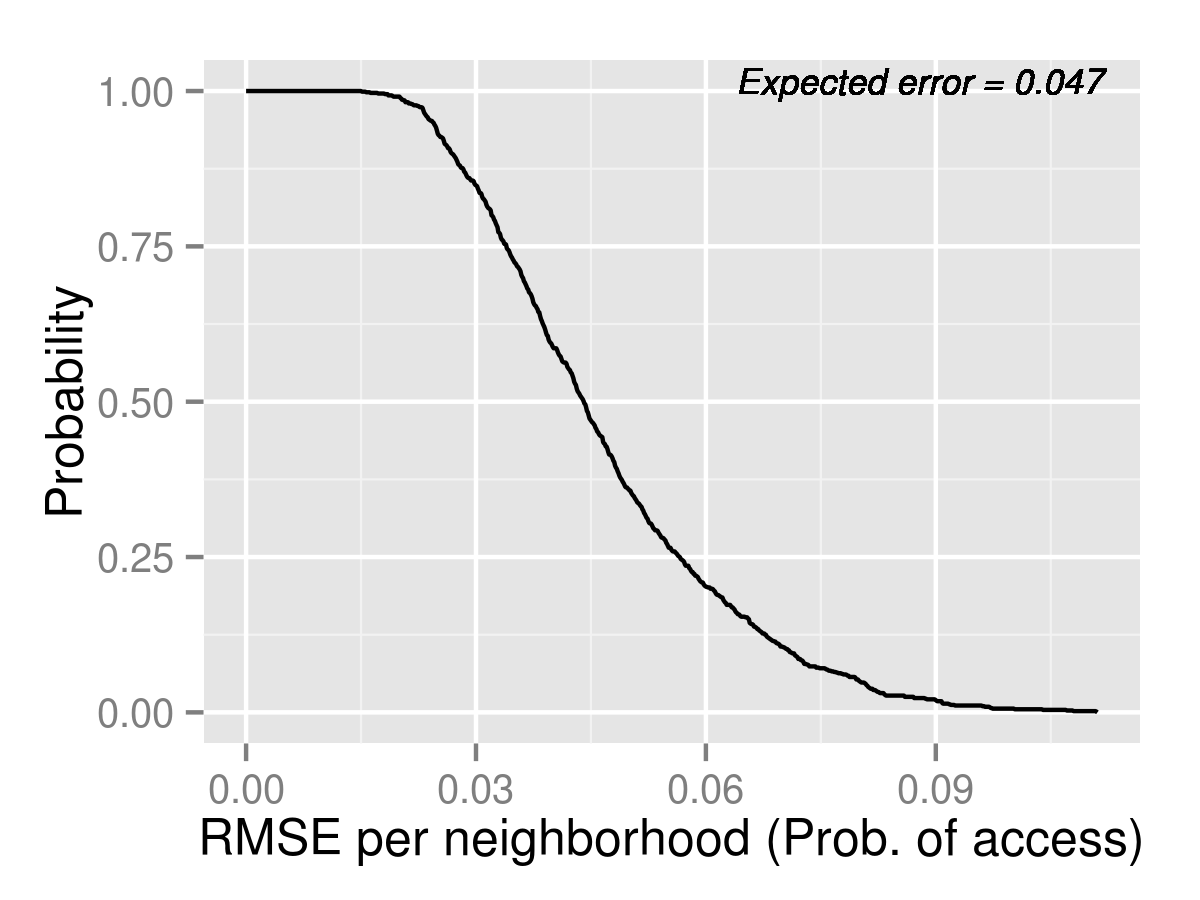}
}
\subfloat[][Access to Quality (MLogit)]{
	\includegraphics[width=0.33\textwidth]{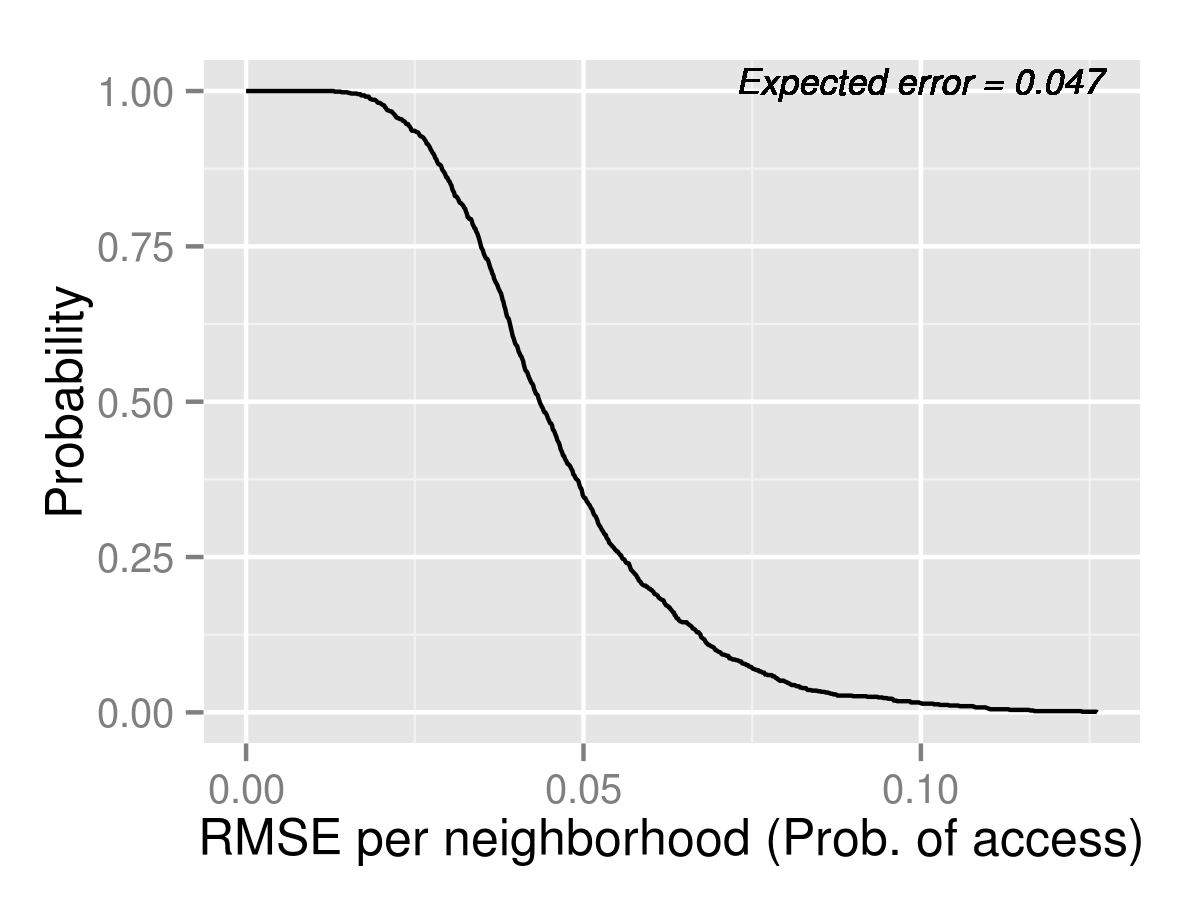}
}\\

\subfloat[][Distance (Naive)]{
 \includegraphics[width=0.33\textwidth]{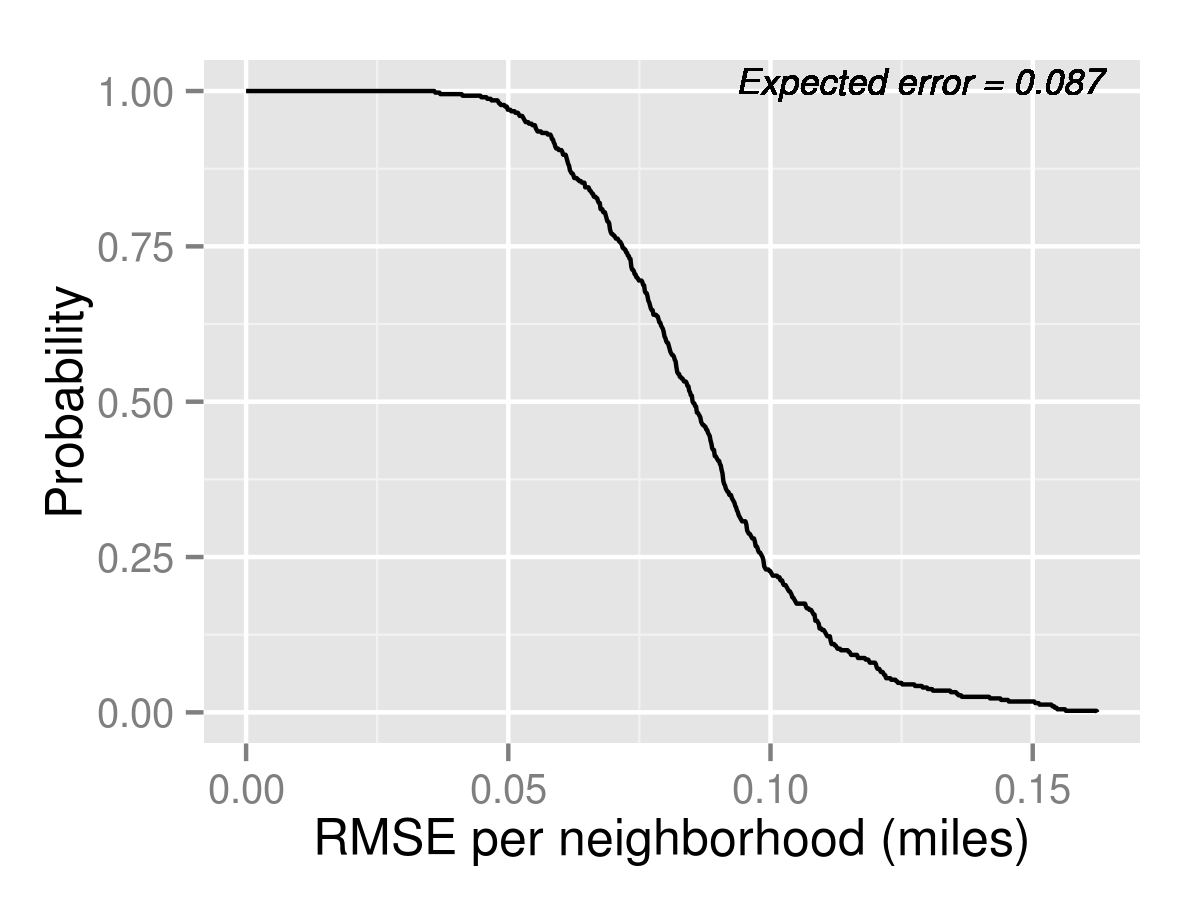}
}
\subfloat[][Distance (Logit)]{
	\includegraphics[width=0.33\textwidth]{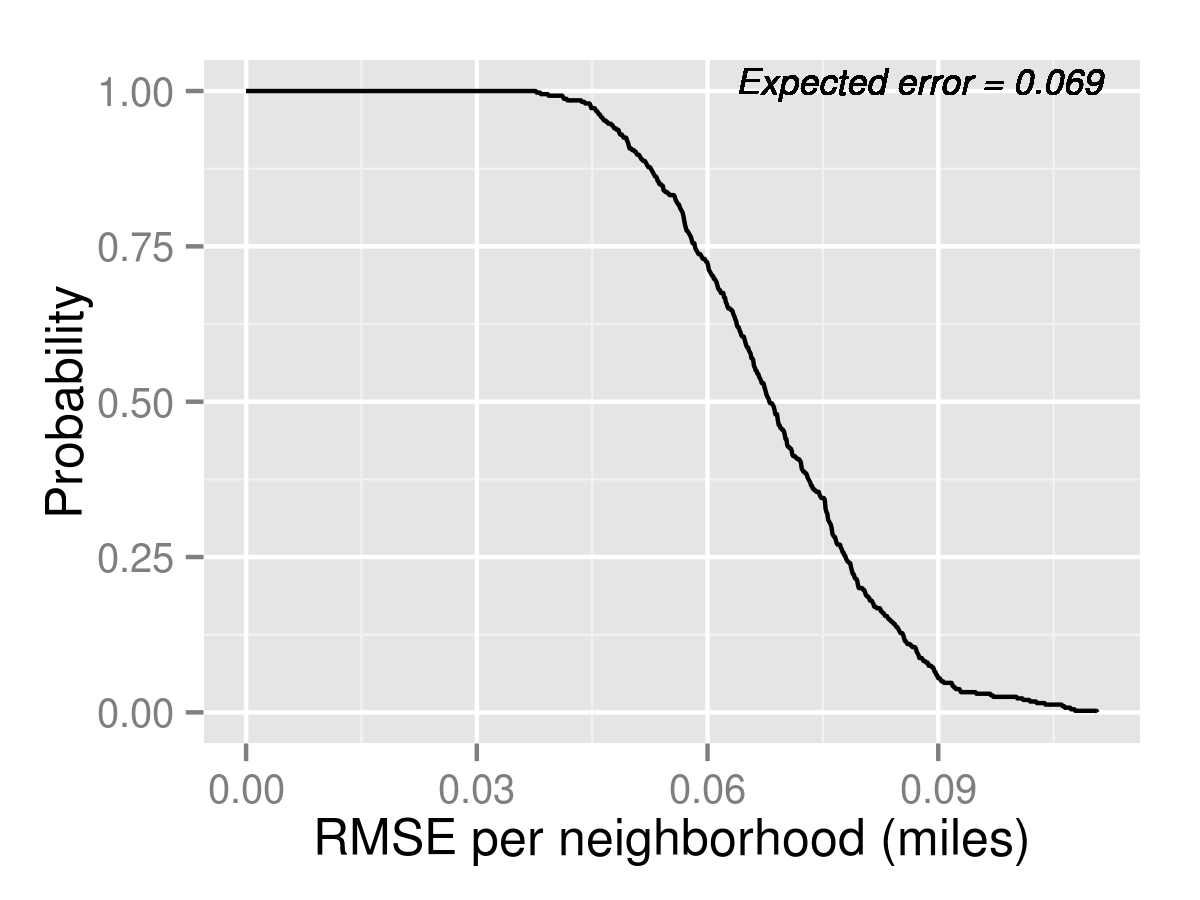}
}
\subfloat[][Distance (MLogit)]{
	\includegraphics[width=0.33\textwidth]{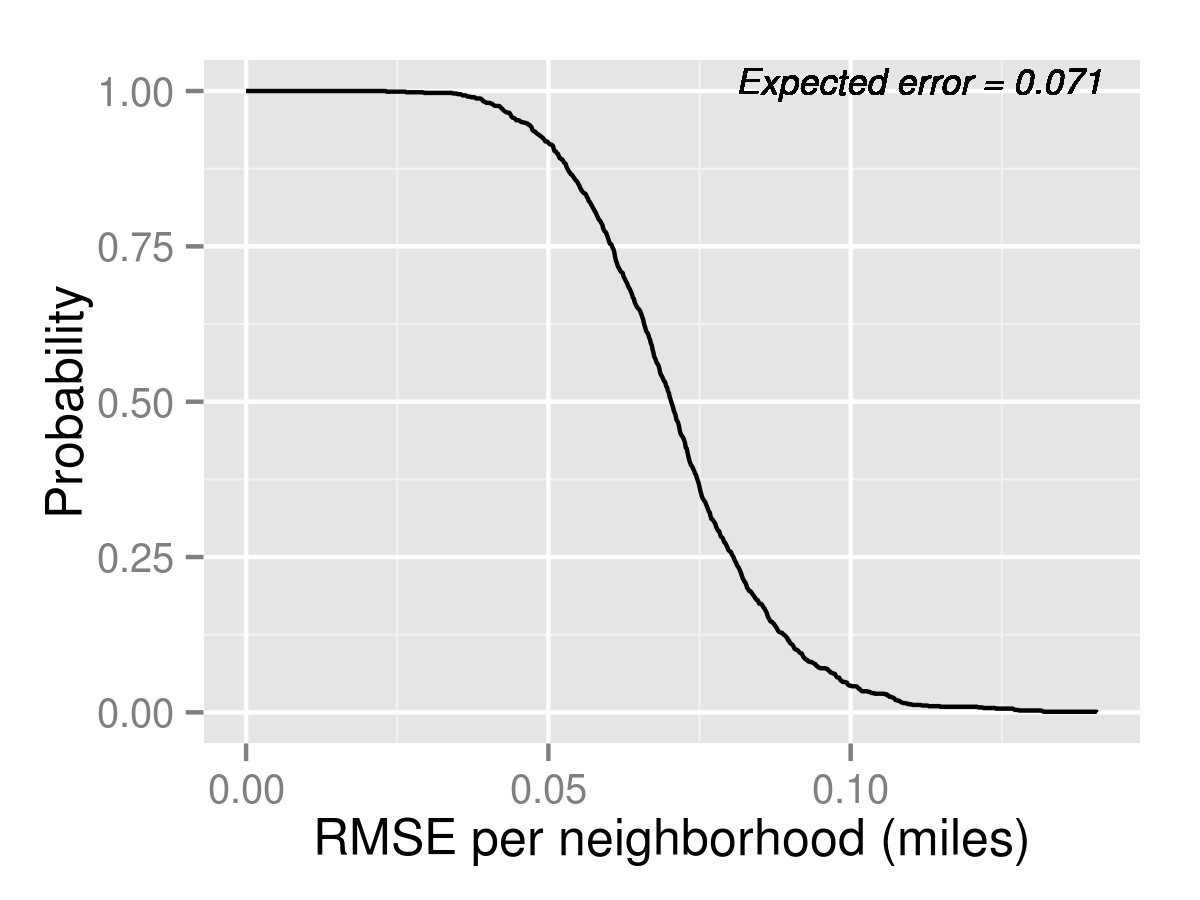}
}\\

\end{figure}

\begin{figure}[h!]
\centering
\caption{Forecasts for market shares for 2014 K2. Tail distribution plots. \label{fig:marketShare2014K2}}

\subfloat[][Top 1 (Naive)]{
 \includegraphics[width=0.33\textwidth]{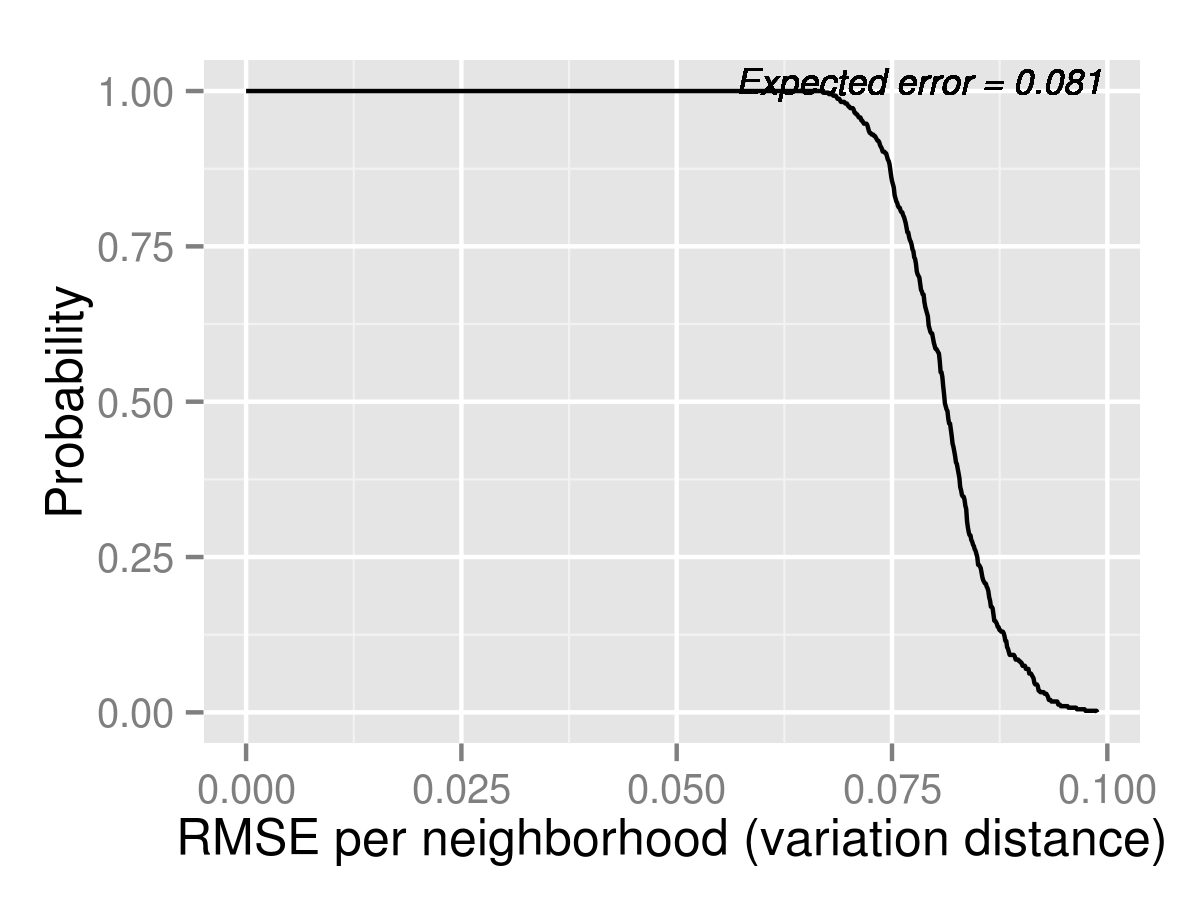}
}
\subfloat[][Top 1 (Logit)]{
	\includegraphics[width=0.33\textwidth]{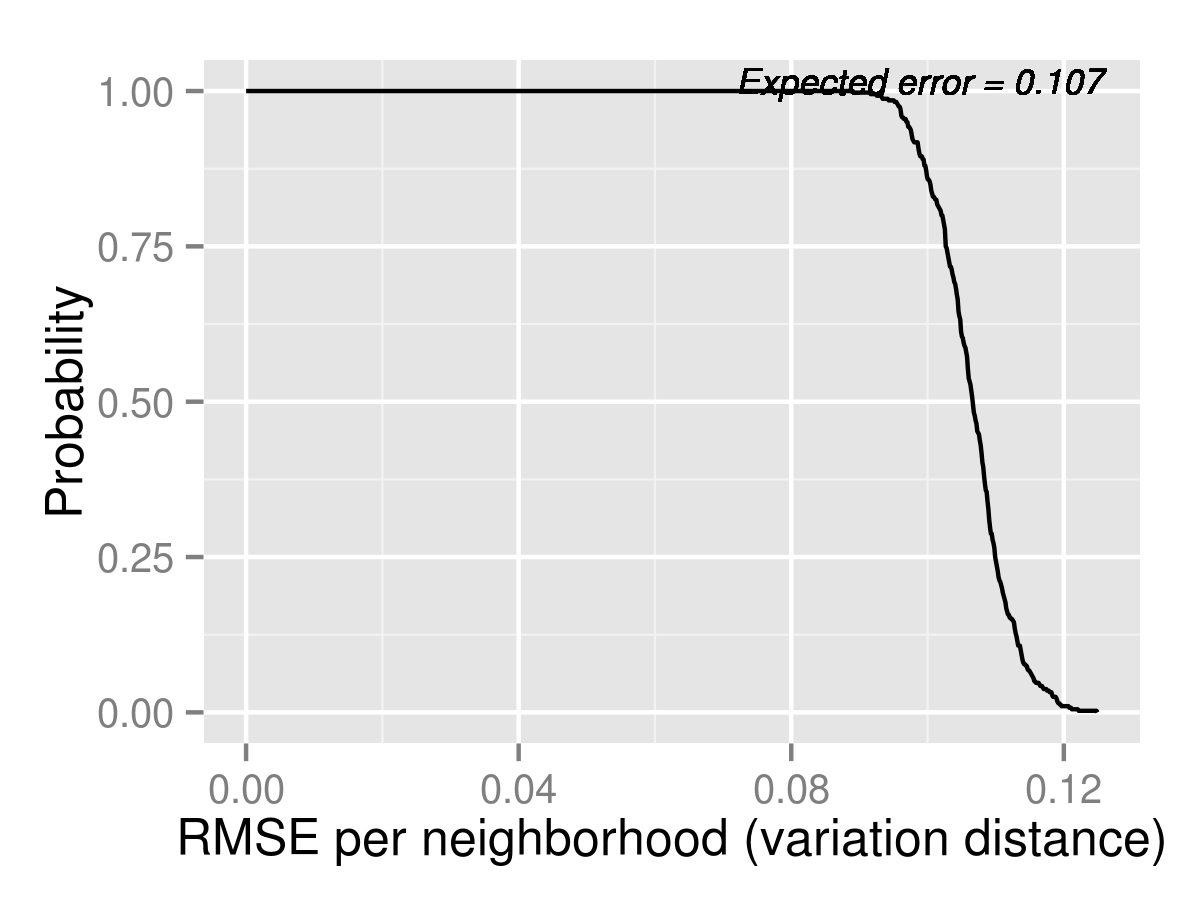}
}
\subfloat[][Top 1 (MLogit)]{
	\includegraphics[width=0.33\textwidth]{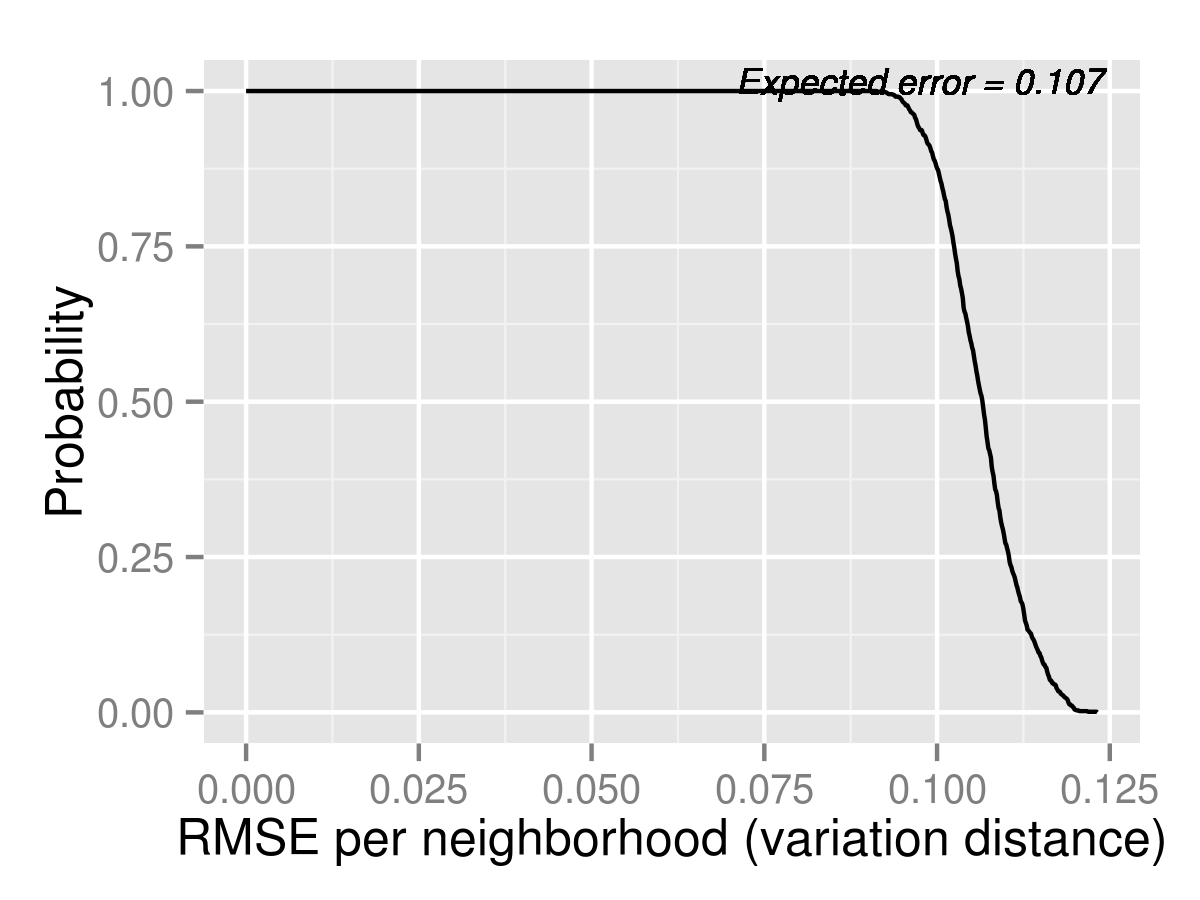}
}\\

\subfloat[][Top 2 (Naive)]{
 \includegraphics[width=0.33\textwidth]{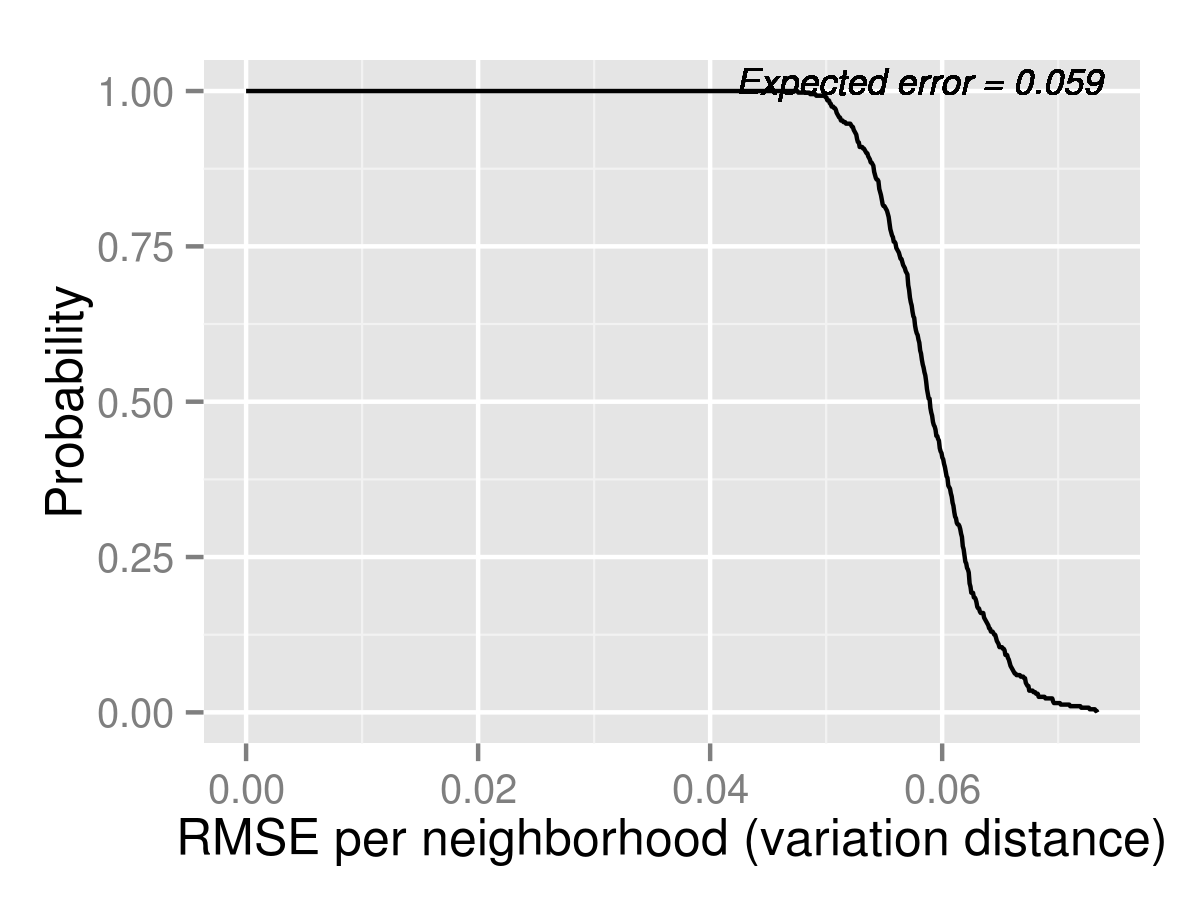}
}
\subfloat[][Top 2 (Logit)]{
	\includegraphics[width=0.33\textwidth]{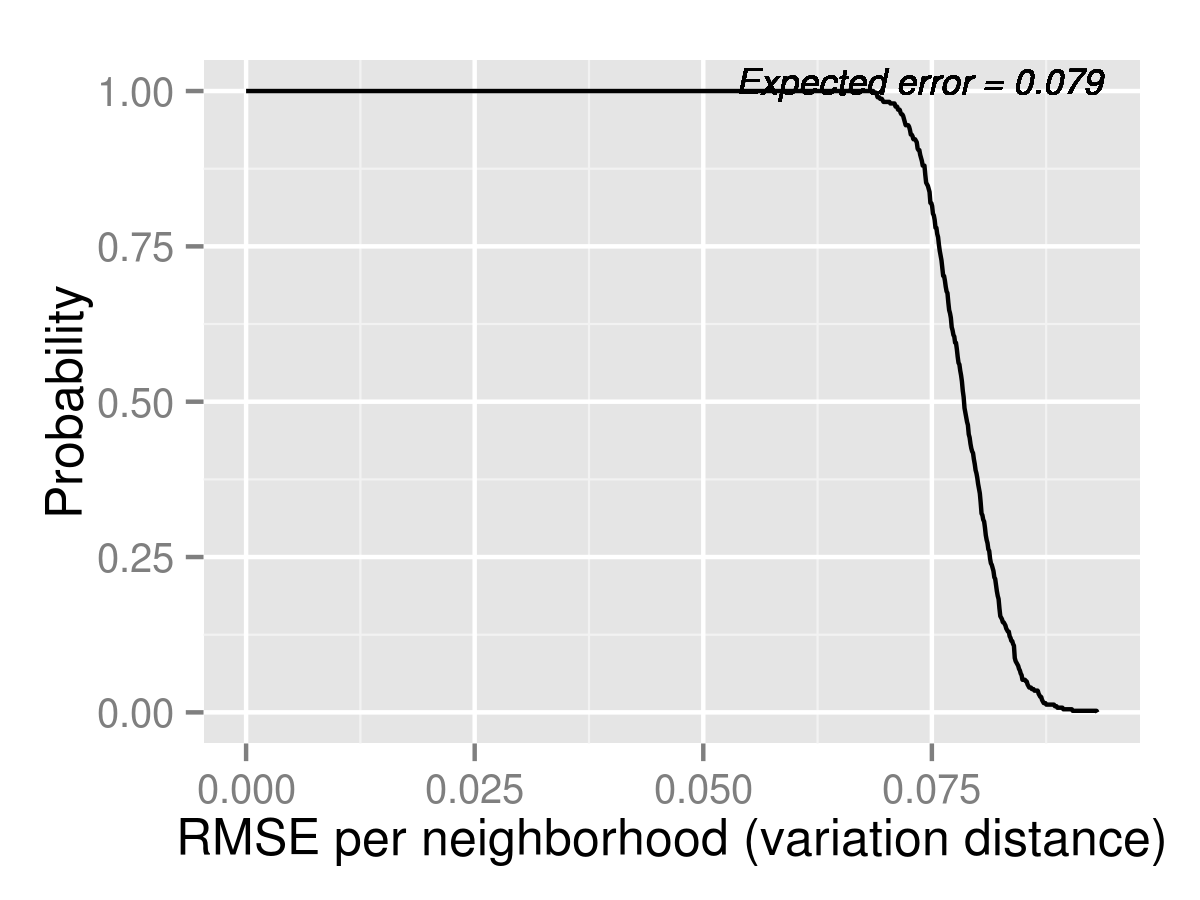}
}
\subfloat[][Top 2 (MLogit)]{
	\includegraphics[width=0.33\textwidth]{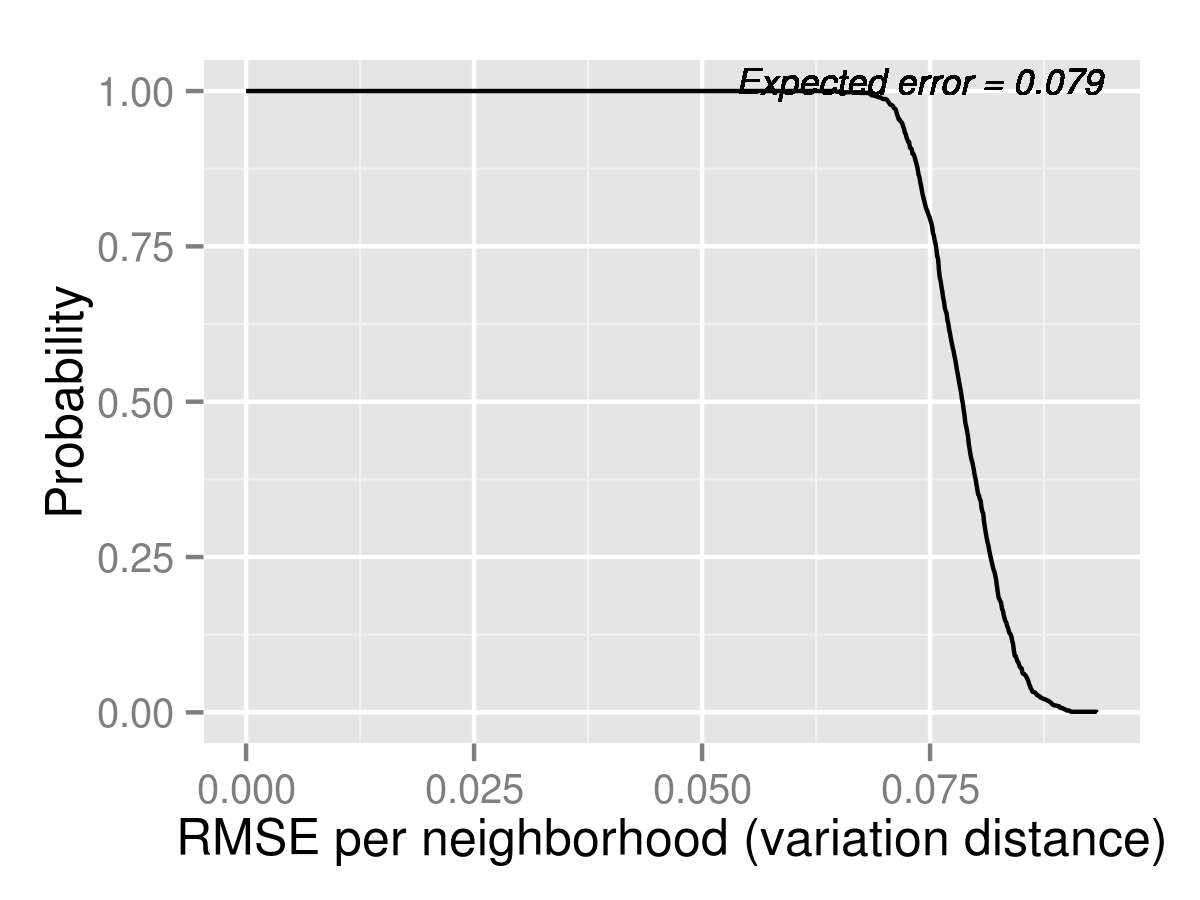}
}\\

\subfloat[][Top 3 (Naive)]{
 \includegraphics[width=0.33\textwidth]{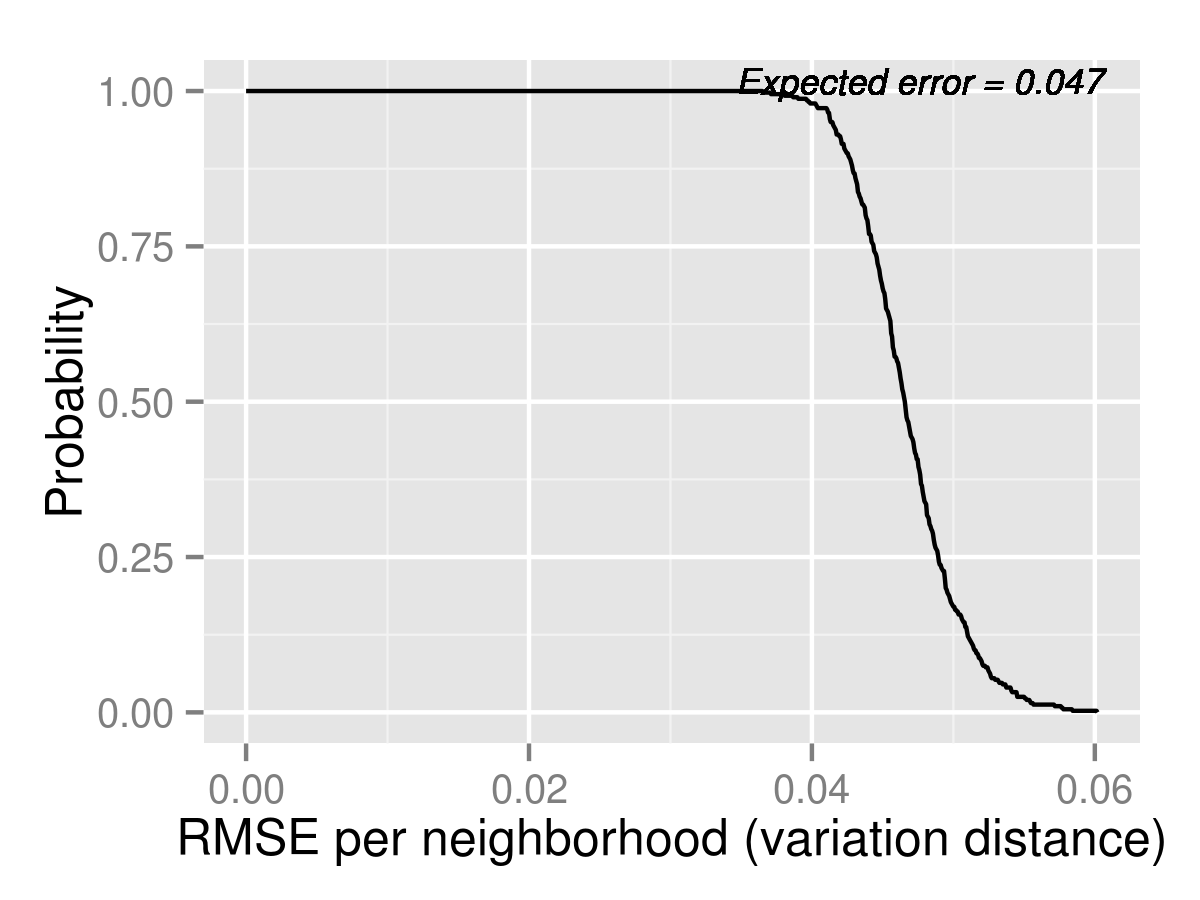}
}
\subfloat[][Top 3 (Logit)]{
	\includegraphics[width=0.33\textwidth]{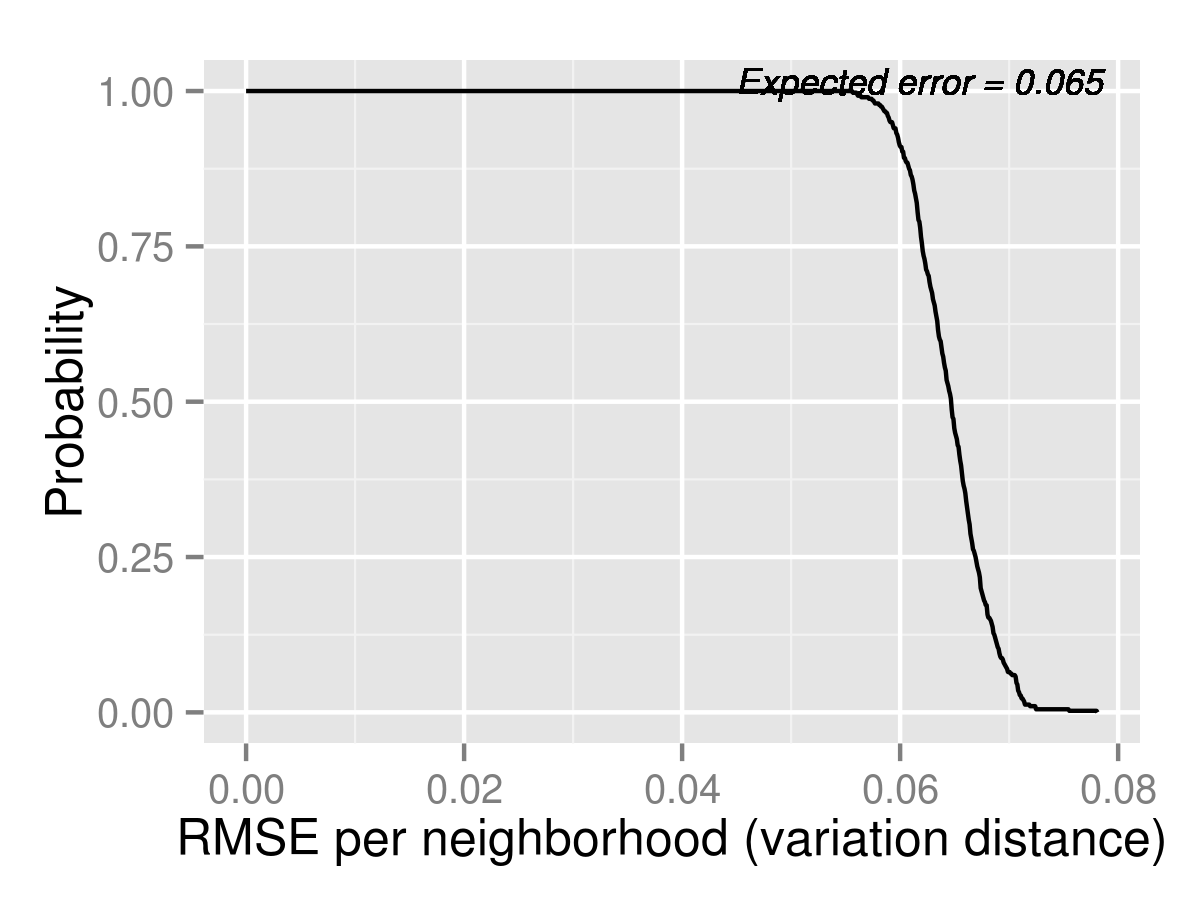}
}
\subfloat[][Top 3 (MLogit)]{
	\includegraphics[width=0.33\textwidth]{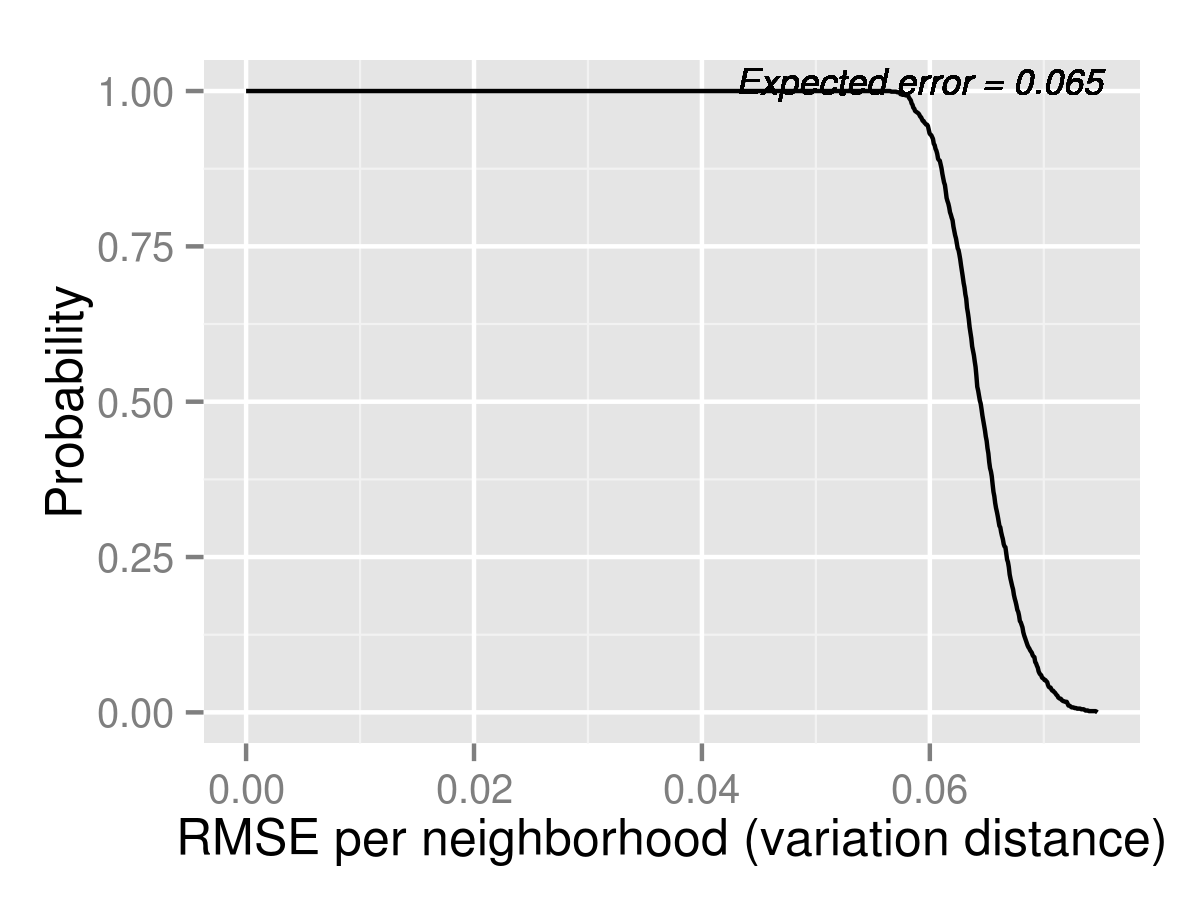}
}\\

\end{figure}

We show the predicted outcomes for 2014 K2 and the associated expected error distributions. The tail distributions of aggregate errors are in figures~\ref{fig:outcomes2014K2} and \ref{fig:marketShare2014K2}. The neighborhood by neighborhood estimates are in tables~\ref{tab:unassigned2014K2}, \ref{tab:aoq2014K2}, and \ref{tab:distance2014K2}. The K1 estimates are in the appendix. All estimates for 2014 are done using 1,000 simulations.

Suppose that families choices are driven by framing instead of underlying preferences, then Naive would do better than in 2013, and might even perform the best out of the three models. If this were the case, then it would suggest that we should re-think using demand modeling for analyzing counterfactuals, because details of how counterfactuals are presented may matter more than what we can learn from past data using demand models.  On the other hand, if framing, although possibly significant, is not crucial to the outcome, then we expect the Logit model to do equally well in 2014, reasonably accurately predicting the outcome measures of unassigned, access to quality, and distance for most neighborhoods. We would also expect MixedLogit to perform better, since it is more flexible in capturing substitution patterns. However, judging by the small differences in the back testing results for 2013, we expect the improvements to be small.

\begin{table}[!htbp]
 \centering

 \caption{Unassigned predictions for 2014 K2.}
 \label{tab:unassigned2014K2}
 \footnotesize
 \begin{tabular}{l c c c c c c}
Neighborhood & Naive & (95 \% C.I.) & Logit & (95 \% C.I.) & MLogit & (95 \% C.I.)\\ \hline  
Allston-Brighton & 4.07 & (0.00,14.00) & 1.28 & (0.00,9.00) & 1.18 & (0.00,10.00)\\ 
Charlestown & 0.68 & (0.00,8.00) & 1.25 & (0.00,10.00) & 1.38 & (0.00,11.00)\\ 
Downtown & 1.24 & (0.00,5.00) & 1.09 & (0.00,7.00) & 1.36 & (0.00,9.00)\\ 
East Boston & 47.22 & (2.00,96.00) & 49.22 & (5.00,107.00) & 49.00 & (6.00,111.03)\\ 
Hyde Park & 2.90 & (0.00,16.00) & 5.09 & (0.00,18.00) & 5.12 & (0.00,18.00)\\ 
Jamaica Plain & 18.15 & (6.00,34.00) & 1.68 & (0.00,7.00) & 2.07 & (0.00,8.00)\\ 
Mattapan & 17.73 & (3.00,37.02) & 3.86 & (0.00,18.00) & 5.17 & (0.00,20.00)\\ 
North Dorchester & 18.79 & (7.00,34.00) & 2.07 & (0.00,8.00) & 3.02 & (0.00,11.00)\\ 
Roslindale & 4.66 & (0.00,22.00) & 3.69 & (0.00,20.00) & 4.27 & (0.00,22.00)\\ 
Roxbury & 42.07 & (20.98,67.03) & 4.24 & (0.00,15.00) & 5.97 & (0.00,18.00)\\ 
South Boston & 8.93 & (1.00,20.00) & 0.35 & (0.00,4.00) & 0.63 & (0.00,5.00)\\ 
South Dorchester & 19.80 & (3.00,49.00) & 10.52 & (0.00,36.00) & 11.18 & (0.00,39.00)\\ 
South End & 9.30 & (0.00,22.00) & 4.03 & (0.00,14.00) & 4.70 & (0.00,15.00)\\ 
West Roxbury & 5.43 & (0.00,25.00) & 8.78 & (0.00,30.00) & 8.42 & (0.00,28.02)\\ 
\end{tabular}
 \end{table}

\begin{table}[!htbp]
 \centering

 \caption{Access to quality predictions for 2014 K2.}
 \label{tab:aoq2014K2}
 \footnotesize
 \begin{tabular}{l c c c c c c}
Neighborhood & Naive & (95 \% C.I.) & Logit & (95 \% C.I.) & MLogit & (95 \% C.I.)\\ \hline  
Allston-Brighton & 0.80 & (0.70,0.94) & 0.98 & (0.87,1.00) & 0.98 & (0.88,1.00)\\ 
Charlestown & 0.94 & (0.83,1.00) & 0.98 & (0.91,1.00) & 0.99 & (0.90,1.00)\\ 
Downtown & 0.85 & (0.76,0.93) & 0.92 & (0.85,0.97) & 0.91 & (0.84,0.96)\\ 
East Boston & 0.80 & (0.72,0.90) & 0.87 & (0.77,0.98) & 0.88 & (0.77,0.98)\\ 
Hyde Park & 0.69 & (0.59,0.81) & 0.86 & (0.76,0.94) & 0.86 & (0.76,0.94)\\ 
Jamaica Plain & 0.68 & (0.60,0.78) & 0.89 & (0.78,1.00) & 0.90 & (0.77,1.00)\\ 
Mattapan & 0.59 & (0.50,0.69) & 0.97 & (0.85,1.00) & 0.96 & (0.83,1.00)\\ 
North Dorchester & 0.50 & (0.42,0.58) & 0.77 & (0.63,0.94) & 0.77 & (0.63,0.93)\\ 
Roslindale & 0.74 & (0.66,0.84) & 0.98 & (0.89,1.00) & 0.98 & (0.88,1.00)\\ 
Roxbury & 0.56 & (0.50,0.63) & 0.83 & (0.72,0.93) & 0.83 & (0.72,0.92)\\ 
South Boston & 0.48 & (0.40,0.57) & 0.72 & (0.61,0.86) & 0.70 & (0.58,0.82)\\ 
South Dorchester & 0.60 & (0.52,0.68) & 0.87 & (0.76,0.97) & 0.86 & (0.75,0.96)\\ 
South End & 0.67 & (0.59,0.74) & 0.82 & (0.73,0.92) & 0.80 & (0.72,0.89)\\ 
West Roxbury & 0.78 & (0.70,0.88) & 0.89 & (0.81,0.95) & 0.89 & (0.81,0.95)\\ 
\end{tabular}
 \end{table}

 \begin{table}[!htbp]
 \centering
  \caption{Distance predictions for 2014 K2.}
  \label{tab:distance2014K2}
 \footnotesize
 \begin{tabular}{l c c c c c c}
Neighborhood & Naive & (95 \% C.I.) & Logit & (95 \% C.I.) & MLogit & (95 \% C.I.)\\ \hline  
Allston-Brighton & 1.44 & (1.23,1.66) & 1.27 & (1.10,1.47) & 1.28 & (1.09,1.50)\\ 
Charlestown & 0.97 & (0.70,1.26) & 0.94 & (0.76,1.12) & 0.95 & (0.78,1.14)\\ 
Downtown & 1.30 & (1.06,1.64) & 1.23 & (1.08,1.40) & 1.23 & (1.08,1.39)\\ 
East Boston & 1.75 & (1.60,1.93) & 1.24 & (1.05,1.44) & 1.27 & (1.06,1.51)\\ 
Hyde Park & 2.04 & (1.89,2.20) & 1.80 & (1.67,1.94) & 1.78 & (1.64,1.92)\\ 
Jamaica Plain & 1.29 & (1.19,1.39) & 1.17 & (1.07,1.26) & 1.16 & (1.07,1.26)\\ 
Mattapan & 1.80 & (1.69,1.93) & 1.71 & (1.61,1.83) & 1.71 & (1.60,1.84)\\ 
North Dorchester & 1.29 & (1.17,1.42) & 1.17 & (1.08,1.27) & 1.17 & (1.07,1.27)\\ 
Roslindale & 1.73 & (1.62,1.82) & 1.53 & (1.44,1.63) & 1.52 & (1.43,1.62)\\ 
Roxbury & 1.37 & (1.28,1.46) & 1.21 & (1.15,1.28) & 1.21 & (1.15,1.28)\\ 
South Boston & 1.46 & (1.26,1.69) & 1.21 & (1.06,1.37) & 1.20 & (1.04,1.35)\\ 
South Dorchester & 1.52 & (1.43,1.61) & 1.43 & (1.35,1.51) & 1.43 & (1.35,1.51)\\ 
South End & 1.41 & (1.25,1.58) & 1.30 & (1.15,1.44) & 1.28 & (1.14,1.43)\\ 
West Roxbury & 1.89 & (1.76,2.02) & 1.70 & (1.57,1.83) & 1.69 & (1.56,1.82)\\ 
\end{tabular}
 \end{table}

\clearpage

\section{Conclusion}

This paper reports on ex ante forecasts using a discrete choice model of demand of a large-scale
change to choice menus occurring in Boston's school choice plan in 2014.  It also reports on a simpler
statistical forecast that is not based on an underlying random utility model.
The methodology and target outcomes are described before information on new preferences
is available to avoid any scope for post-analysis bias.
Part II of this report will
revisit these forecasts using data from the new system to assess the strengths and limitations of discrete choice
models of demand in our context.

\clearpage
\newpage

\appendix

\section{Detailed Participation Forecasts}
\label{app:populationForecast}
\begin{table}[!htbp]
 \centering
 \caption{Forecast for total number of new applicants in 2013.}
 \label{}
 \footnotesize
 \begin{tabular}{l c c c}
Year & Predicted Mean & Standard Deviation & Actual\\ \hline
2013 & 5893.30 & 382.50 & 5255\\
\end{tabular}
 \end{table}

 \begin{table}[!htbp]
 \centering
 \label{}
 \caption{Forecast for proportion of the new applicants distributed to each grade and neighborhood in 2013.}
 \footnotesize
 \begin{tabular}{l c c c| c c c}
 & Predicted Mean & Standard Dev. & Actual & Predicted Mean & Standard Dev. & Actual\\ \hline
Neighborhood & \multicolumn{3}{c|}{K1}   & \multicolumn{3}{c}{K2} \\

Allston-Brighton & 0.0184 & 0.0010 & 0.0156 & 0.0148 & 0.0027 & 0.0156\\
Charlestown & 0.0078 & 0.0019 & 0.0118 & 0.0196 & 0.0002 & 0.0171\\
Downtown & 0.0219 & 0.0001 & 0.0156 & 0.0135 & 0.0020 & 0.0179\\
East Boston & 0.0543 & 0.0053 & 0.0504 & 0.0603 & 0.0053 & 0.0500\\
Hyde Park & 0.0313 & 0.0025 & 0.0322 & 0.0205 & 0.0010 & 0.0207\\
Jamaica Plain & 0.0310 & 0.0020 & 0.0379 & 0.0207 & 0.0002 & 0.0221\\
Mattapan & 0.0311 & 0.0028 & 0.0297 & 0.0258 & 0.0024 & 0.0272\\
North Dorchester & 0.0244 & 0.0038 & 0.0245 & 0.0187 & 0.0038 & 0.0154\\
Roslindale & 0.0452 & 0.0032 & 0.0447 & 0.0260 & 0.0033 & 0.0287\\
Roxbury & 0.0569 & 0.0048 & 0.0556 & 0.0521 & 0.0033 & 0.0525\\
South Boston & 0.0137 & 0.0012 & 0.0131 & 0.0139 & 0.0015 & 0.0135\\
South Dorchester & 0.0617 & 0.0040 & 0.0668 & 0.0453 & 0.0040 & 0.0481\\
South End & 0.0187 & 0.0023 & 0.0198 & 0.0207 & 0.0001 & 0.0206\\
West Roxbury & 0.0420 & 0.0033 & 0.0373 & 0.0226 & 0.0028 & 0.0217\\
\end{tabular}
 \end{table}

\begin{table}[!htbp]
 \centering
 \label{}
 \caption{Forecast for continuing ratios for each grade and neighborhood in 2013.}
 \footnotesize
 \begin{tabular}{l c c c| c c c}
& Predicted Mean & Standard Dev. & Actual & Predicted Mean & Standard Dev. & Actual\\ \hline
Neighborhood & \multicolumn{3}{c|}{K1}   & \multicolumn{3}{c}{K2} \\
Allston-Brighton & 0.83 & 0.02 & 0.76 & 0.78 & 0.08 & 0.76\\
Charlestown & 0.88 & 0.18 & 0.40 & 0.76 & 0.12 & 0.77\\
Downtown & 0.63 & 0.32 & 0.17 & 0.71 & 0.02 & 0.74\\
East Boston & 0.87 & 0.09 & 0.78 & 0.85 & 0.04 & 0.86\\
Hyde Park & 0.77 & 0.18 & 0.55 & 0.80 & 0.05 & 0.77\\
Jamaica Plain & 1 & 0 & 0.31 & 0.91 & 0.00 & 0.73\\
Mattapan & 0.92 & 0.02 & 0.58 & 0.79 & 0.02 & 0.81\\
North Dorchester & 0.76 & 0.23 & 0.59 & 0.77 & 0.14 & 0.79\\
Roslindale & 0.70 & 0.34 & 0.40 & 0.80 & 0.04 & 0.83\\
Roxbury & 1 & 0 & 0.38 & 0.78 & 0.03 & 0.81\\
South Boston & 0.75 & 0.35 & 0.08 & 0.78 & 0.09 & 0.79\\
South Dorchester & 0.70 & 0.05 & 0.61 & 0.79 & 0.04 & 0.81\\
South End & 0.65 & 0.32 & 0.80 & 0.83 & 0.02 & 0.71\\
West Roxbury & 1 & 0 & 0.22 & 0.79 & 0.05 & 0.84\\
\end{tabular}
 \end{table}

 \begin{table}[!htbp]
 \centering
 \label{}
 \caption{Forecast for total number of new applicants in 2014.}
 \footnotesize
 \begin{tabular}{l c c }
 & Predicted Mean & Standard Deviation\\ \hline
Total New Applicants & 5798.5 & 366.4\\
\end{tabular}
 \end{table}

 \begin{table}[!htbp]
 \centering
 \label{}
 \caption{Forecast for proportion of the new applicants distributed to each grade and neighborhood in 2014.}
 \footnotesize
 \begin{tabular}{l c c| c c}
  & Predicted Mean & Standard Deviation & Predicted Mean & Standard Deviation\\ \hline
Neighborhood & \multicolumn{2}{c|}{K1}  & \multicolumn{2}{c|}{K2} \\
Allston-Brighton & 0.0177 & 0.0016 & 0.0150 & 0.0022\\
Charlestown & 0.0136 & 0.0006 & 0.0190 & 0.0013\\
Downtown & 0.0155 & 0.0027 & 0.0198 & 0.0006\\
East Boston & 0.0533 & 0.0047 & 0.0577 & 0.0067\\
Hyde Park & 0.0315 & 0.0021 & 0.0206 & 0.0008\\
Jamaica Plain & 0.0327 & 0.0038 & 0.0210 & 0.0007\\
Mattapan & 0.0308 & 0.0024 & 0.0262 & 0.0021\\
North Dorchester & 0.0245 & 0.0031 & 0.0179 & 0.0035\\
Roslindale & 0.0451 & 0.0027 & 0.0267 & 0.0030\\
Roxbury & 0.0566 & 0.0040 & 0.0522 & 0.0027\\
South Boston & 0.0136 & 0.0010 & 0.0138 & 0.0013\\
South Dorchester & 0.0630 & 0.0041 & 0.0460 & 0.0036\\
South End & 0.0190 & 0.0020 & 0.0207 & 0.0001\\
West Roxbury & 0.0408 & 0.0036 & 0.0223 & 0.0023\\
\end{tabular}
 \end{table}

 \begin{table}[!htbp]
 \centering
 \label{}
 \caption{Forecast for continuing ratios for each grade and neighborhood in 2014.}
 \footnotesize
 \begin{tabular}{l c c| c c}
& Predicted Mean & Standard Deviation & Predicted Mean & Standard Deviation\\ \hline
Neighborhood & \multicolumn{2}{c|}{K1}  & \multicolumn{2}{c|}{K2} \\
Allston-Brighton & 0.81 & 0.04 & 0.78 & 0.07\\
Charlestown & 0.72 & 0.30 & 0.76 & 0.10\\
Downtown & 0.52 & 0.35 & 0.72 & 0.02\\
East Boston & 0.85 & 0.09 & 0.85 & 0.03\\
Hyde Park & 0.72 & 0.19 & 0.79 & 0.04\\
Jamaica Plain & 0.77 & 0.40 & 0.78 & 0.06\\
Mattapan & 0.83 & 0.17 & 0.80 & 0.02\\
North Dorchester & 0.71 & 0.20 & 0.77 & 0.11\\
Roslindale & 0.63 & 0.32 & 0.81 & 0.04\\
Roxbury & 0.84 & 0.31 & 0.79 & 0.03\\
South Boston & 0.53 & 0.46 & 0.78 & 0.08\\
South Dorchester & 0.68 & 0.06 & 0.79 & 0.03\\
South End & 0.69 & 0.27 & 0.80 & 0.06\\
West Roxbury & 0.81 & 0.39 & 0.80 & 0.05\\
\end{tabular}
 \end{table}

\section{Estimates for K1.}
\label{app:resultsK1}

\subsection{Back Testing Forecasts for 2013 K1.}
\label{sec:results2013K1}

\begin{figure}[h!]
\centering
\caption{Back testing predictions for assignment outcomes for 2013 K1. Tail distribution plots. \label{fig:outcomes2013K1}}

\subfloat[][Unassigned (Naive)]{
 \includegraphics[width=0.33\textwidth]{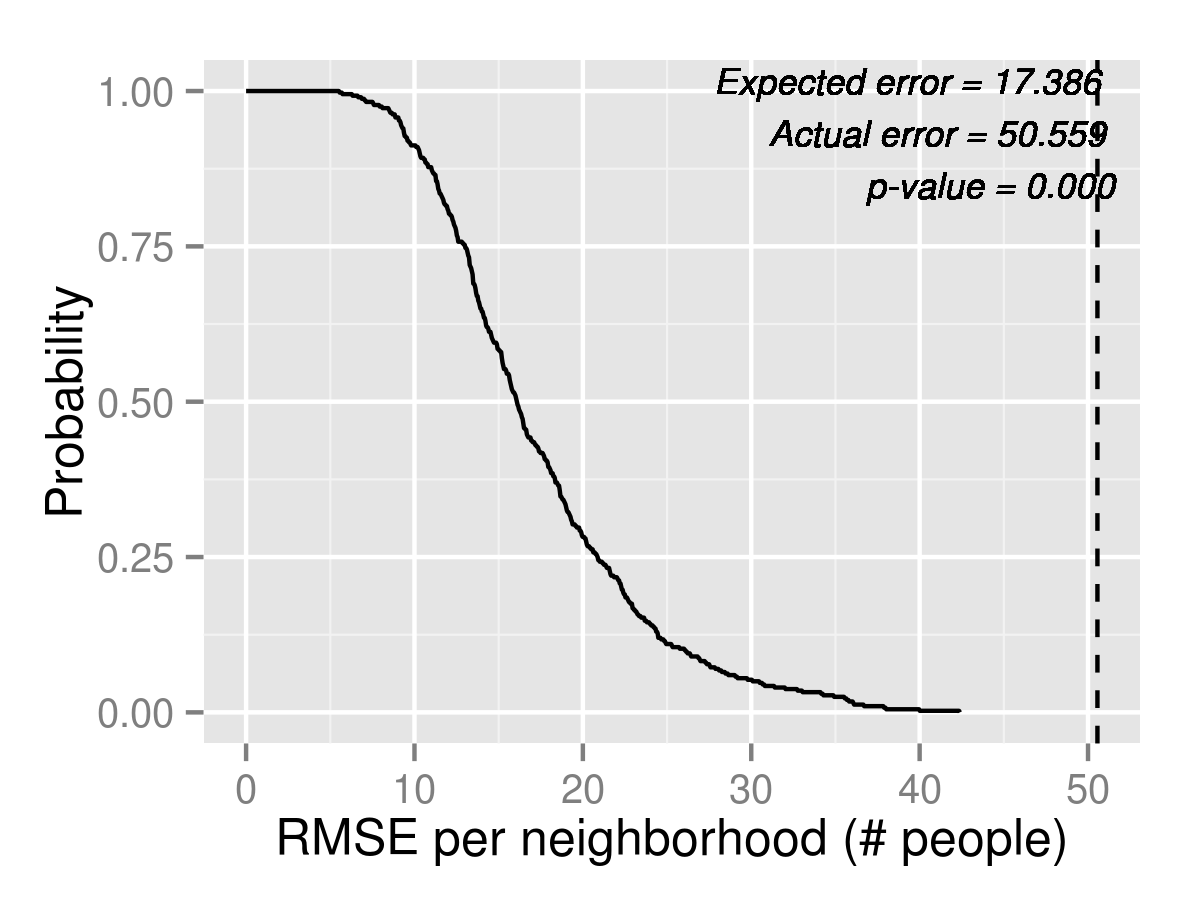}
}
\subfloat[][Unassigned (Logit)]{
	\includegraphics[width=0.33\textwidth]{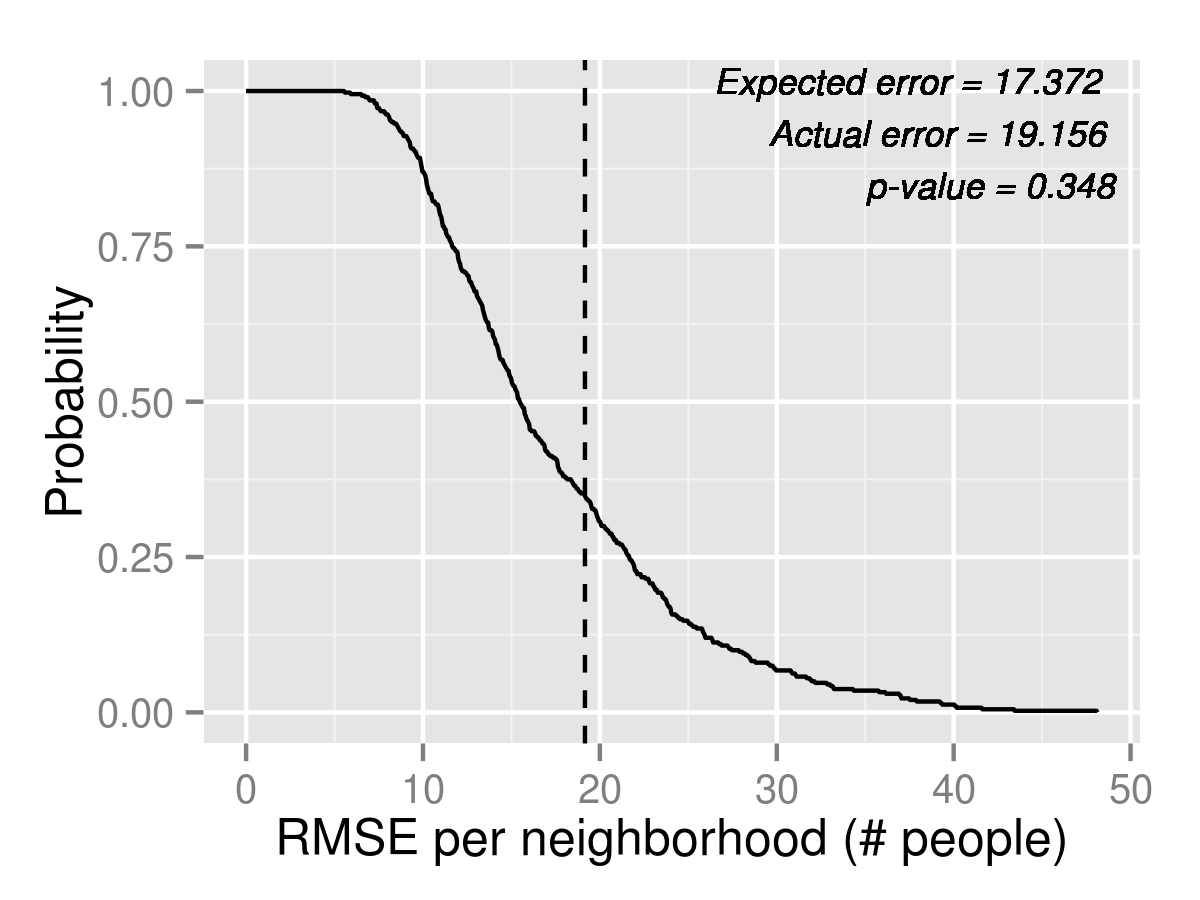}
}
\subfloat[][Unassigned (MLogit)]{
	\includegraphics[width=0.33\textwidth]{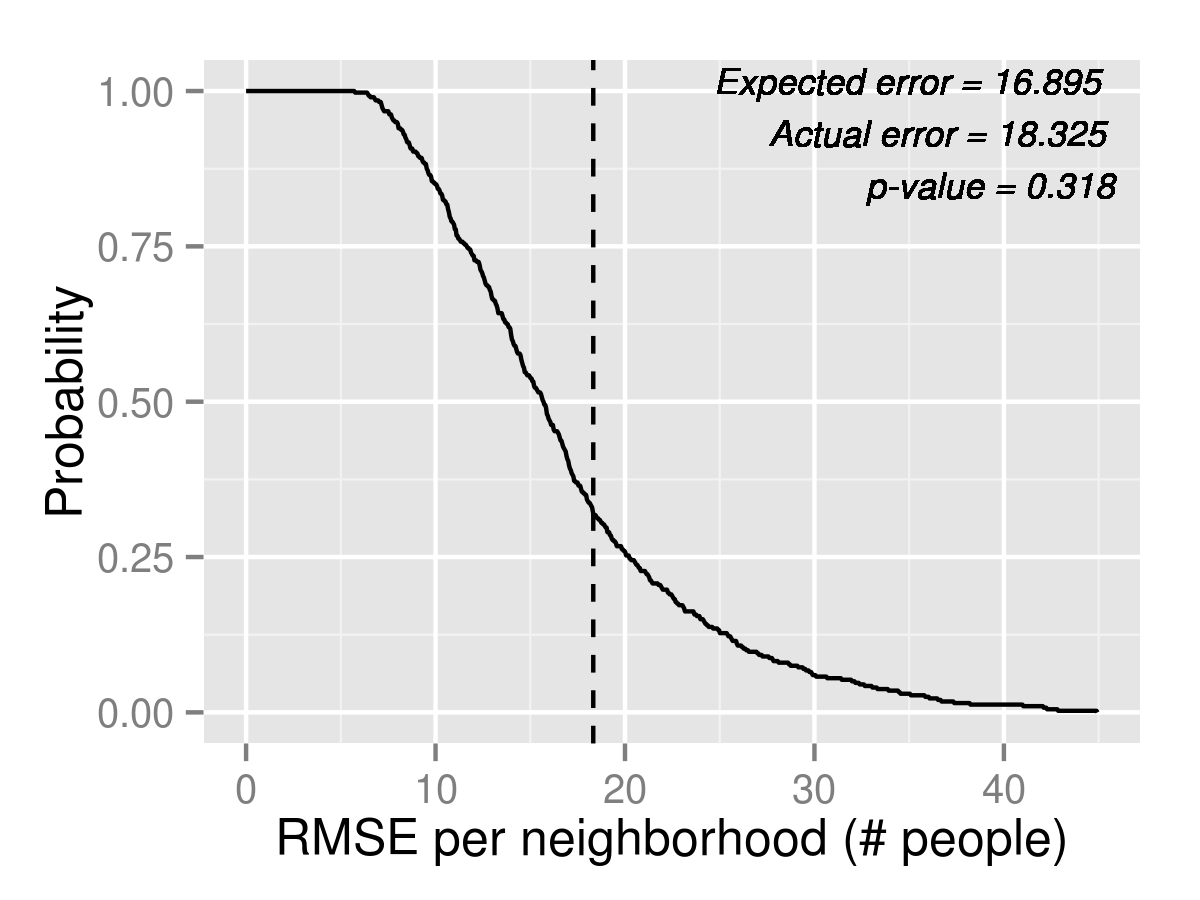}
}\\

\subfloat[][Access to Quality (Naive)]{
 \includegraphics[width=0.33\textwidth]{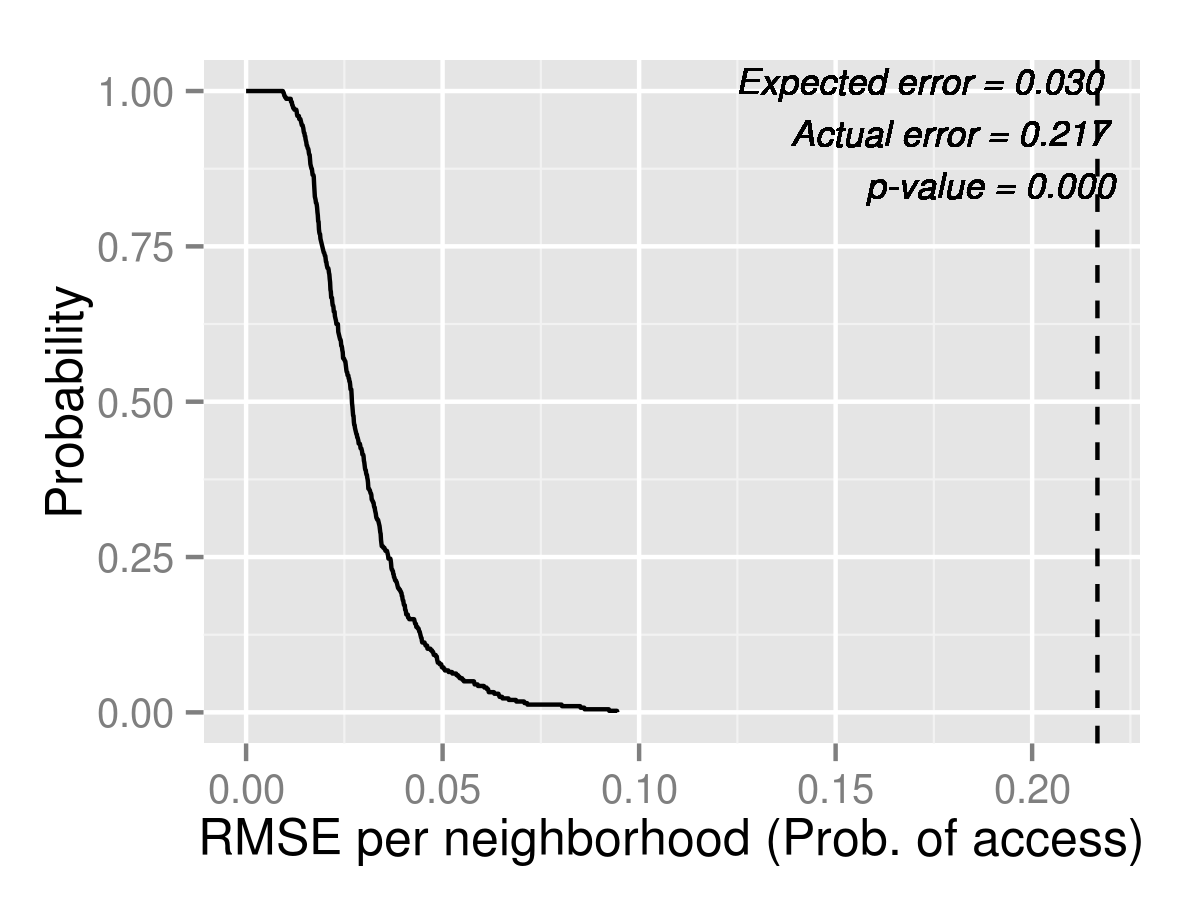}
}
\subfloat[][Access to Quality (Logit)]{
	\includegraphics[width=0.33\textwidth]{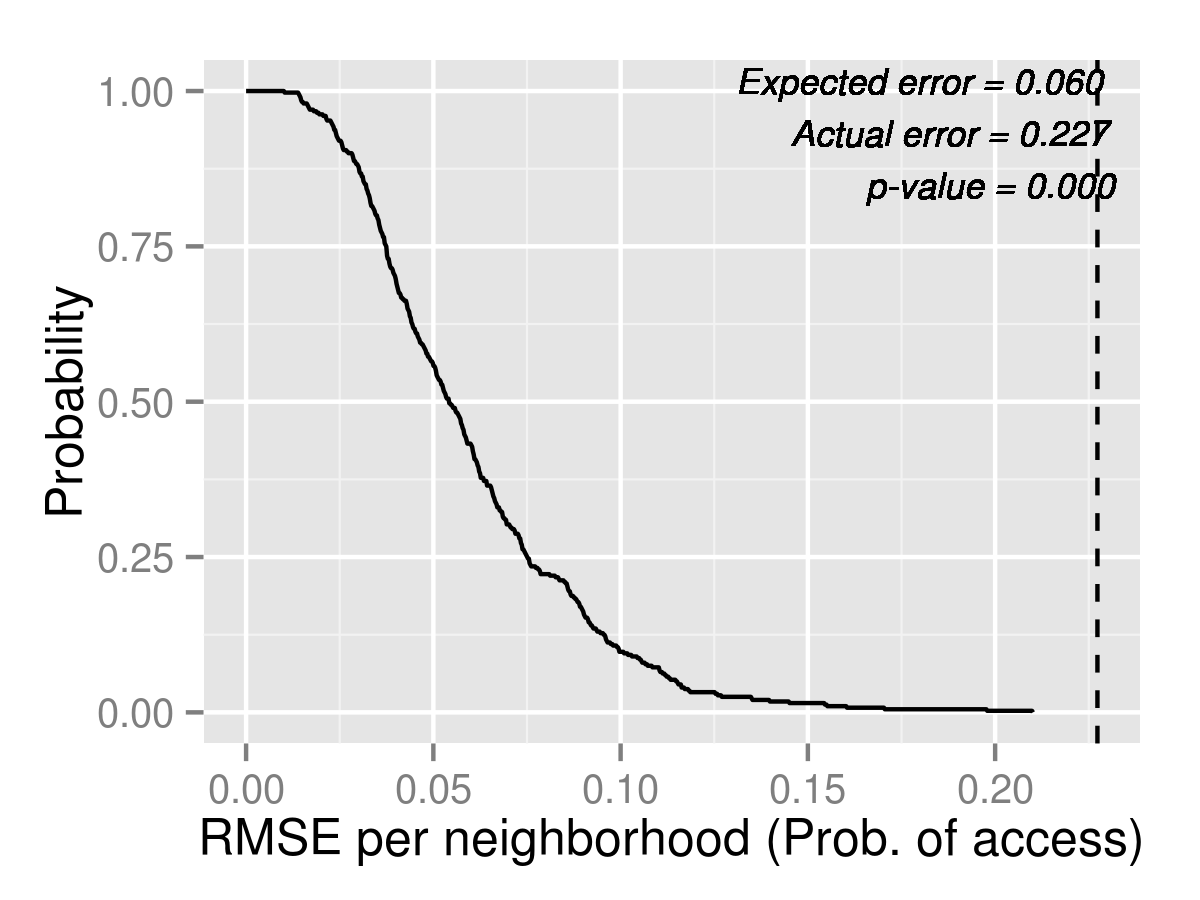}
}
\subfloat[][Access to Quality (MLogit)]{
	\includegraphics[width=0.33\textwidth]{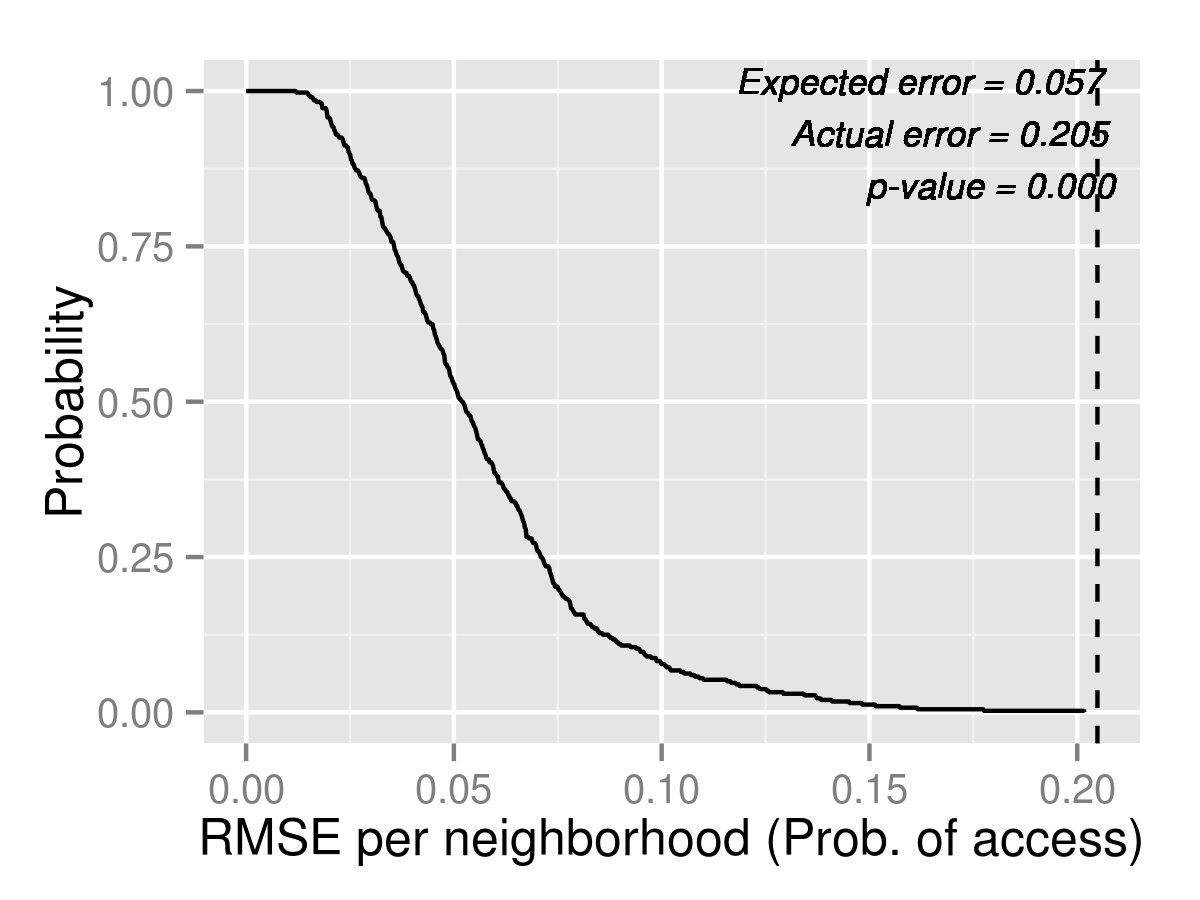}
}\\

\subfloat[][Distance (Naive)]{
 \includegraphics[width=0.33\textwidth]{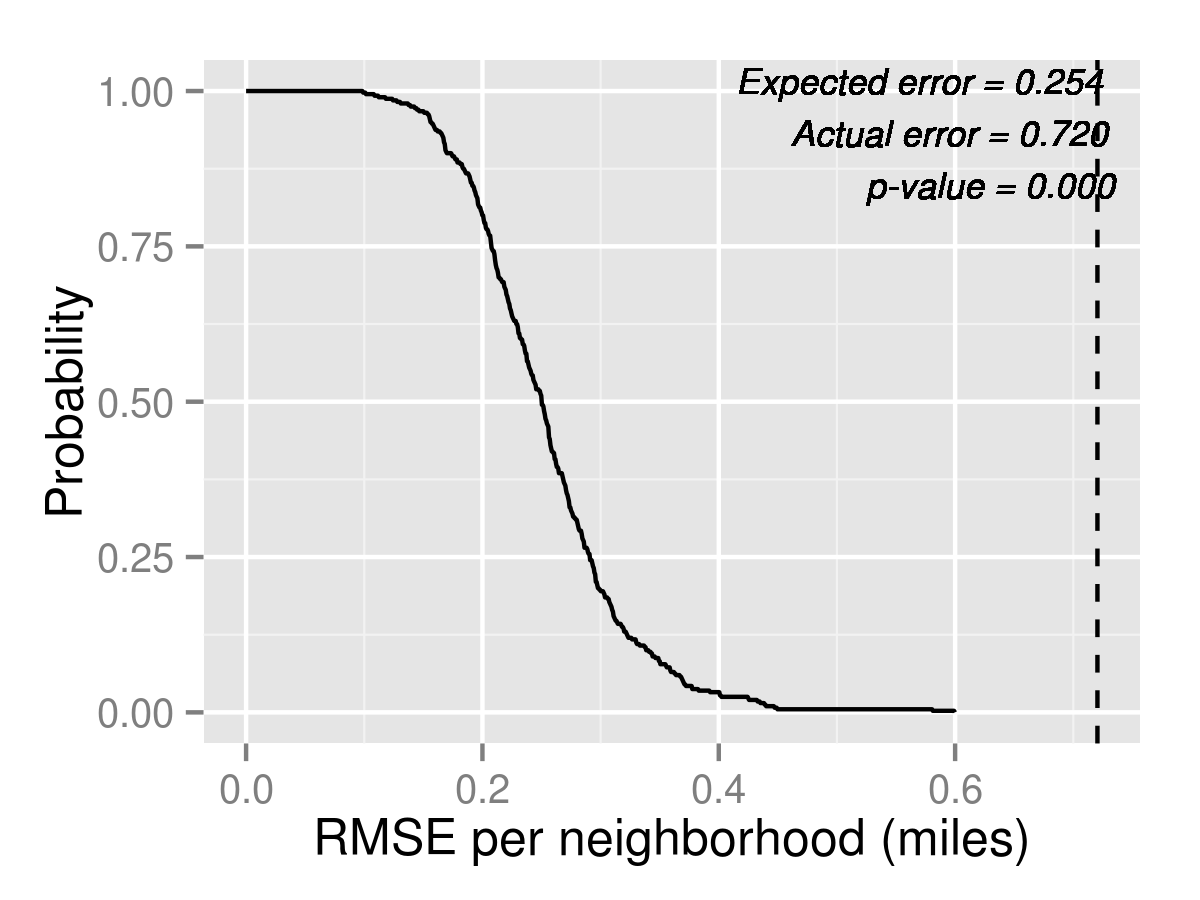}
}
\subfloat[][Distance (Logit)]{
	\includegraphics[width=0.33\textwidth]{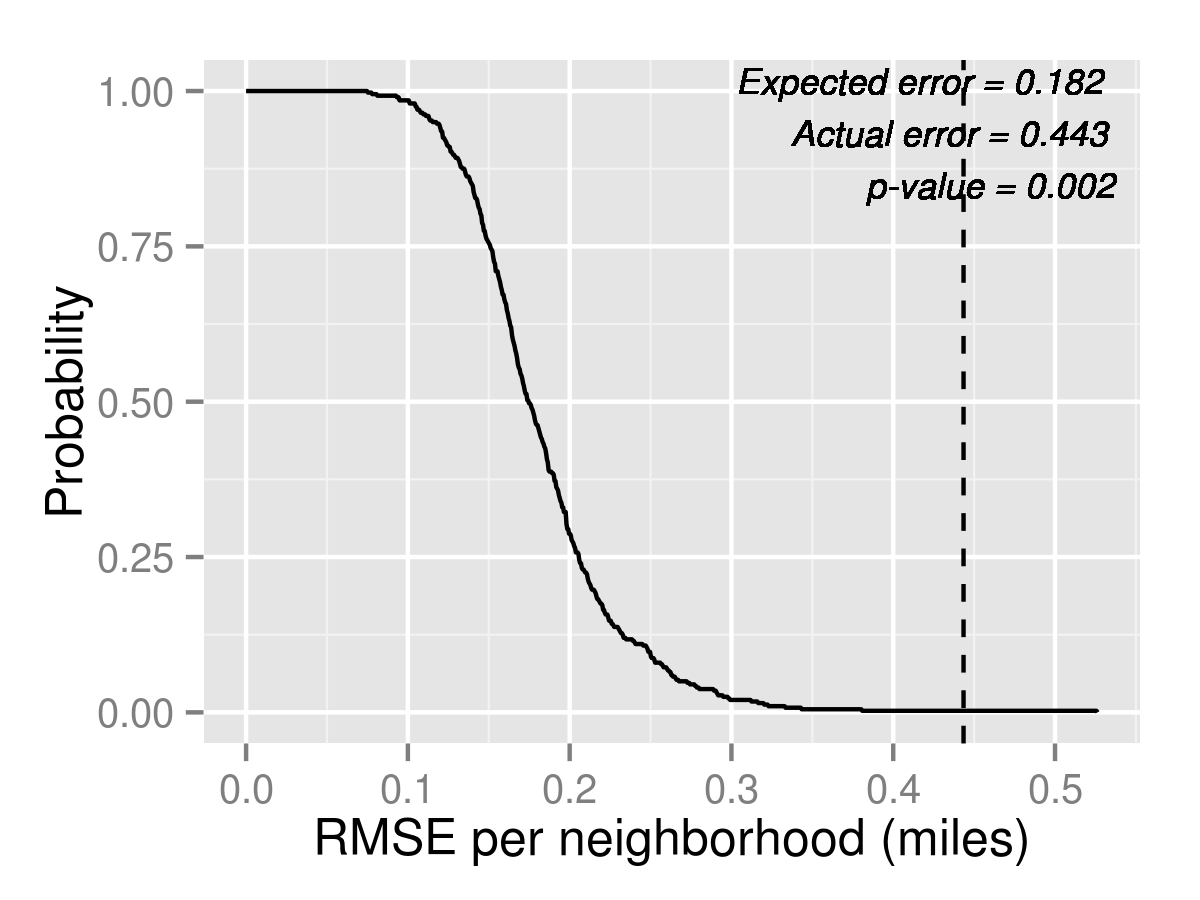}
}
\subfloat[][Distance (MLogit)]{
	\includegraphics[width=0.33\textwidth]{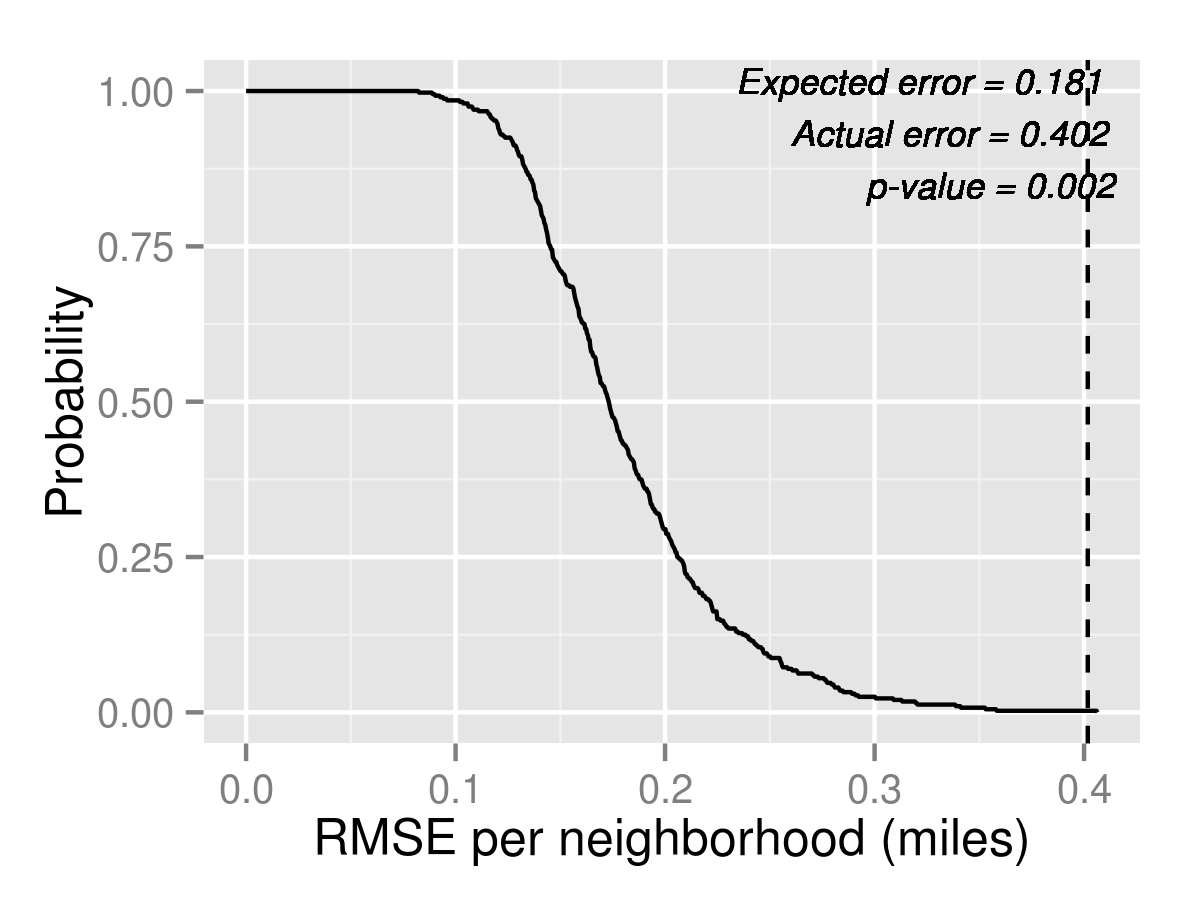}
}\\

\end{figure}

As with K2, Table~\ref{tab:unassigned2013K1} shows that the Logit model predicts unassigned for most neighborhoods better than the Naive model. MixedLogit predicts similarly as Logit but judging by the Root Mean Square Error (RMSE) in Figure~\ref{fig:outcomes2013K1}, MixedLogit performs slightly better.

For access to quality, shown in Table~\ref{tab:aoq2013K1}, Logit and MixedLogit do worse relative to how they did for grade K2. It turns out that in the actual data, one of the zones, the North zone, had left-over seats in a Tier 1 or 2 school, which makes access to quality equal to one for all neighborhoods in that zone. This is not predicted by Logit or MixedLogit, which fill every Tier 1 or 2 seat. However, the naive model predicts this correctly, while it under-predicts the access in all the other neighborhoods.
For distance, shown in Table~\ref{tab:distance2013K1}, the predictions for Naive are uniformly too high, while the predictions for Logit and MixedLogit are reasonably close, except that they over-predict distance for Allston-Brighton, Charlestown, Jamaica Plain, Roxbury, and West Roxbury.

For market shares, all models again fail to explain the data, although the error is 2-3 times smaller in Logit and MixedLogit compared to Naive. MixedLogit slightly decreases the actual RMSE error over Logit in all cases.

\begin{figure}[h!]
\centering
\caption{Back testing predictions for market shares for 2013 K1. Tail distribution plots. \label{fig:marketShare2013K1}}

\subfloat[][Top 1 (Naive)]{
 \includegraphics[width=0.33\textwidth]{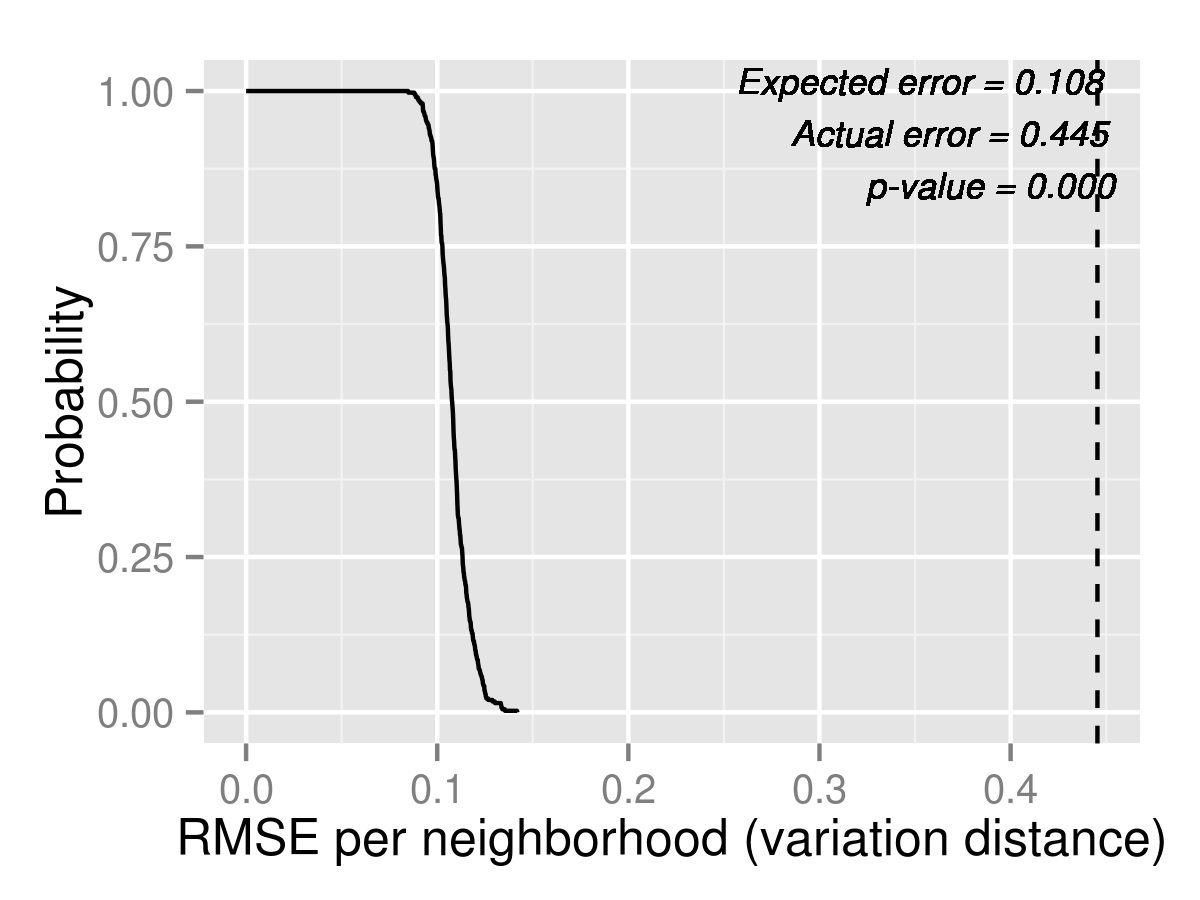}
}
\subfloat[][Top 1 (Logit)]{
	\includegraphics[width=0.33\textwidth]{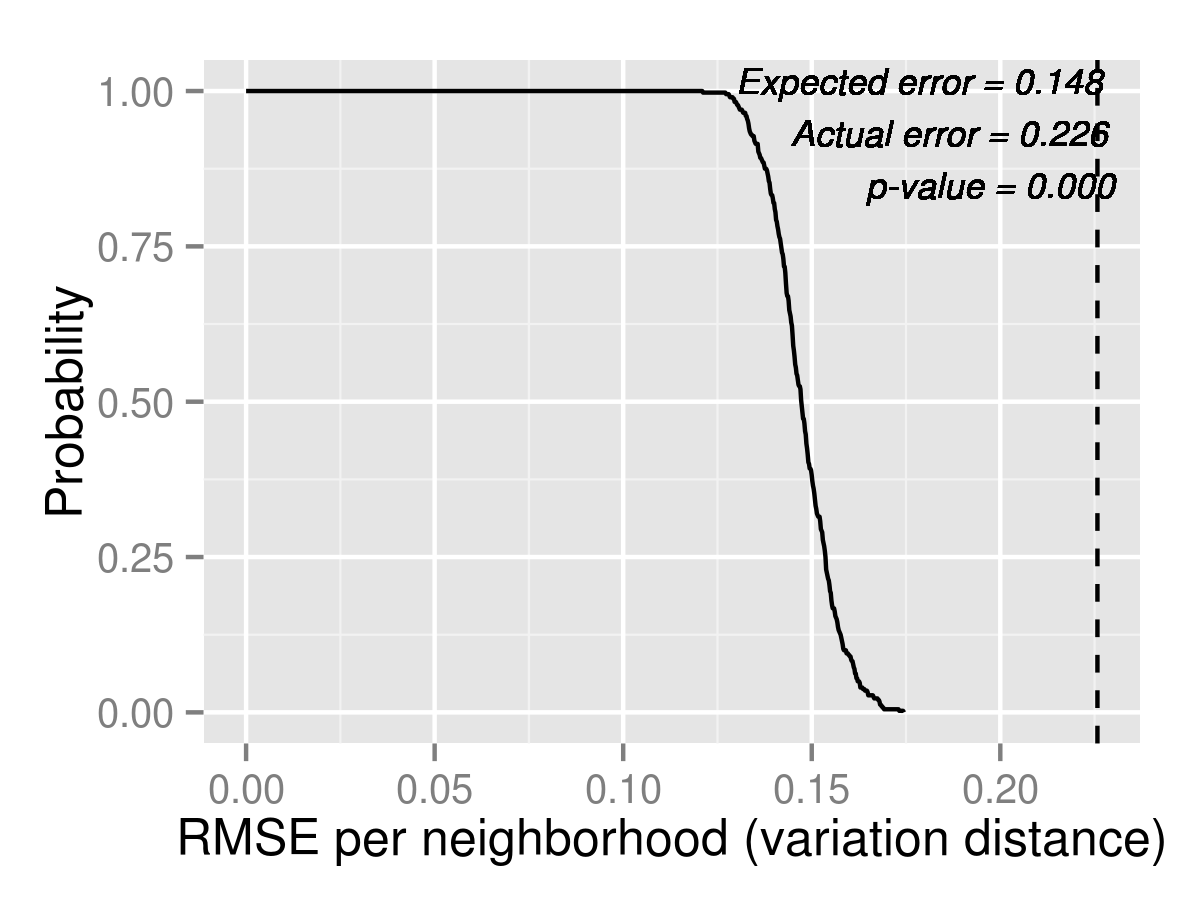}
}
\subfloat[][Top 1 (MLogit)]{
	\includegraphics[width=0.33\textwidth]{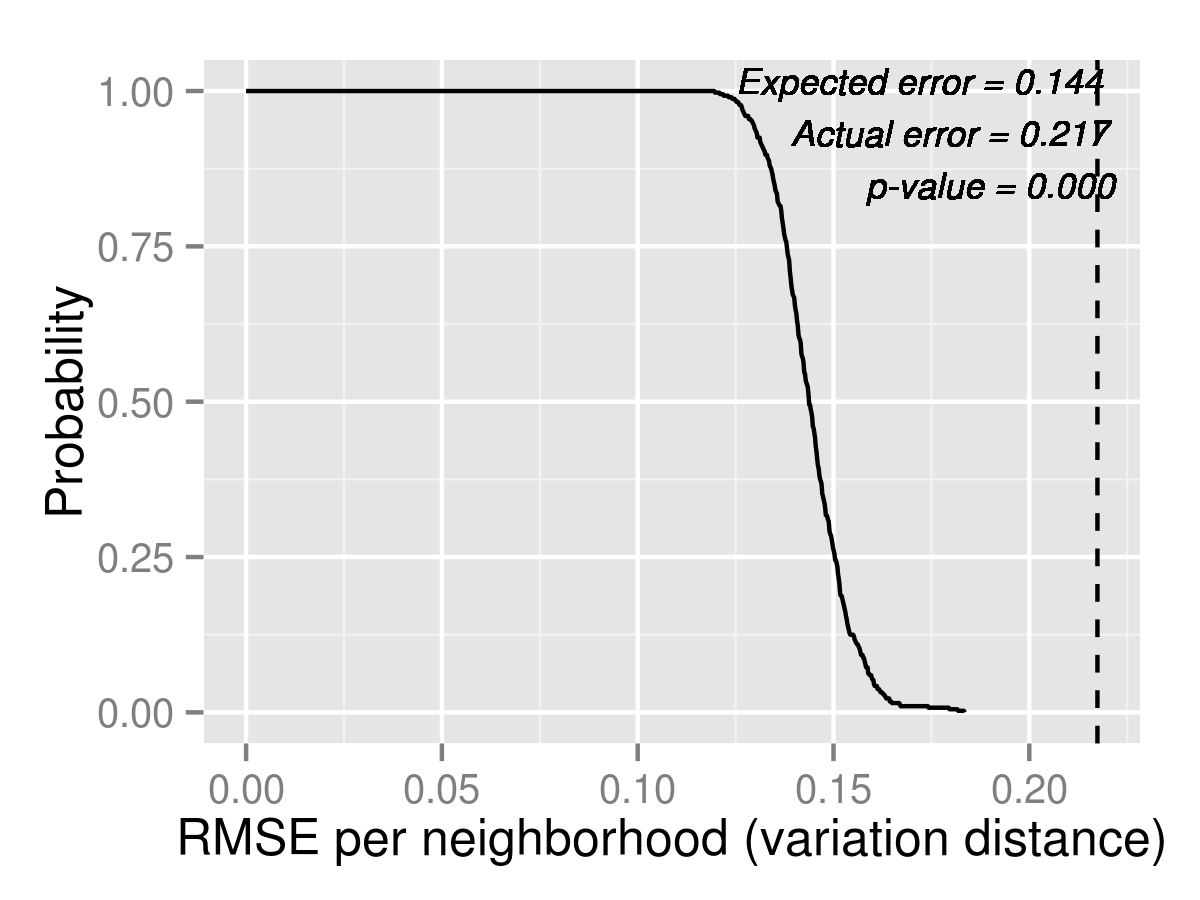}
}\\

\subfloat[][Top 2 (Naive)]{
 \includegraphics[width=0.33\textwidth]{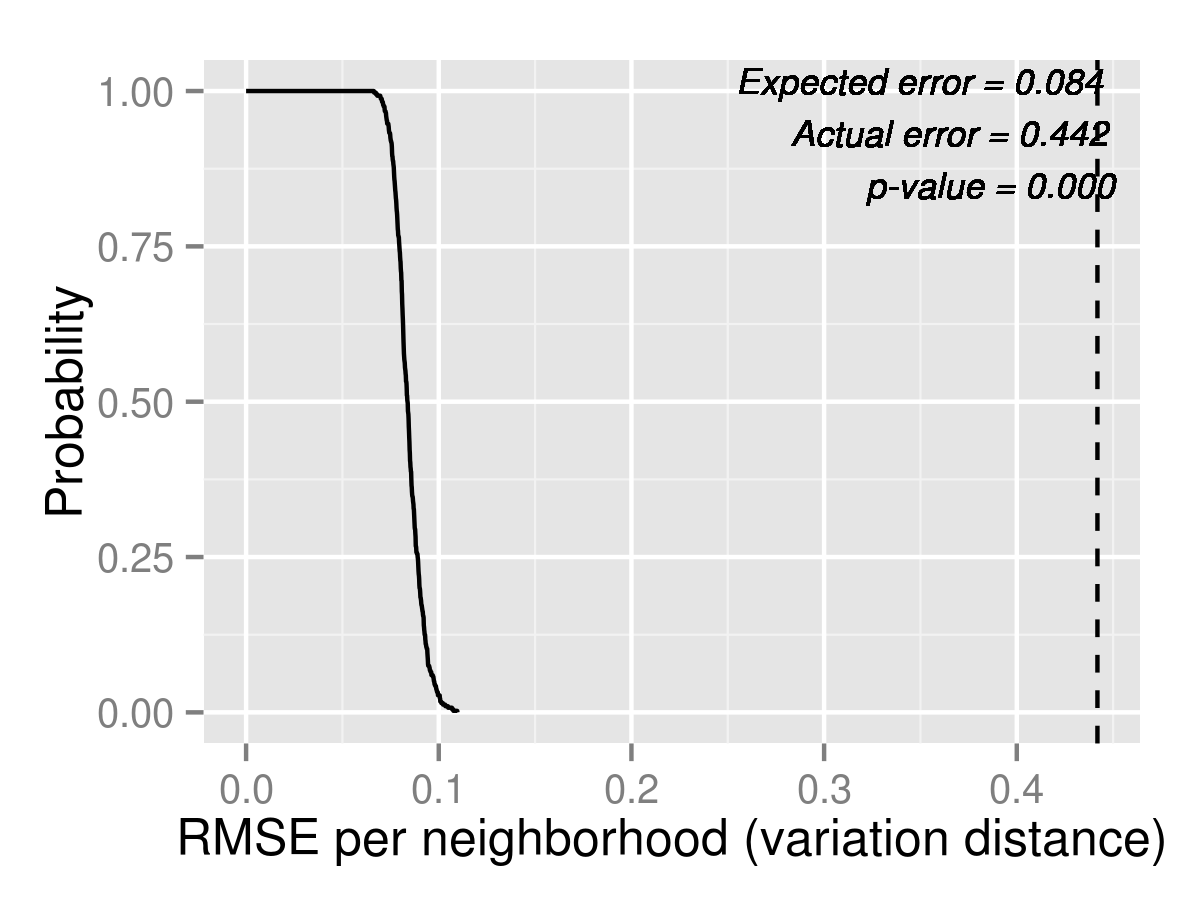}
}
\subfloat[][Top 2 (Logit)]{
	\includegraphics[width=0.33\textwidth]{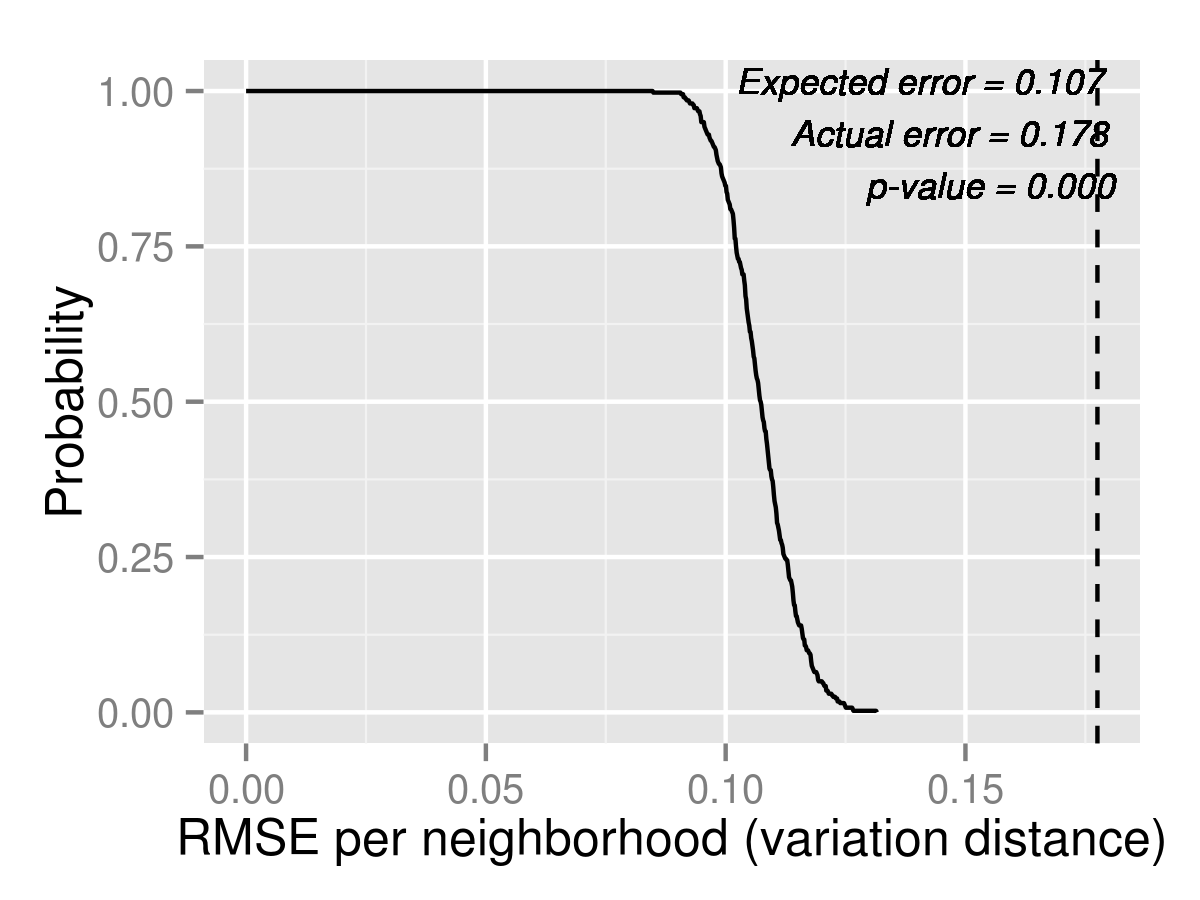}
}
\subfloat[][Top 2 (MLogit)]{
	\includegraphics[width=0.33\textwidth]{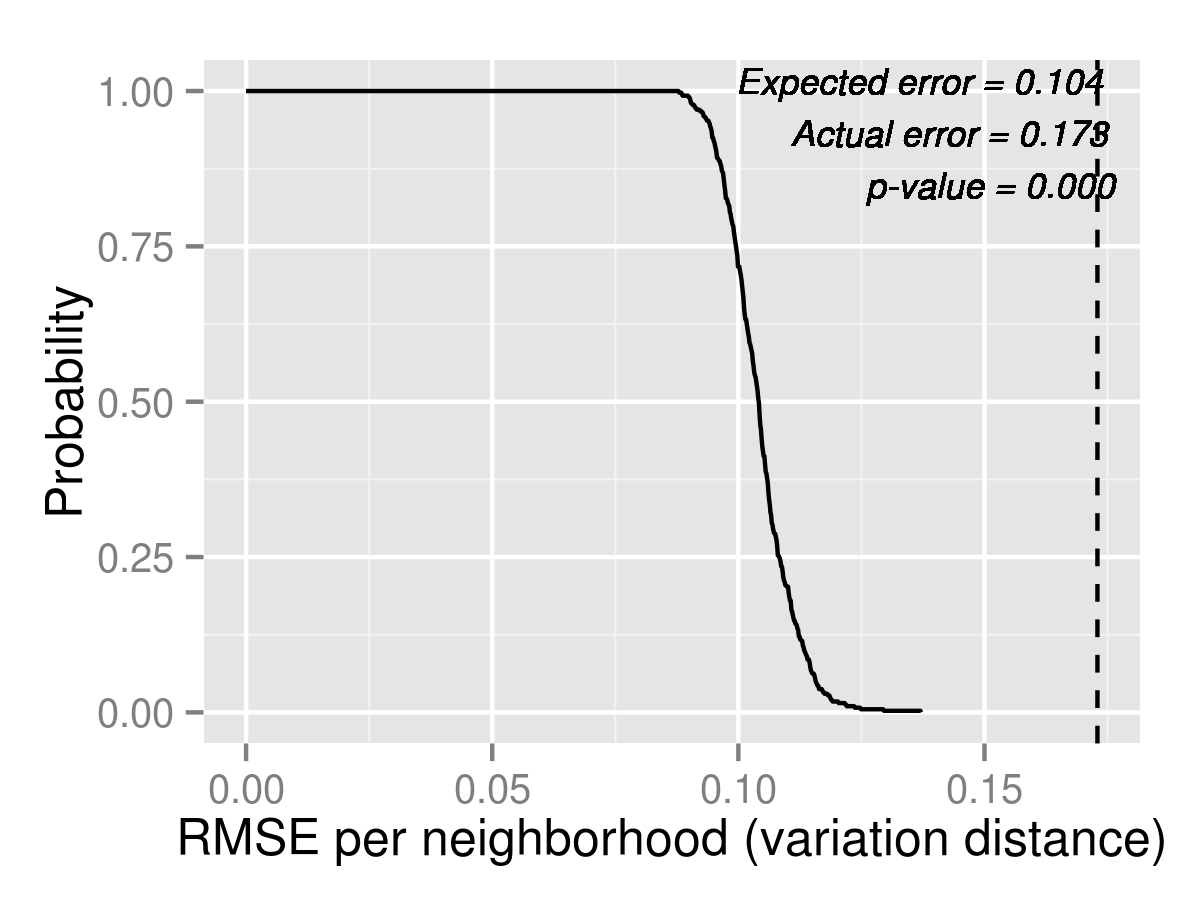}
}\\

\subfloat[][Top 3 (Naive)]{
 \includegraphics[width=0.33\textwidth]{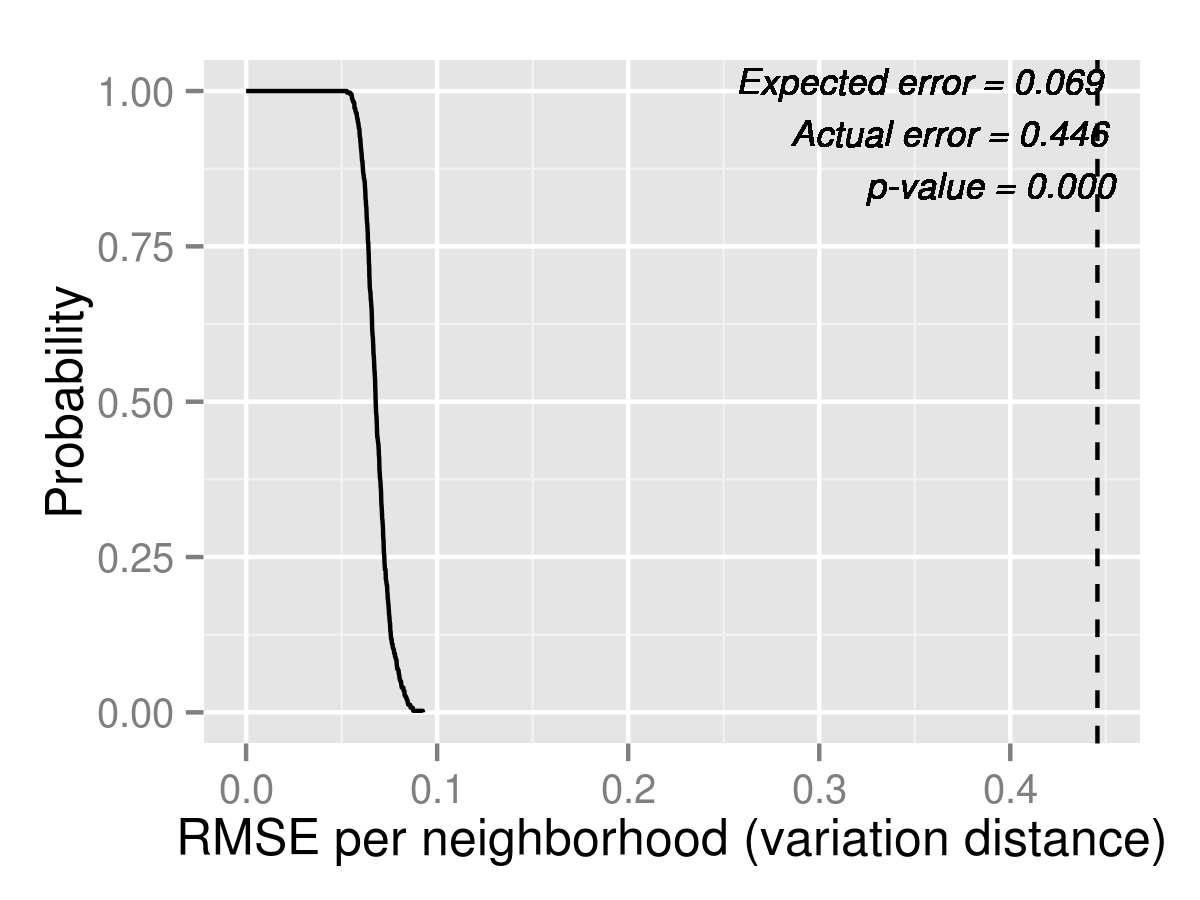}
}
\subfloat[][Top 3 (Logit)]{
	\includegraphics[width=0.33\textwidth]{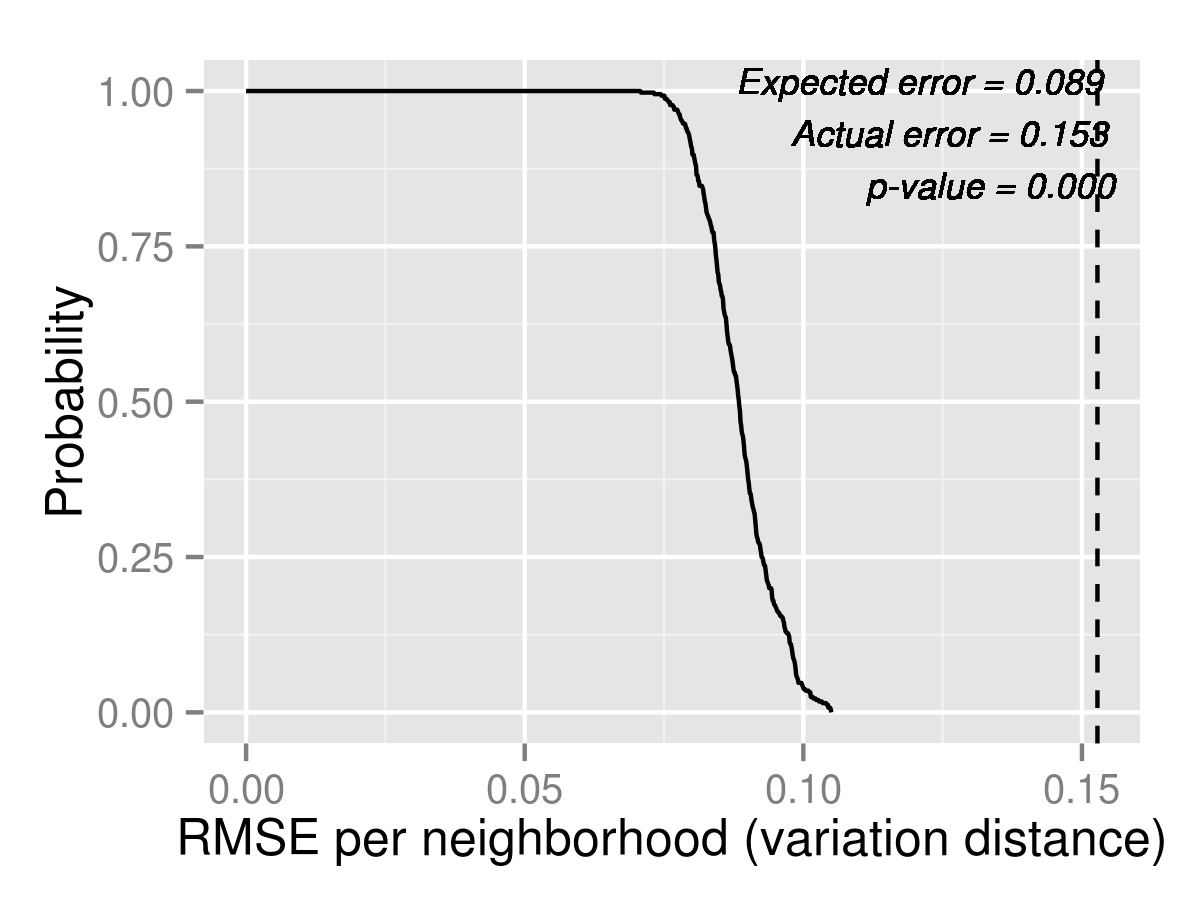}
}
\subfloat[][Top 3 (MLogit)]{
	\includegraphics[width=0.33\textwidth]{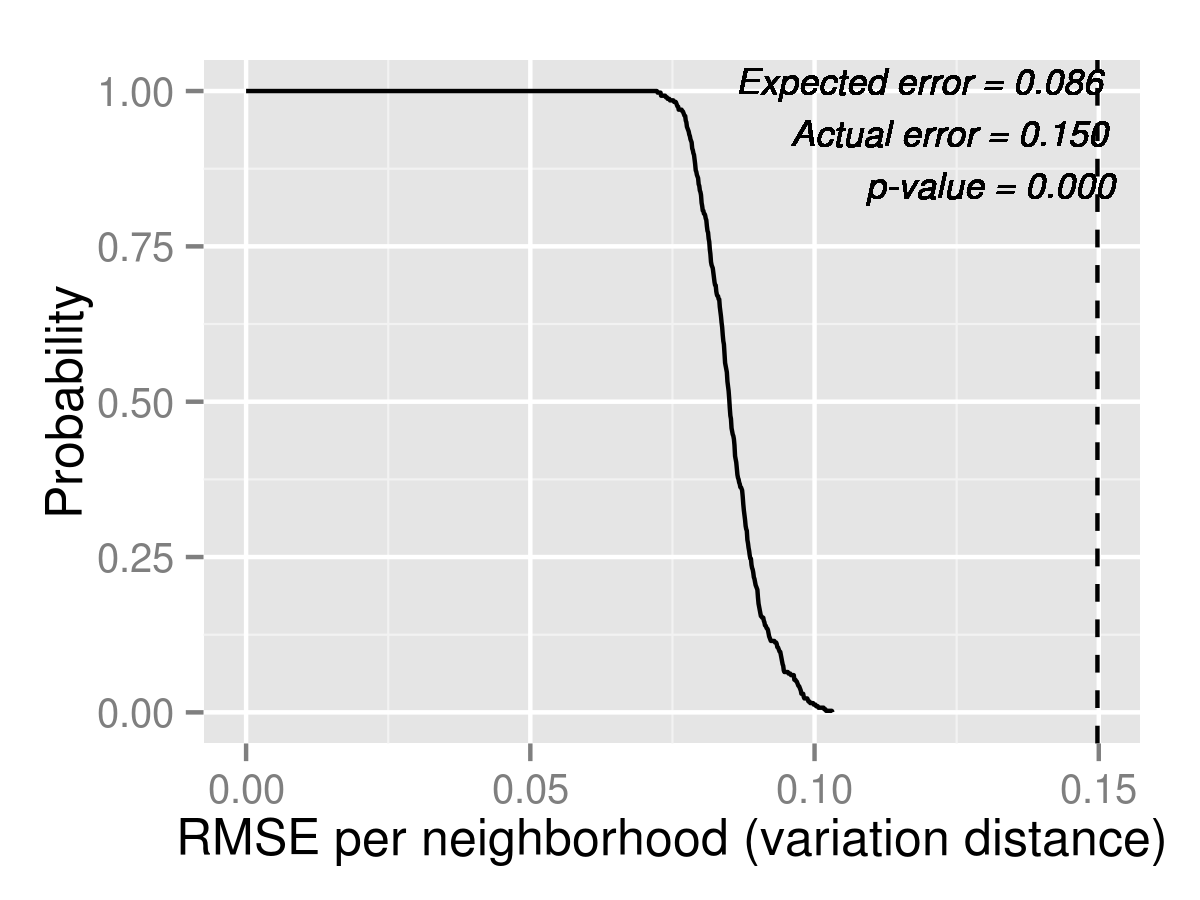}
}\\

\end{figure}

\begin{table}[!htbp]
 \centering

 \caption{Back testing unassigned predictions for 2013 K1.}
 \label{tab:unassigned2013K1}
 \footnotesize
 \begin{tabular}{l c c c c c c c}
Neighborhood & Naive & (95 \% C.I.) & Logit & (95 \% C.I.) & MLogit & (95 \% C.I.) & Actual\\ \hline  
Allston-Brighton & 62.74 & (46.00,81.03) & 34.59 & (16.98,54.00) & 31.83 & (16.00,50.00) & 7\\ 
Charlestown & 29.34 & (13.97,50.00) & 20.06 & (7.00,40.00) & 19.42 & (6.00,36.00) & 24\\ 
Downtown & 75.83 & (59.00,95.03) & 47.12 & (26.00,69.00) & 45.13 & (25.98,69.03) & 36\\ 
East Boston & 173.83 & (118.00,242.03) & 129.75 & (71.97,202.03) & 126.91 & (65.00,188.12) & 83\\ 
Hyde Park & 99.48 & (71.97,130.00) & 69.54 & (40.00,102.00) & 71.66 & (43.98,105.00) & 62\\ 
Jamaica Plain & 109.90 & (78.00,141.00) & 72.85 & (41.98,101.03) & 71.31 & (41.98,104.03) & 63\\ 
Mattapan & 92.30 & (62.00,125.00) & 63.19 & (34.00,94.00) & 63.76 & (36.00,94.00) & 56\\ 
North Dorchester & 78.19 & (47.98,113.00) & 47.66 & (22.00,80.03) & 47.94 & (23.00,78.00) & 31\\ 
Roslindale & 130.11 & (92.00,174.03) & 101.83 & (64.97,146.03) & 104.45 & (66.00,151.03) & 95\\ 
Roxbury & 136.74 & (93.97,187.03) & 90.94 & (43.95,145.05) & 88.58 & (42.00,143.05) & 67\\ 
South Boston & 41.67 & (27.95,59.00) & 24.58 & (11.00,40.00) & 23.52 & (10.00,38.00) & 18\\ 
South Dorchester & 197.29 & (146.00,260.05) & 129.50 & (74.00,196.03) & 129.65 & (74.00,192.03) & 122\\ 
South End & 57.55 & (36.00,82.03) & 33.85 & (12.00,57.02) & 31.79 & (14.97,51.10) & 31\\ 
West Roxbury & 135.33 & (92.85,185.03) & 112.99 & (73.97,156.03) & 113.97 & (75.00,155.05) & 84\\ 
\end{tabular}
 \end{table}

\begin{table}[!htbp]
 \centering

 \caption{Back testing access to quality predictions for 2013 K1.}
 \label{tab:aoq2013K1}
 \footnotesize
 \begin{tabular}{l c c c c c c c}
Neighborhood & Naive & (95 \% C.I.) & Logit & (95 \% C.I.) & MLogit & (95 \% C.I.) & Actual\\ \hline  
Allston-Brighton & 1 & (1.00,1.00) & 0.66 & (0.52,0.81) & 0.69 & (0.56,0.83) & 1\\ 
Charlestown & 1 & (1.00,1.00) & 0.65 & (0.50,0.80) & 0.69 & (0.55,0.83) & 1\\ 
Downtown & 1 & (1.00,1.00) & 0.68 & (0.54,0.83) & 0.71 & (0.58,0.85) & 1\\ 
East Boston & 1 & (1.00,1.00) & 0.66 & (0.54,0.80) & 0.70 & (0.58,0.83) & 1\\ 
Hyde Park & 0.36 & (0.30,0.43) & 0.56 & (0.45,0.69) & 0.55 & (0.45,0.68) & 0.60\\ 
Jamaica Plain & 0.53 & (0.45,0.62) & 0.59 & (0.47,0.74) & 0.61 & (0.50,0.75) & 0.80\\ 
Mattapan & 0.33 & (0.26,0.41) & 0.56 & (0.44,0.69) & 0.54 & (0.44,0.67) & 0.57\\ 
North Dorchester & 0.45 & (0.38,0.52) & 0.57 & (0.46,0.70) & 0.56 & (0.46,0.69) & 0.67\\ 
Roslindale & 0.46 & (0.37,0.55) & 0.61 & (0.50,0.74) & 0.63 & (0.51,0.77) & 0.79\\ 
Roxbury & 0.60 & (0.52,0.69) & 0.61 & (0.51,0.73) & 0.63 & (0.52,0.74) & 0.81\\ 
South Boston & 0.30 & (0.23,0.39) & 0.53 & (0.41,0.67) & 0.52 & (0.40,0.68) & 0.57\\ 
South Dorchester & 0.35 & (0.29,0.42) & 0.56 & (0.44,0.68) & 0.54 & (0.43,0.67) & 0.60\\ 
South End & 1 & (1.00,1.00) & 0.70 & (0.57,0.83) & 0.73 & (0.61,0.86) & 1\\ 
West Roxbury & 0.46 & (0.38,0.55) & 0.61 & (0.51,0.74) & 0.63 & (0.52,0.76) & 0.80\\ 
\end{tabular}
 \end{table}

\begin{table}[!htbp]
 \centering
 \caption{Back testing distance predictions for 2013 K1.}
 \label{tab:distance2013K1}
 \footnotesize
 \begin{tabular}{l c c c c c c c}
Neighborhood & Naive & (95 \% C.I.) & Logit & (95 \% C.I.) & MLogit & (95 \% C.I.) & Actual\\ \hline  
Allston-Brighton & 2.69 & (2.17,3.25) & 1.73 & (1.41,2.06) & 1.76 & (1.45,2.10) & 1.17\\ 
Charlestown & 2.65 & (1.75,3.78) & 2.88 & (2.04,3.71) & 2.72 & (1.86,3.43) & 1.69\\ 
Downtown & 2.03 & (1.41,2.70) & 2.00 & (1.64,2.38) & 1.97 & (1.63,2.35) & 1.95\\ 
East Boston & 3.20 & (2.73,3.71) & 2.12 & (1.69,2.52) & 2.28 & (1.90,2.66) & 2.44\\ 
Hyde Park & 3.15 & (2.80,3.50) & 2.68 & (2.40,3.04) & 2.69 & (2.35,3.03) & 2.43\\ 
Jamaica Plain & 2.09 & (1.72,2.48) & 1.81 & (1.58,2.06) & 1.80 & (1.57,2.04) & 1.26\\ 
Mattapan & 2.37 & (2.08,2.66) & 2.21 & (1.97,2.44) & 2.17 & (1.94,2.44) & 2.08\\ 
North Dorchester & 1.84 & (1.37,2.34) & 1.58 & (1.30,1.87) & 1.62 & (1.31,1.95) & 1.41\\ 
Roslindale & 1.87 & (1.63,2.18) & 2.10 & (1.84,2.39) & 2.07 & (1.83,2.35) & 1.95\\ 
Roxbury & 2.15 & (1.89,2.39) & 1.76 & (1.56,1.97) & 1.74 & (1.55,1.96) & 1.47\\ 
South Boston & 2.31 & (1.74,2.95) & 1.71 & (1.31,2.14) & 1.74 & (1.31,2.20) & 1.76\\ 
South Dorchester & 1.68 & (1.47,1.88) & 1.72 & (1.57,1.86) & 1.69 & (1.56,1.83) & 1.72\\ 
South End & 2.57 & (1.94,3.21) & 1.81 & (1.50,2.19) & 1.81 & (1.49,2.19) & 1.53\\ 
West Roxbury & 2.08 & (1.78,2.40) & 2.33 & (2.03,2.62) & 2.27 & (1.99,2.56) & 1.76\\ 
\end{tabular}
 \end{table}

 \subsection{Forecasts for 2014 K1}

\begin{figure}[h!]
\centering
\caption{Forecasts for assignment outcomes for 2014 K1. Tail distribution plots. \label{fig:outcomes2014K1}}

\subfloat[][Unassigned (Naive)]{
 \includegraphics[width=0.33\textwidth]{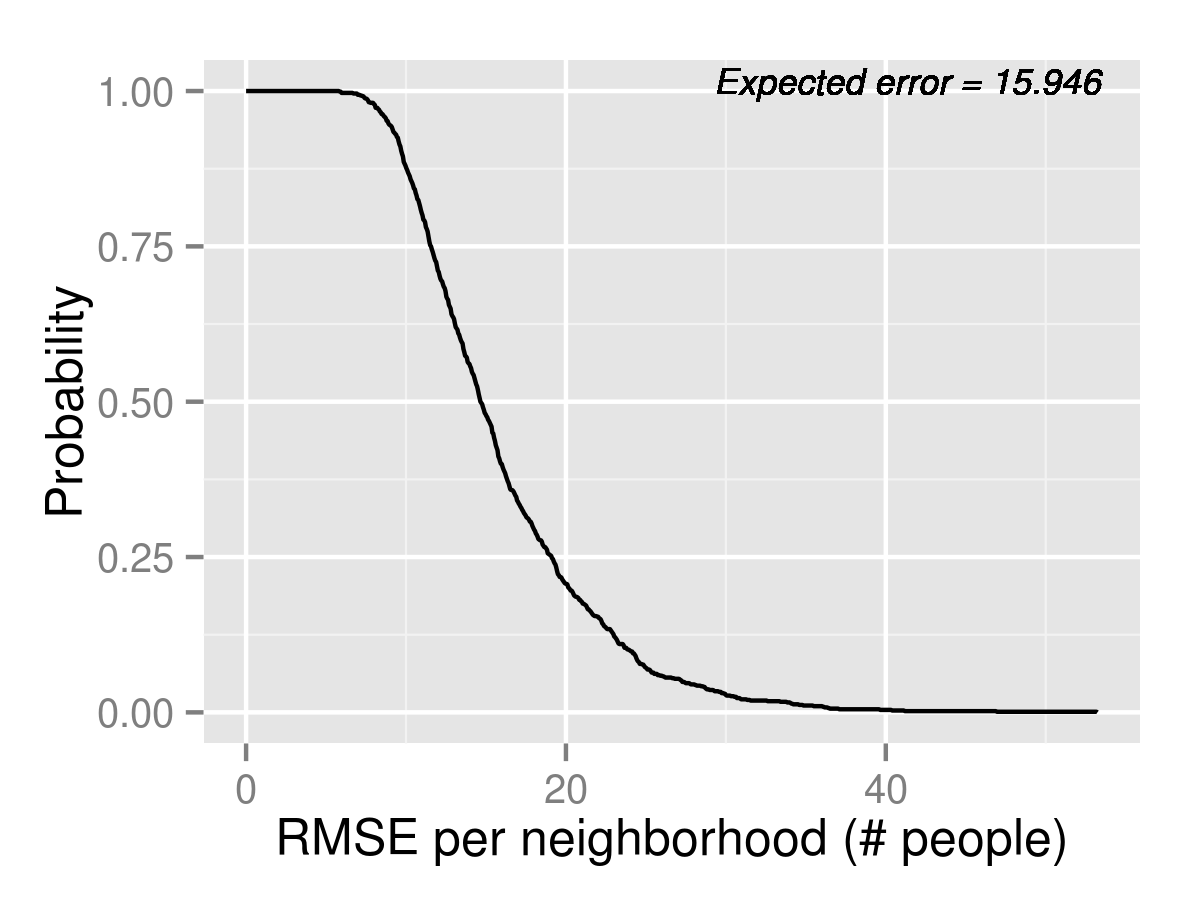}
}
\subfloat[][Unassigned (Logit)]{
	\includegraphics[width=0.33\textwidth]{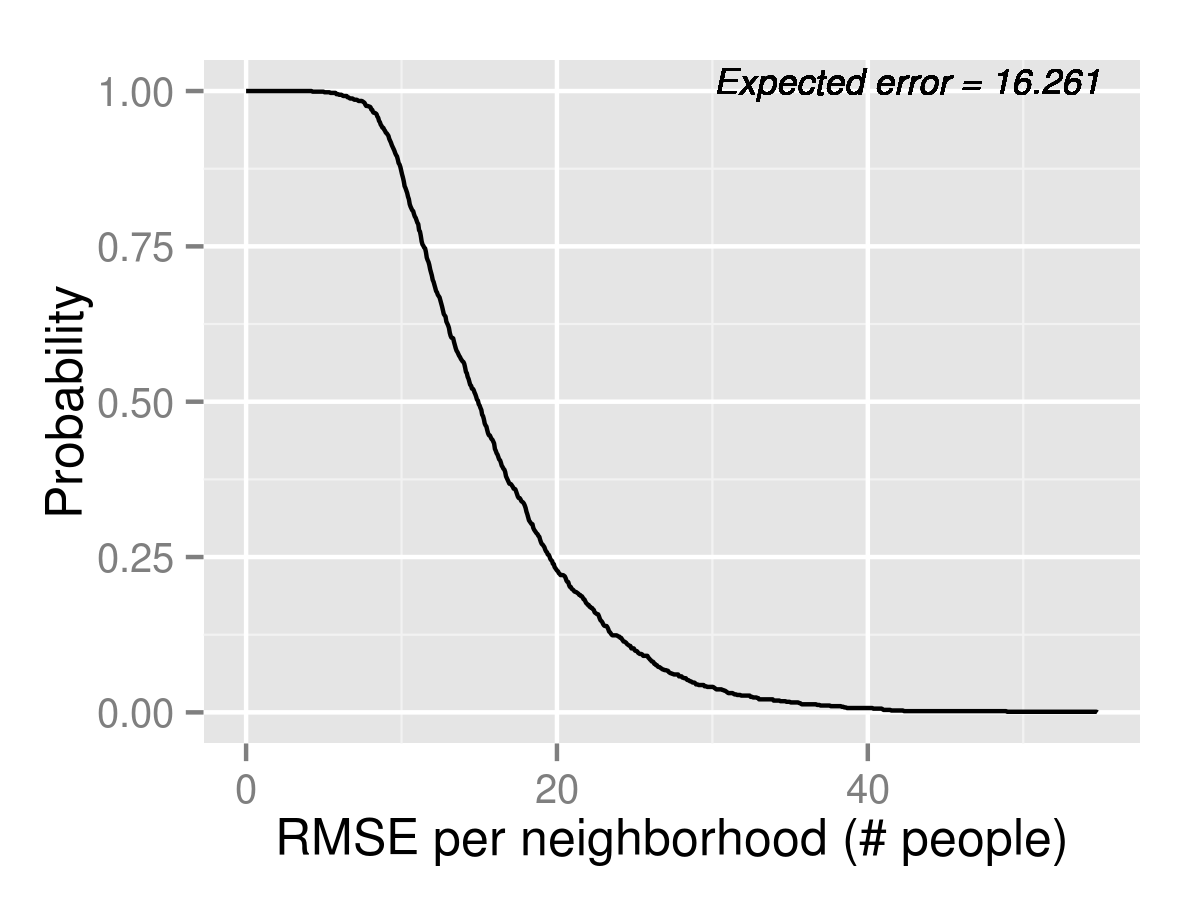}
}
\subfloat[][Unassigned (MLogit)]{
	\includegraphics[width=0.33\textwidth]{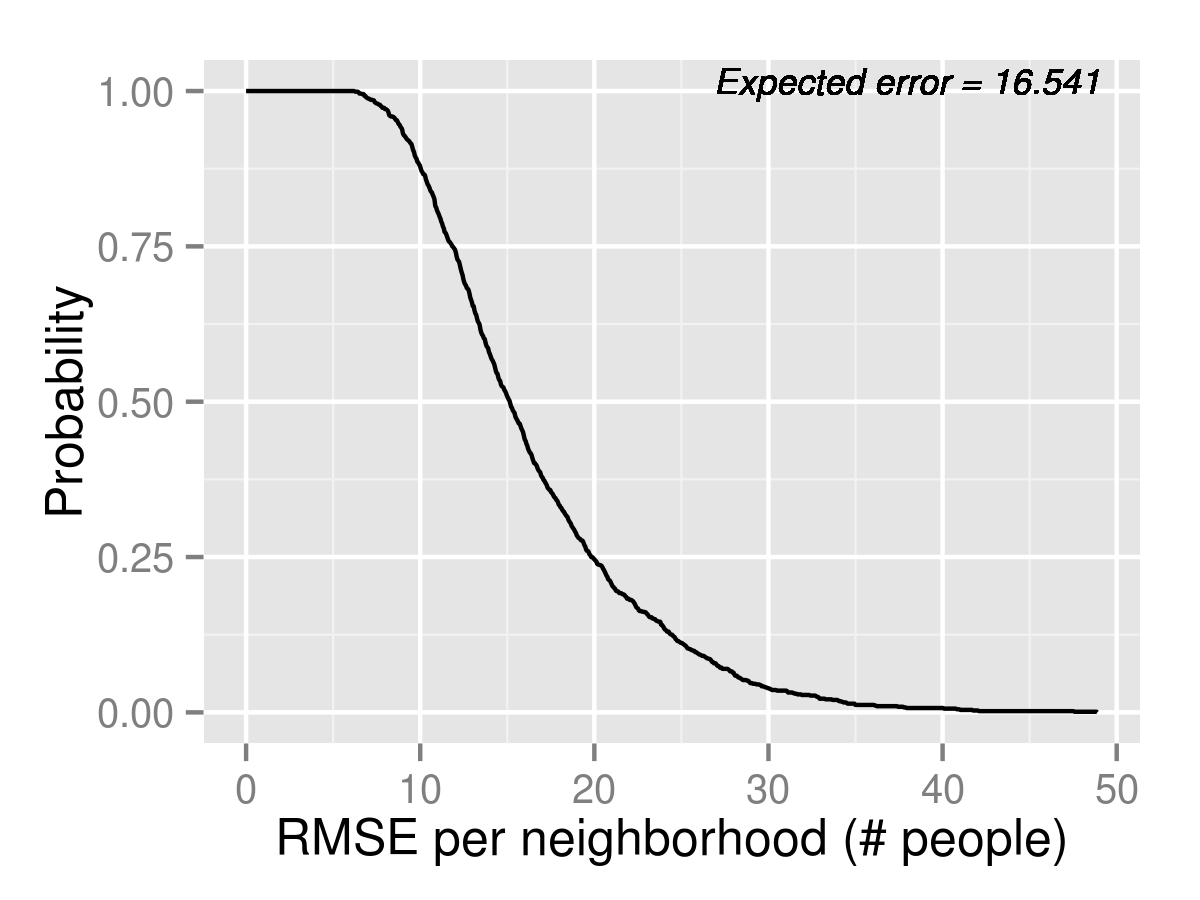}
}
\\

\subfloat[][Access to Quality (Naive)]{
 \includegraphics[width=0.33\textwidth]{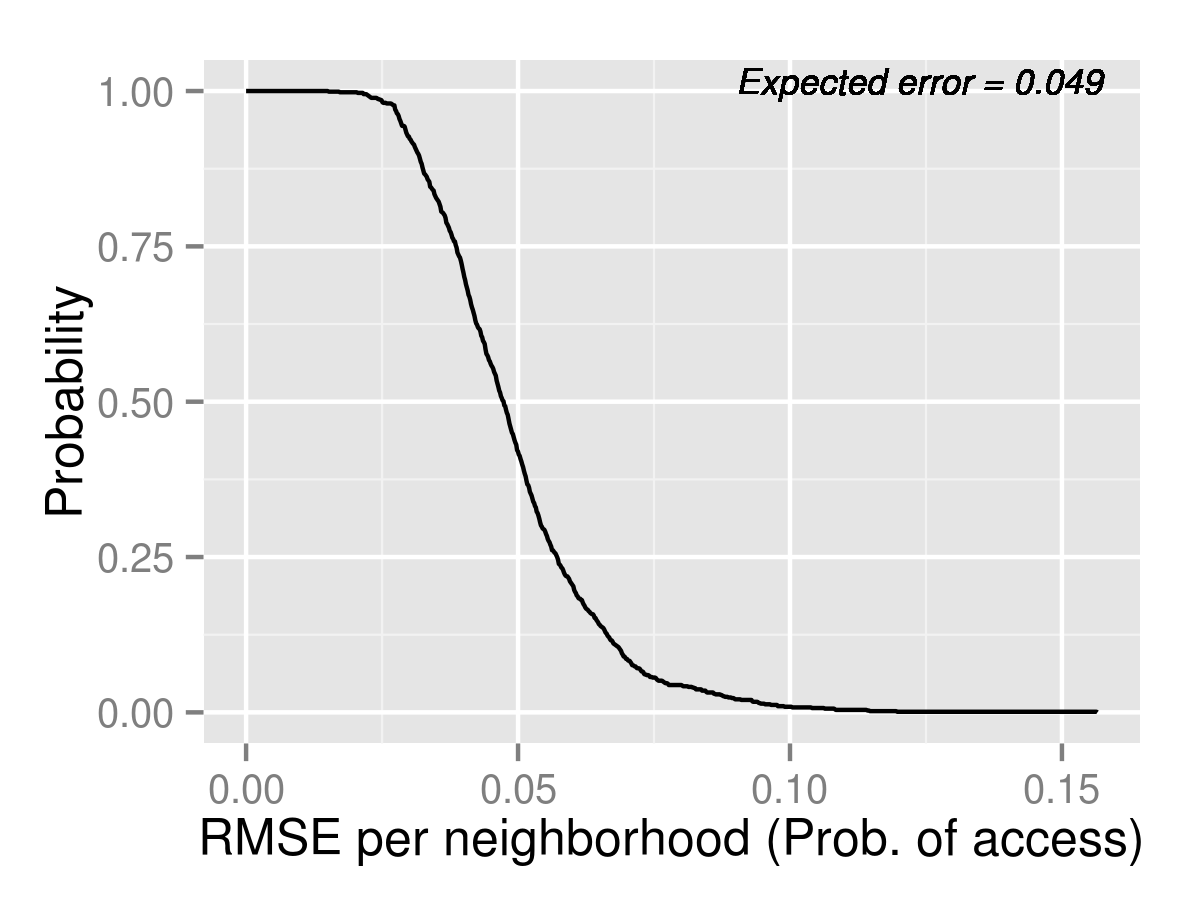}
}
\subfloat[][Access to Quality (Logit)]{
	\includegraphics[width=0.33\textwidth]{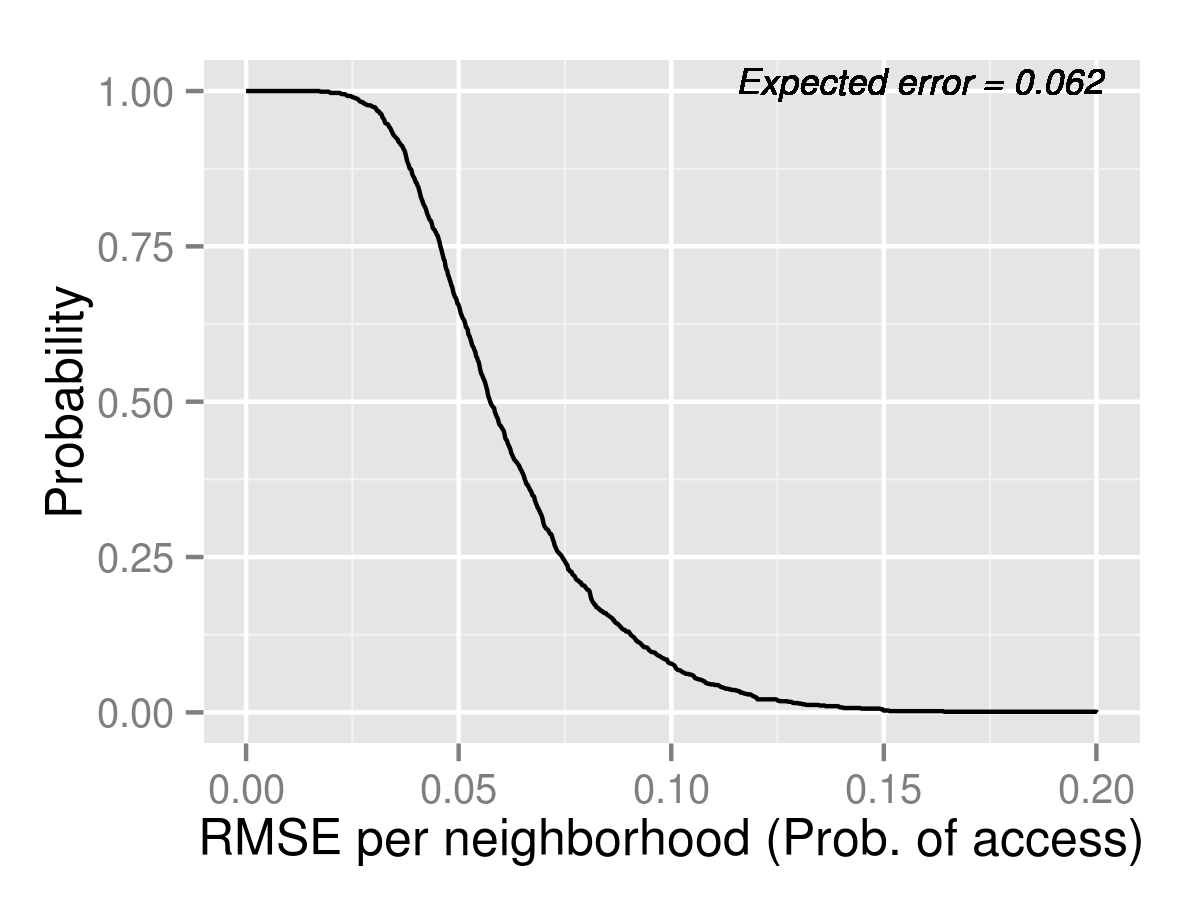}
}
\subfloat[][Access to Quality (MLogit)]{
	\includegraphics[width=0.33\textwidth]{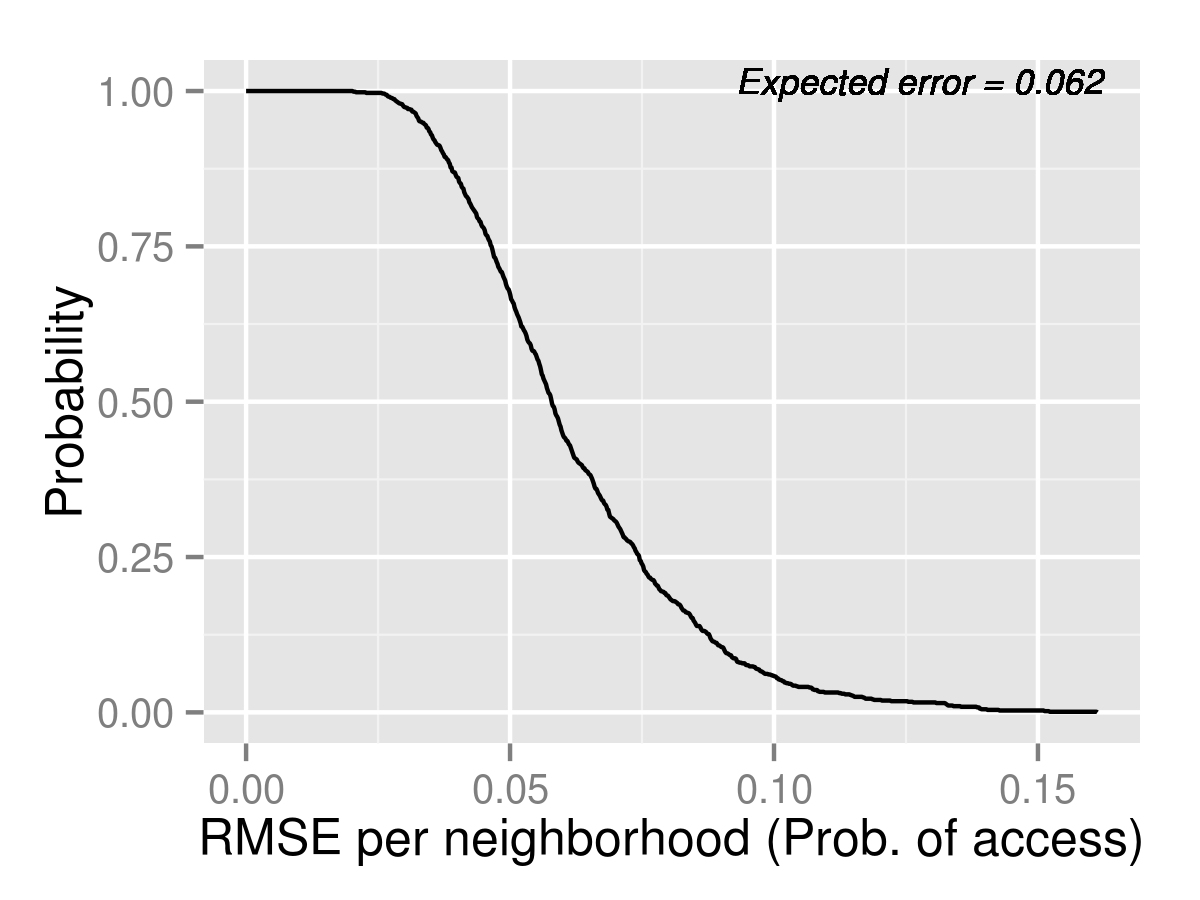}
}\\

\subfloat[][Distance (Naive)]{
 \includegraphics[width=0.33\textwidth]{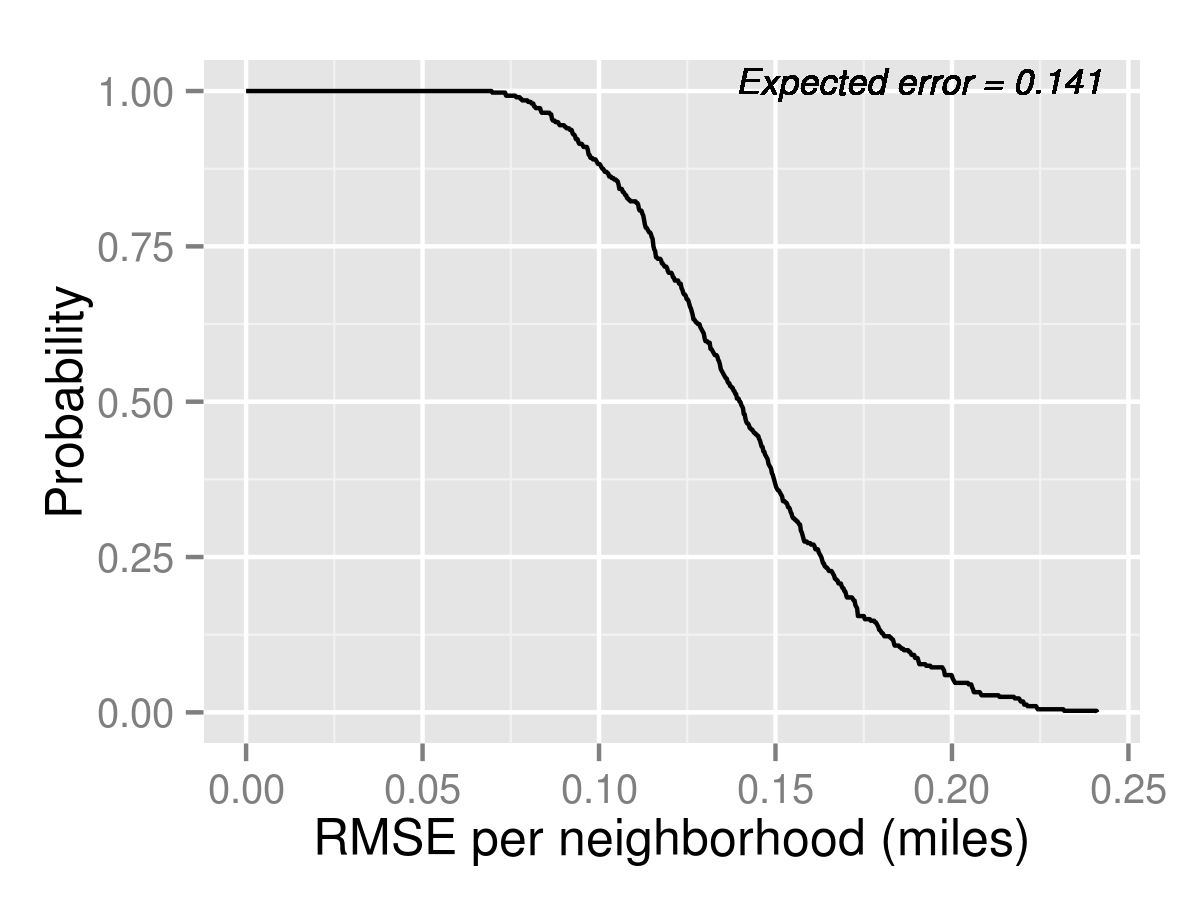}
}
\subfloat[][Distance (Logit)]{
	\includegraphics[width=0.33\textwidth]{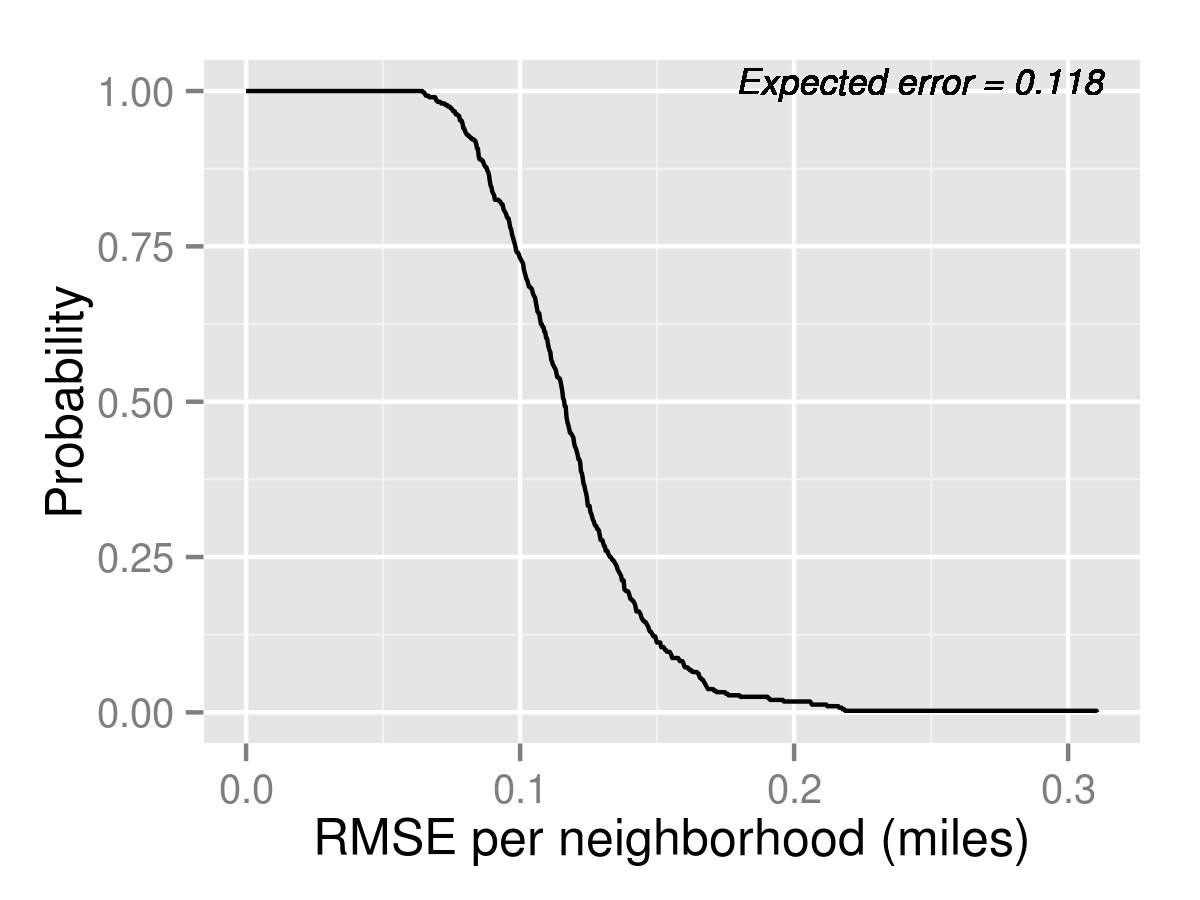}
}
\subfloat[][Distance (MLogit)]{
	\includegraphics[width=0.33\textwidth]{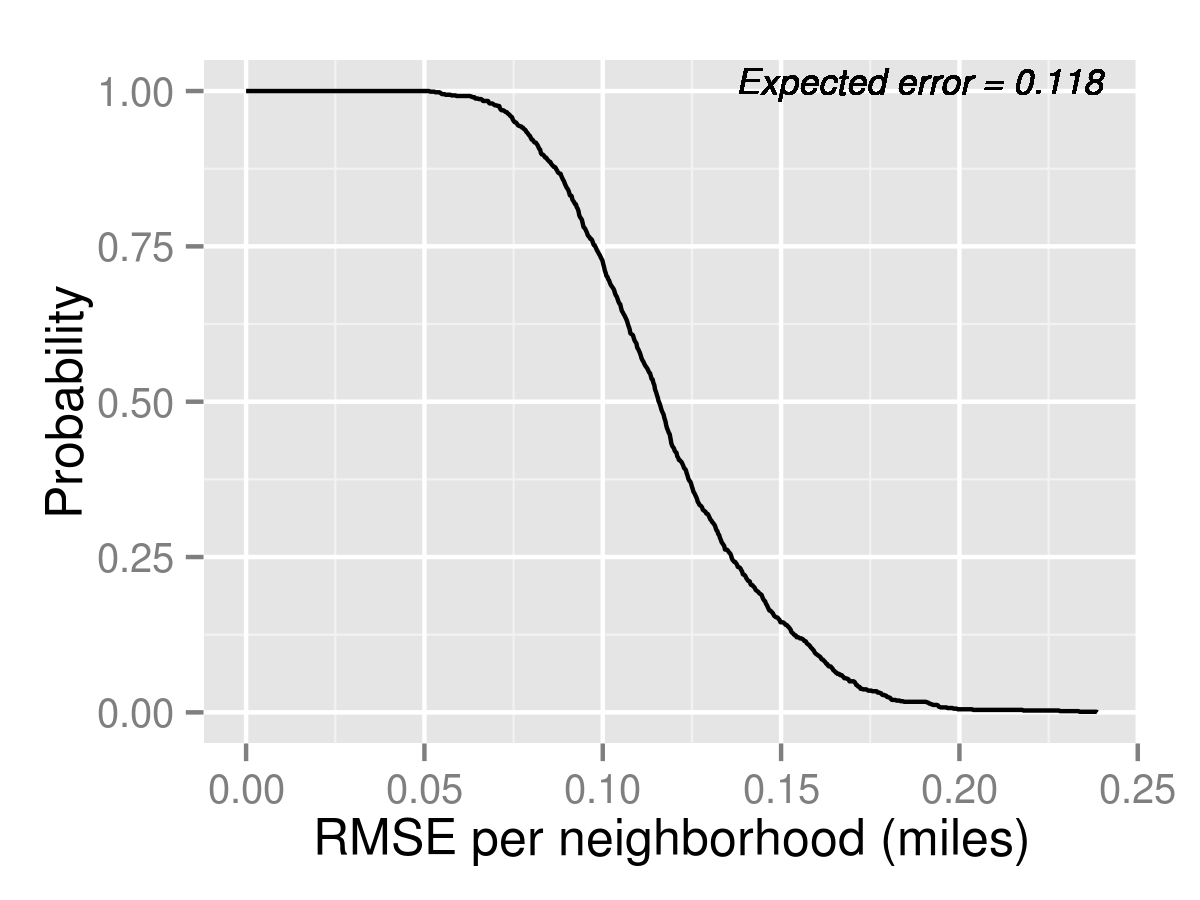}
}\\

\end{figure}

\begin{figure}[h!]
\centering
\caption{Forecasts for market shares for 2014 K1. Tail distribution plots. \label{fig:marketShare2014K1}}

\subfloat[][Top 1 (Naive)]{
 \includegraphics[width=0.33\textwidth]{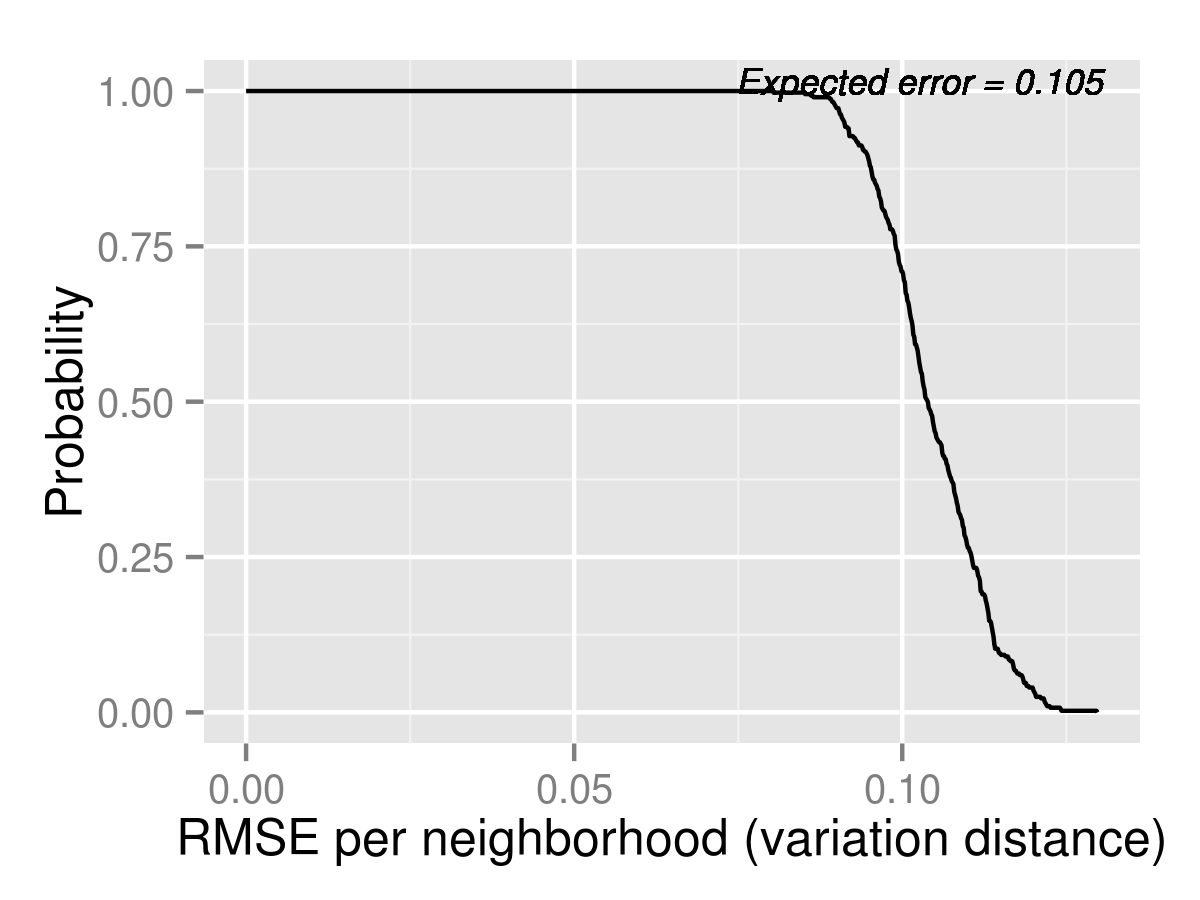}
}
\subfloat[][Top 1 (Logit)]{
	\includegraphics[width=0.33\textwidth]{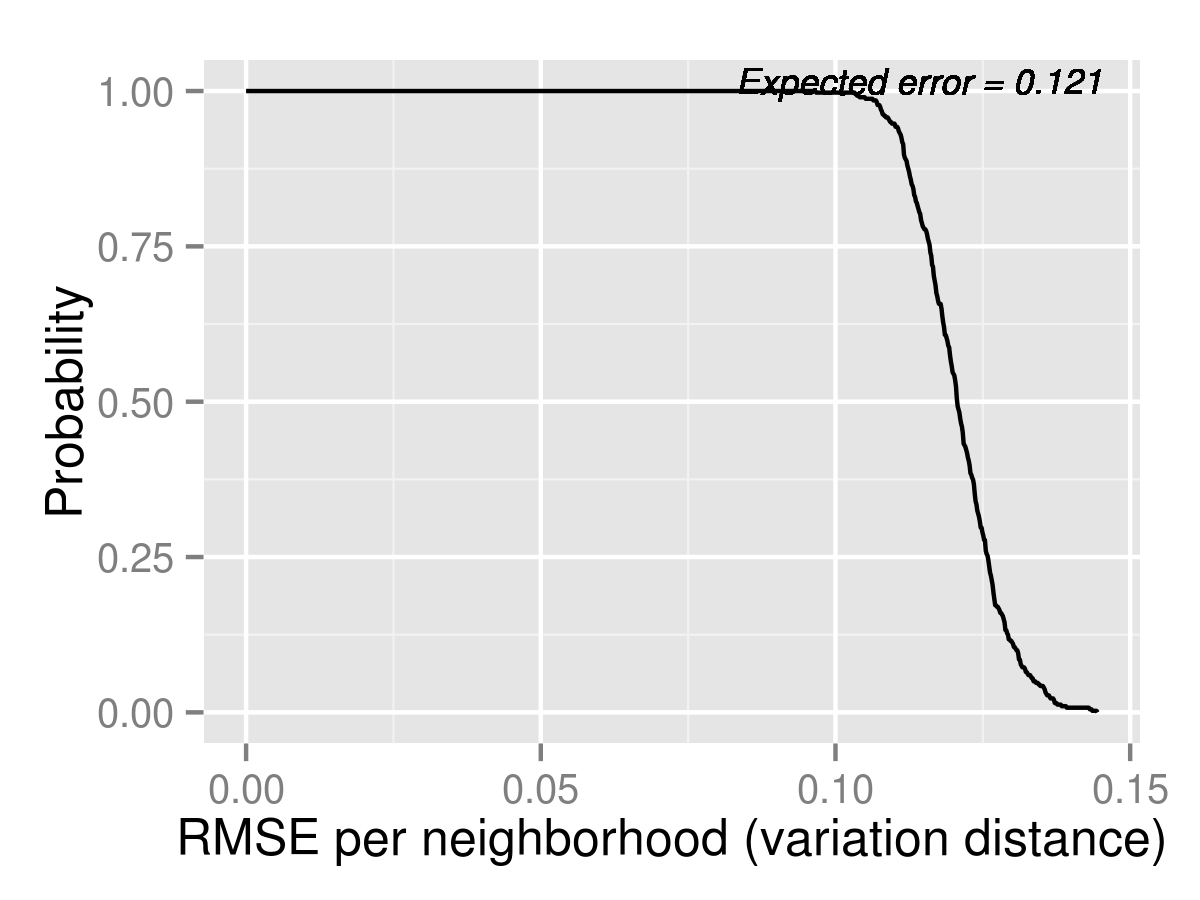}
}
\subfloat[][Top 1 (MLogit)]{
	\includegraphics[width=0.33\textwidth]{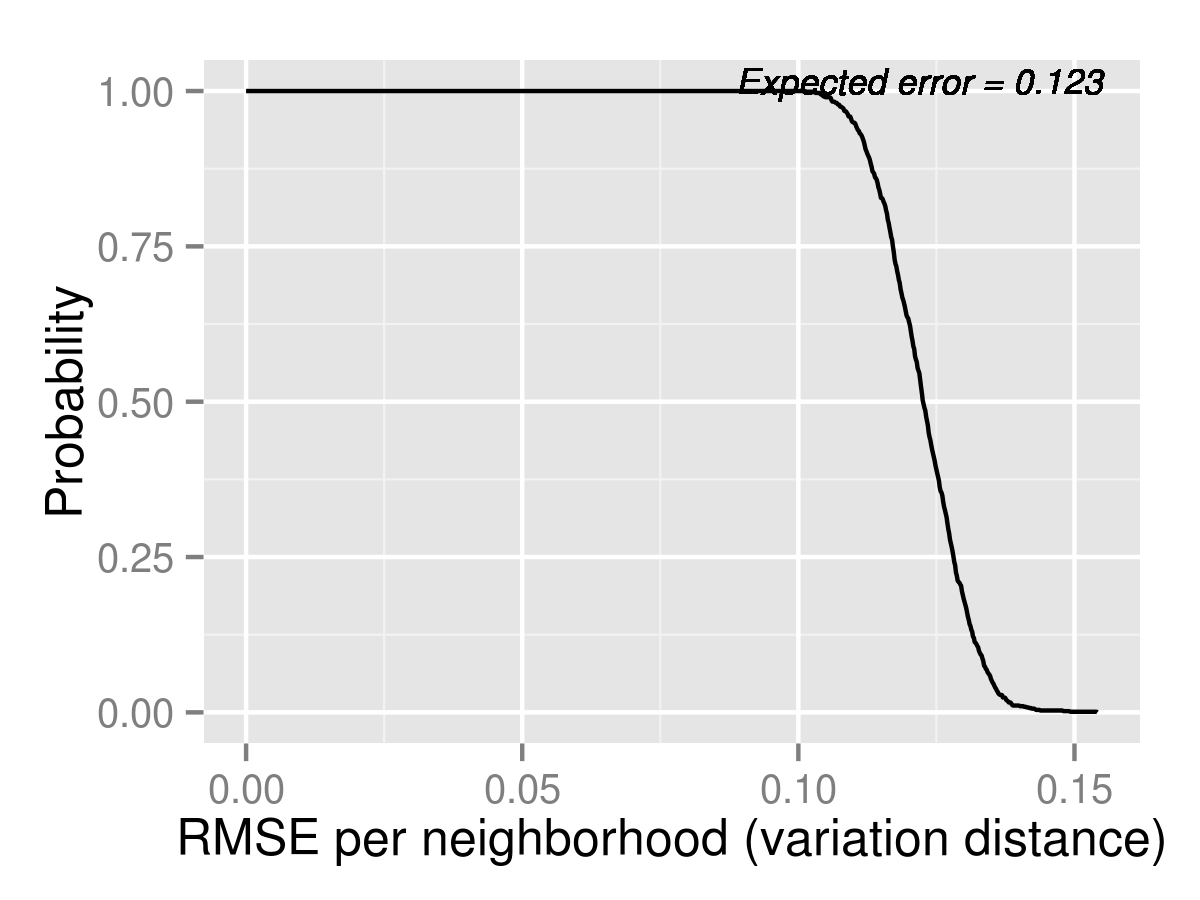}
}\\

\subfloat[][Top 2 (Naive)]{
 \includegraphics[width=0.33\textwidth]{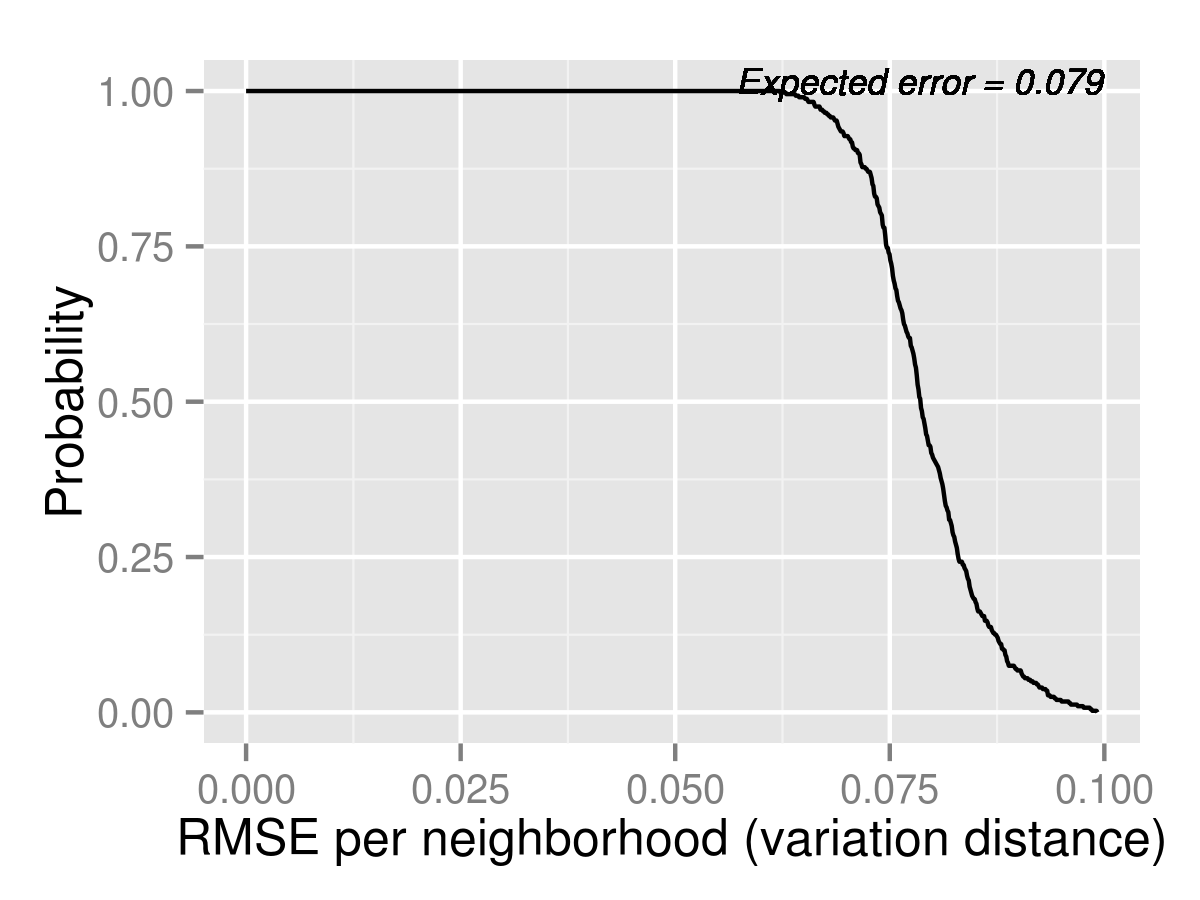}
}
\subfloat[][Top 2 (Logit)]{
	\includegraphics[width=0.33\textwidth]{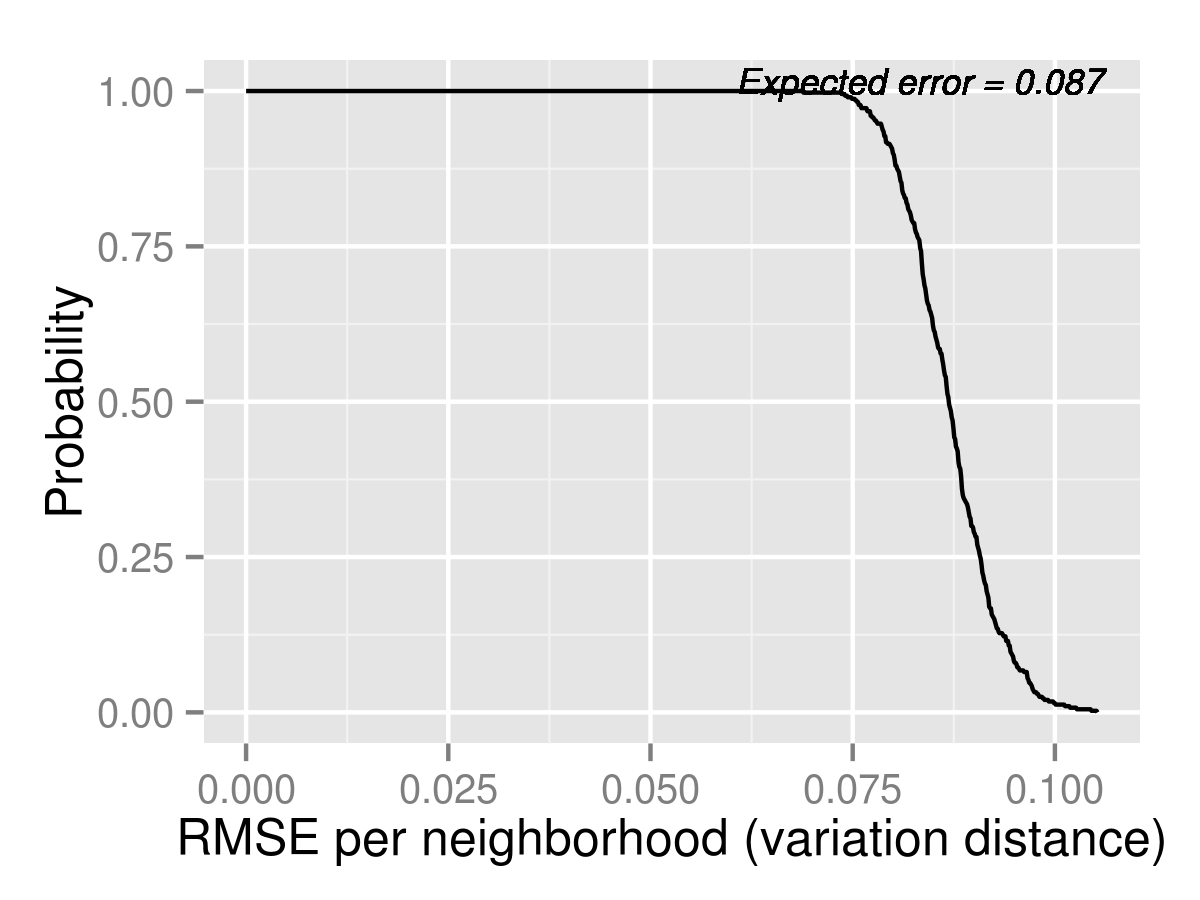}
}
\subfloat[][Top 2 (MLogit)]{
	\includegraphics[width=0.33\textwidth]{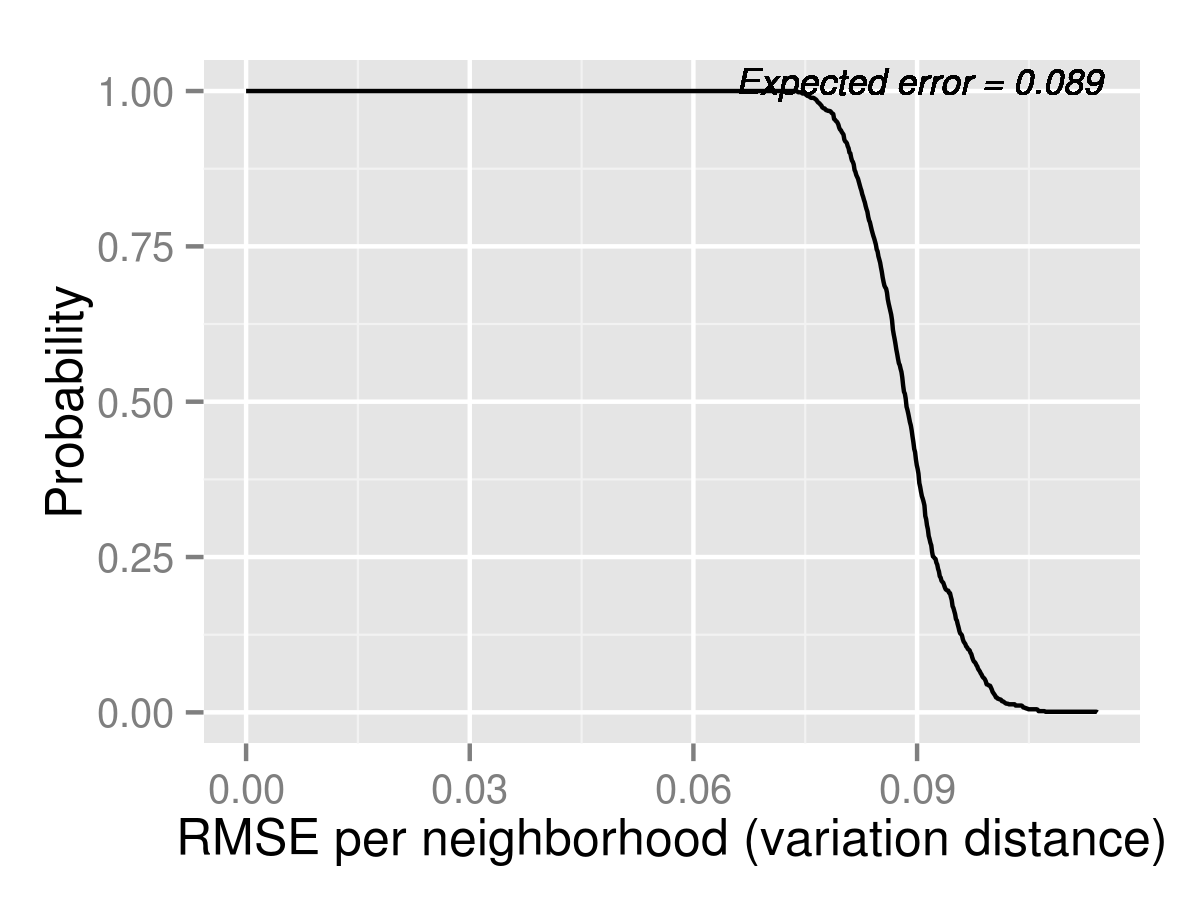}
}\\

\subfloat[][Top 3 (Naive)]{
 \includegraphics[width=0.33\textwidth]{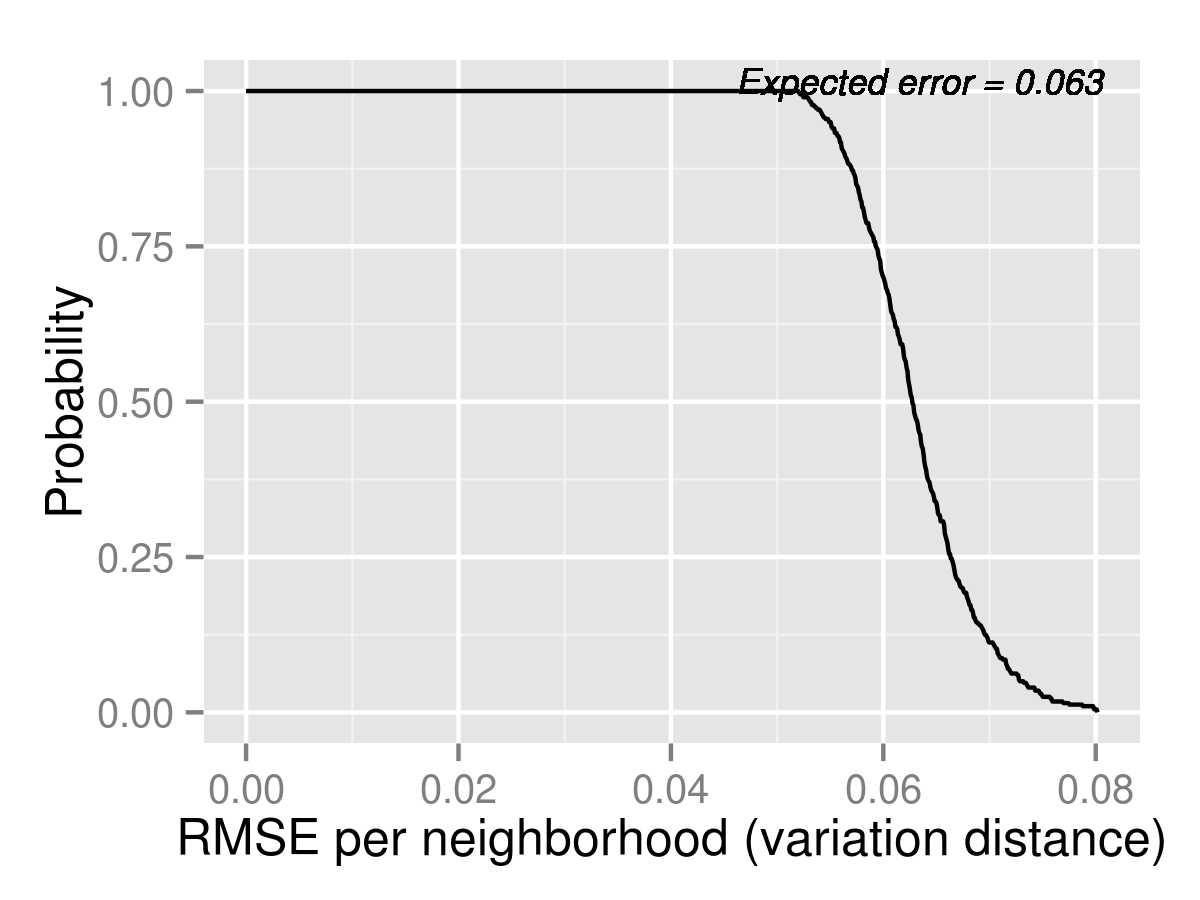}
}
\subfloat[][Top 3 (Logit)]{
	\includegraphics[width=0.33\textwidth]{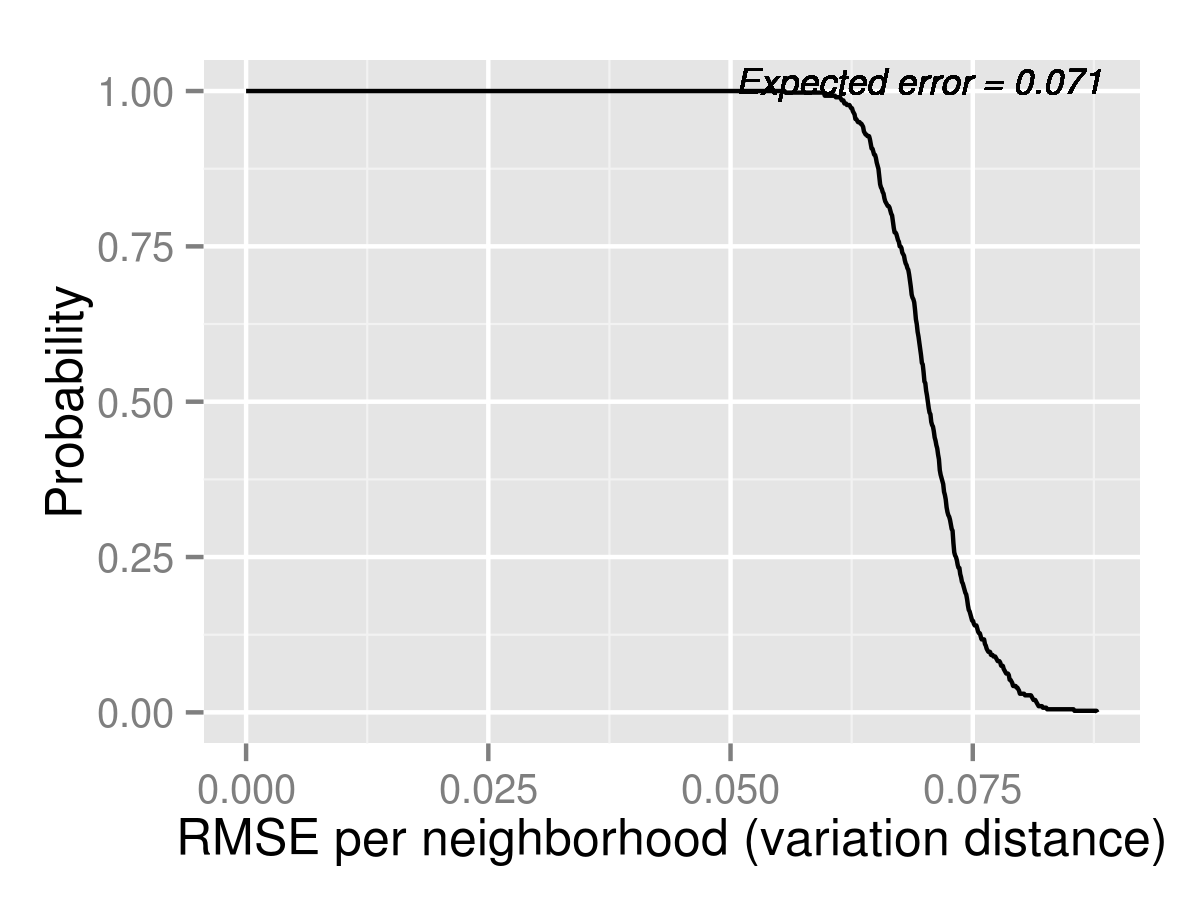}
}
\subfloat[][Top 3 (MLogit)]{
	\includegraphics[width=0.33\textwidth]{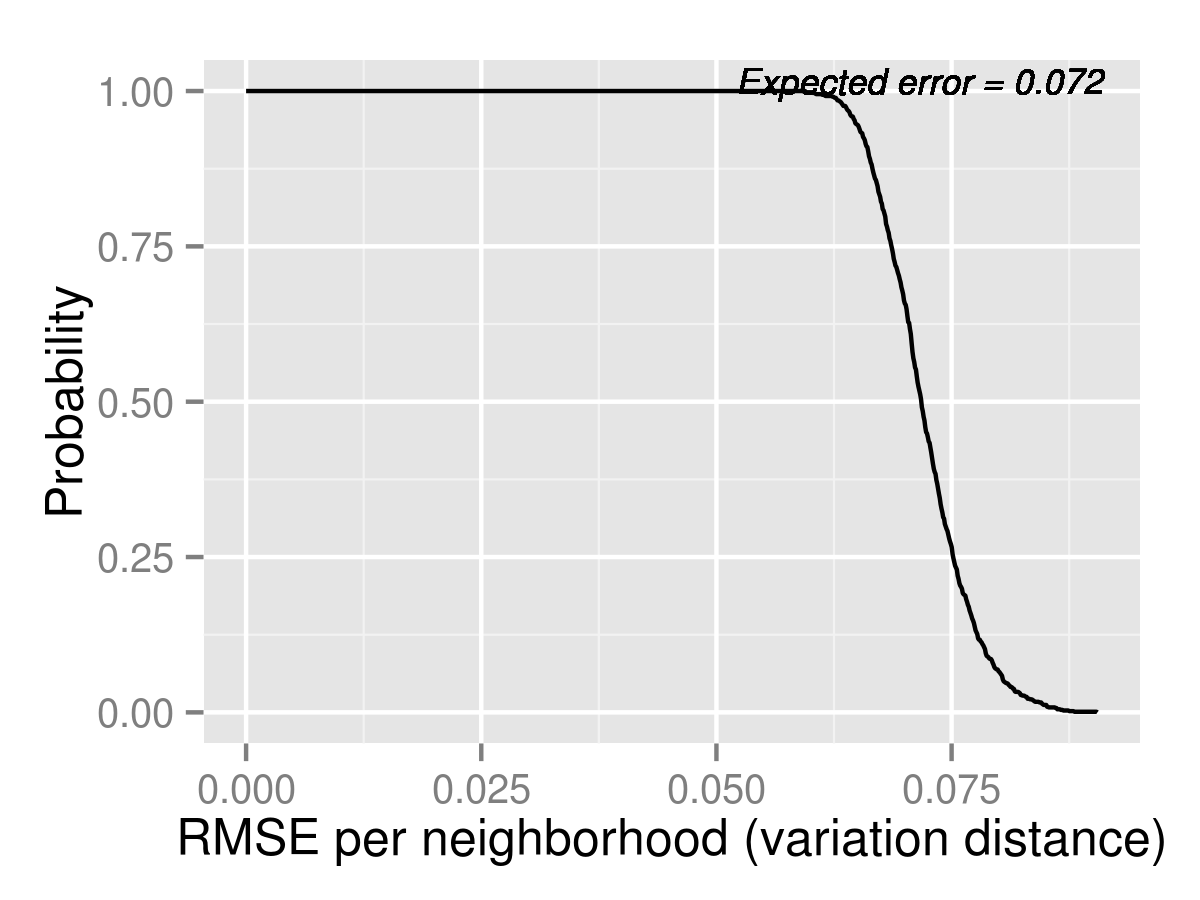}
}\\

\end{figure}

 As in section~\ref{sec:future}, we compute forecasts after simulating 1000 times. The tail distribution plots are in Figures~\ref{fig:outcomes2014K1} and \ref{fig:marketShare2014K1}. The neighborhood by neighborhood predictions are in Tables~\ref{tab:unassigned2014K1}, \ref{tab:aoq2014K1}, and \ref{tab:distance2014K1}. Again, Logit and MixedLogit make very similar predictions in all cases. 

\begin{table}[!htbp]
 \centering
  \caption{Unassigned predictions for 2014 K1.}
  \label{tab:unassigned2014K1}
 \footnotesize
 \begin{tabular}{l c c c c c c}
Neighborhood & Naive & (95 \% C.I.) & Logit & (95 \% C.I.) & MLogit & (95 \% C.I.)\\ \hline  
Allston-Brighton & 21.36 & (3.00,41.02) & 14.75 & (0.00,35.02) & 11.96 & (0.00,33.00)\\ 
Charlestown & 36.05 & (21.00,50.00) & 40.67 & (27.00,55.00) & 40.18 & (27.00,54.00)\\ 
Downtown & 30.25 & (12.00,54.00) & 31.91 & (15.00,55.00) & 31.91 & (14.00,54.00)\\ 
East Boston & 123.49 & (69.97,186.00) & 132.59 & (71.00,195.03) & 132.58 & (72.00,198.00)\\ 
Hyde Park & 80.39 & (51.00,111.00) & 82.77 & (54.00,114.03) & 81.96 & (54.00,114.03)\\ 
Jamaica Plain & 81.56 & (46.00,122.03) & 60.26 & (27.00,101.00) & 56.88 & (25.00,94.00)\\ 
Mattapan & 84.71 & (58.00,115.00) & 70.77 & (45.00,100.00) & 69.33 & (43.00,98.00)\\ 
North Dorchester & 55.55 & (30.00,84.03) & 28.06 & (8.00,52.00) & 27.34 & (8.00,51.00)\\ 
Roslindale & 108.77 & (71.00,151.03) & 114.96 & (79.00,156.00) & 114.25 & (78.00,153.00)\\ 
Roxbury & 81.77 & (50.98,118.00) & 55.48 & (15.00,102.03) & 54.89 & (15.00,103.03)\\ 
South Boston & 17.26 & (6.00,31.00) & 15.72 & (3.00,31.00) & 15.85 & (4.00,31.00)\\ 
South Dorchester & 142.72 & (98.00,194.03) & 109.65 & (62.00,160.00) & 109.08 & (60.00,163.03)\\ 
South End & 34.00 & (16.98,54.02) & 23.71 & (8.00,43.00) & 23.00 & (8.00,43.00)\\ 
West Roxbury & 81.46 & (47.98,121.00) & 93.77 & (59.00,134.03) & 92.67 & (54.00,134.00)\\ 
\end{tabular}
 \end{table}

\begin{table}[!htbp]
 \centering
 \caption{Access to quality predictions for 2014 K1.}
 \label{tab:aoq2014K1}
 \footnotesize
 \begin{tabular}{l c c c c c c}
Neighborhood & Naive & (95 \% C.I.) & Logit & (95 \% C.I.) & MLogit & (95 \% C.I.)\\ \hline  
Allston-Brighton & 0.49 & (0.35,0.64) & 0.82 & (0.62,1.00) & 0.86 & (0.65,1.00)\\ 
Charlestown & 0.37 & (0.25,0.52) & 0.41 & (0.29,0.55) & 0.41 & (0.29,0.55)\\ 
Downtown & 0.43 & (0.34,0.53) & 0.58 & (0.46,0.71) & 0.59 & (0.47,0.72)\\ 
East Boston & 0.53 & (0.39,0.70) & 0.50 & (0.37,0.65) & 0.50 & (0.37,0.65)\\ 
Hyde Park & 0.36 & (0.27,0.45) & 0.39 & (0.30,0.50) & 0.40 & (0.30,0.51)\\ 
Jamaica Plain & 0.41 & (0.32,0.50) & 0.62 & (0.49,0.76) & 0.63 & (0.49,0.76)\\ 
Mattapan & 0.27 & (0.20,0.34) & 0.46 & (0.35,0.59) & 0.46 & (0.35,0.58)\\ 
North Dorchester & 0.39 & (0.31,0.47) & 0.58 & (0.45,0.72) & 0.57 & (0.45,0.70)\\ 
Roslindale & 0.38 & (0.32,0.45) & 0.44 & (0.36,0.53) & 0.44 & (0.35,0.53)\\ 
Roxbury & 0.43 & (0.35,0.51) & 0.64 & (0.52,0.77) & 0.64 & (0.53,0.78)\\ 
South Boston & 0.34 & (0.25,0.44) & 0.51 & (0.38,0.67) & 0.50 & (0.38,0.64)\\ 
South Dorchester & 0.39 & (0.33,0.47) & 0.58 & (0.47,0.72) & 0.58 & (0.46,0.70)\\ 
South End & 0.50 & (0.41,0.60) & 0.67 & (0.56,0.80) & 0.68 & (0.56,0.79)\\ 
West Roxbury & 0.53 & (0.44,0.63) & 0.55 & (0.46,0.65) & 0.56 & (0.46,0.66)\\ 
\end{tabular}
 \end{table}

\begin{table}[!htbp]
 \centering
 \caption{Distance predictions for 2014 K1.}
 \label{tab:distance2014K1}
 \footnotesize
 \begin{tabular}{l c c c c c c}
Neighborhood & Naive & (95 \% C.I.) & Logit & (95 \% C.I.) & MLogit & (95 \% C.I.)\\ \hline  
Allston-Brighton & 1.23 & (1.02,1.45) & 1.22 & (1.02,1.43) & 1.23 & (1.05,1.45)\\ 
Charlestown & 1.82 & (1.40,2.29) & 1.41 & (0.97,1.89) & 1.38 & (0.94,1.89)\\ 
Downtown & 1.57 & (1.28,1.93) & 1.43 & (1.17,1.69) & 1.42 & (1.19,1.69)\\ 
East Boston & 1.97 & (1.66,2.29) & 1.44 & (1.21,1.72) & 1.38 & (1.13,1.65)\\ 
Hyde Park & 2.05 & (1.79,2.34) & 1.90 & (1.68,2.13) & 1.87 & (1.63,2.12)\\ 
Jamaica Plain & 1.31 & (1.14,1.50) & 1.27 & (1.15,1.41) & 1.27 & (1.15,1.40)\\ 
Mattapan & 2.10 & (1.77,2.47) & 1.77 & (1.54,2.00) & 1.76 & (1.53,2.01)\\ 
North Dorchester & 1.47 & (1.21,1.75) & 1.28 & (1.13,1.44) & 1.29 & (1.11,1.48)\\ 
Roslindale & 1.81 & (1.60,2.06) & 1.62 & (1.45,1.79) & 1.60 & (1.43,1.79)\\ 
Roxbury & 1.43 & (1.28,1.61) & 1.29 & (1.15,1.43) & 1.26 & (1.14,1.40)\\ 
South Boston & 1.24 & (0.98,1.55) & 1.12 & (0.92,1.33) & 1.12 & (0.92,1.34)\\ 
South Dorchester & 1.58 & (1.42,1.72) & 1.48 & (1.36,1.60) & 1.49 & (1.36,1.62)\\ 
South End & 1.50 & (1.21,1.83) & 1.35 & (1.08,1.63) & 1.33 & (1.09,1.61)\\ 
West Roxbury & 2.08 & (1.85,2.31) & 1.76 & (1.54,2.00) & 1.76 & (1.52,1.96)\\ 
\end{tabular}
 \end{table}

 \clearpage
 \newpage

 \section{Calibrating the Mixed Logit Model using MCMC}
 \label{app:mixedlogit-mcmc}
 
 Unlike in the simple Logit model, the log likelihood function associated with the mixed logit model is difficult to evaluate directly, involving many multi-dimensional integrals. Hence, we calibrate it using Markov Chain Monte Carlo (MCMC) instead of maximum likelihood.
 
 The basic framework to calibrate mixed logit models using MCMC have been established in previous works such as~\citeasnoun{train:03}, and is based on Gibbs sampling and the Metropolis-Hasting algorithm. However, the examples there do not have so many fixed coefficients as we have. (We have a fixed effect for every school and over 80 schools, and since only the relative difference between fixed effects matter for preferences, but not their absolute values, these fixed effects are likely to be highly correlated.) It is known that the simple Metropolis Hasting with random walk proposals does not perform well when estimating many dimensions (see~\citeasnoun{katafygiotis/zuev:2008}), especially if the dimensions are correlated. So we modify the framework to use Metropolis-Within-Gibbs (MWG), which samples blocks of coordinates iteratively and not everything at once, and Hamiltonian Monte Carlo (HMC), which incorporates gradient information to suggest good directions to sample. We will describe these methods in 
Section~\ref{app:mcmc-review}.
 
 \subsection{Specifying the Likelihood Function}
 \label{app:mixedlogit-likelihood}
 
 The first step of applying MCMC techniques is specifying the full likelihood function of observing the data given the model parameters. To do this, we will restate the mixed logit model in a more precise way and clearly define the parameters. 
   
 The model of interest is as follows. For student $i$ and program $j$, let $\mathbf{F}_{ij}$ be a vector in which the corresponding components correspond to the following features for this student-program pair: ``continuing,'' ``sibling,'' ``ell language match,'' ``distance*black/hispanic,'' ``distance*income est.,'' ``mcas*black,'' ``mcas*income est.,'' ``\% white/asian*black/hispanic,'' and ``\% white/asian*income est.'' Let $\mathbf{G}_{ij}$ be a vector with components corresponding to the following features: ``ell match,'' ``walk zone,'' ``distance,'' ``mcas'' and ``\% white/asian.'' The explanations for these features are in Section~\ref{sec:logit}. The features in $\mathbf{F}_{ij}$ are assumed to have the same coefficients for all students, while the features in $\mathbf{G}_{ij}$ correspond to features that have random coefficients. These random coefficients are in turn partitioned into 3 blocks, with the first block being ``ell match,'' the second ``walk zone,'' and the third ``distance,'' ``mcas,'' 
and ``\% white/asian.'' The coefficients within each block are 
arbitrarily correlated, while the coefficients across blocks are independent. There is also a fixed effect for every school, capturing its general, common attractiveness. (Note that this departs from the notation in Section~\ref{sec:mixed-logit} as $\mathbf{F}_{ij}$ there corresponds to not only the above but also all of the features in $\mathbf{G}_{ij}$ and also the school-specific fixed-effects. This current notation makes the following exposition easier.)
  
 More precisely speaking, the utility of student $i$ for program $j$ in school $s(j)$ is 
 \[u_{ij} = \tilde{\alpha}_{s(j)} + \boldsymbol{\beta}\cdot\mathbf{F}_{ij} + \boldsymbol{\gamma_i} \cdot \mathbf{G}_{ij} + \epsilon_{ij},\]
 where vector $\boldsymbol{\tilde{\alpha}} = \spvec{\boldsymbol{\alpha};0}$ corresponds to the school fixed-effects. (The last component is normalized to zero because only relative differences between fixed effects drive utilities.) $\boldsymbol{\beta}$ corresponds to the other fixed coefficients. $\boldsymbol{\gamma_i}$ is the instantiation of the random coefficients for student $i$, and $\epsilon_{ij}$ is an unobservable idiosyncratic taste shock. The terms $\boldsymbol{\gamma_i}$ are i.i.d. and distributed $\mathtt{Normal}(\mathbf{b},\mathbf{W})$, where $\mathbf{b}$ is the mean vector and $\mathbf{W}$ is the covariance matrix. The idiosyncratic shocks $\epsilon_{ij}$ are i.i.d. standard Gumbel distributed. Because of our partition of the random coefficients into 3 blocks, with the blocks having 1, 1 and 3 variables respectively, the covariance matrix can be written in the block diagonal form
 \[\mathbf{W} = \begin{pmatrix}
                 \mathbf{W_1} & & \\
                 & \mathbf{W_2} & \\
                 & & \mathbf{W_3} \\
                \end{pmatrix},\]
where $\mathbf{W_1}$, $\mathbf{W_2}$ and $\mathbf{W_3}$ are $1 \times 1$, $1 \times 1$ and $3 \times 3$ symmetric positive definite matrices. In summary, the parameters to estimate are $\bs{\alpha}$, $\bs{\beta}$, $\mathbf{b}$, $\mathbf{W_1}$, $\mathbf{W_2}$ and $\mathbf{W_3}$. 

The data to fit these parameters are the observed choices of every students along with the observed characteristics $\mathbf{F}_{ij}$ and $\mathbf{G}_{ij}$. Suppose that student $i$ makes $m_i$ choices, and let the chosen programs from best to worst be $y_{i1}, y_{i2}, \cdots, y_{im_i}$. Given the instantiation of $\bs{\gamma_i}$, the likelihood function on $\bs{\alpha}$ and $\bs{\beta}$ is

\begin{equation}\phi_i(\bs{\alpha}, \bs{\beta} | \bs{\gamma_i}) = \prod_{c=1}^{m_i} \frac{ \exp(\tilde{\alpha}_{s(y_{ic})}+ \bs{\beta} \cdot \mathbf{F}_{iy_{ic}} + \bs{\gamma_i} \cdot \mathbf{G}_{iy_{ic}})}{ \sum_{d=c}^{m_i} \exp (\tilde{\alpha}_{s(y_{id})}+ \beta \cdot \mathbf{F}_{iy_{id}} + \bs{\gamma_i} \cdot \mathbf{G}_{iy_{id}})} . \label{eq:phi}\end{equation}

This is the logit likelihood function. (Recall that $\bs{\tilde{\alpha}} = \spvec{\bs{\alpha};0}$ is the fixed effects including the last school that is normalized to zero, and $s(j)$ denotes the school where program $j$ is in.)  

The full likelihood function incorporating all the data is

\begin{equation} \Phi(\bs{\alpha},\bs{\beta},\mathbf{b}, \mathbf{W}) = \prod_{i=1}^n \int_{\mathbb{R}^5} \phi_i(\bs{\alpha},\bs{\beta} | \bs{\gamma_i})\exp(-\frac{1}{2}\mathbf{W}^{-1}\|\bs{\gamma_i}-\mathbf{b}\|^2) d\bs{\gamma_i} .\label{eq:Phi} \end{equation}
Here, $n$ is the number of students; recall that the random coefficients $\bs{\gamma_i}$ each has five dimensions.)

Our estimates will be based on sampling the parameters based on this likelihood function $\Phi$. Because $\Phi$ is complex, we do this by MCMC. As a detour, we will give an overview of MCMC and the specific techniques we use. Readers who are familiar with these techniques can jump to Section~\ref{app:mcmc-our}.

\subsection{Overview of the MCMC}
\label{app:mcmc-review}
The idea behind Markov Chain Monte Carlo (MCMC) is to samples from a distribution by constructing a Markov chain whose unique stationary distribution is the desired distribution of interest. So if the chain is easy to simulate and if it is fast-mixing, meaning that it converges quickly to the stationary distribution, then we can sample by simply simulating the chain. After throwing out a so-called ``burn-in'' period at the beginning, we would have arrived at samples from the desired distribution. 

The work horse of MCMC are Gibbs sampling and Metropolis-Hasting. Gibbs sampling is used when the desired distribution can be factored into several marginal distributions that are easier to sample. For example, to sample from a joint distribution on $x$, $y$ and $z$, one might iteratively sample one variable at a time conditional on the other ones. Precisely speaking, we initialize $x^0$, $y^0$ and $z^0$ arbitrarily. For each $t\ge 1$, sample iteratively from the following conditional distributions:
\[\begin{array}{rcl}
 x^t & | & y^{t-1}, z^{t-1} \\
 y^t & | & x^{t}, z^{t-1} \\
 z^t & | & x^{t}, y^{t}
\end{array}\]

After a sufficient number $S$ of samples, and after throwing out the initial burn-in of $B$ samples, $\{(x^t,y^t,z^t) : B< t \le S\}$ would approximate samples from the original distribution, although successive samples are not independent. One can remove the serial correlation by either sampling independently from this set, or by keeping only samples in which $t$ is a multiple of $\Delta$, where $\Delta$ is a chosen positive integer. 

Metropolis-Hasting is a technique to sample from an arbitrary distribution with given likelihood function $L(x)$. There are many variants, but the common idea is to use a proposal distribution that is easy to sample from and reject certain samples to get the likelihood ratios to be correct. The proposal distribution may depend on the current iterate $x$. Let transition probability density be $T(y|x)$; this is the probability density of proposing $y$ given that the current sample is $x$. In order to obtain the correct likelihoods, we can only accept a fraction of the samples proposed, and reject the others. The probability that we accept proposal $y$ given the previous iterate being $x$ is
\[A(y|x) = \min (1, \frac{L(y) T(x|y)}{L(x) T(y|x)}) .\]

Note that if $T(y|x)$ is proportional to $L(y)$, then the acceptance probability is always $1$ as the proposal distribution already matches the target. Otherwise, the above formula is tuned so that the following identity, called ``detailed balance'' in the literature, holds:
\[L(x)T(y|x)A(y|x) = L(y)T(x|y)A(x|y).\]

This equation guarantees that the desired density $p(x)$ is a stationary distribution of the Markov chain induced by the proposal and acceptance process. Furthermore, if the chain is ergodic, which is true for example if the proposal distribution has full support, then $p(x)$ is the only stationary distribution. 

The sampling procedure is then to initialize $x^0$ arbitrarily, and for each $t\ge 1$
\begin{enumerate}
 \item Draw $y$ according to $T(y|x^{t-1})$.
 \item Set $x^t= \begin{cases}
                  y & \text{with prob. } A(y|x^{t-1}),\\
                  x^{t-1} & \text{otherwise.}
                 \end{cases}$
\end{enumerate}

By iterating this many times and discarding sufficiently many burn-in samples, we would have arrived at the desired distribution. 

Because of the flexibility in the proposal distributions, there are many variants of the above techniques. The goal is to find a proposal distribution that strikes a good balance of being easy to sample from and approximating the target distribution locally. If it is not easy to sample from, then each step would take too long; if it is too far from the target distribution, then the acceptance probabilities would be very low and the chain may get stuck at a certain iterate for a very long time. In the following sections we present the three variants we use: Random Walk Metropolis (RWM), Metropolis-Within-Gibbs (MWG), and Hamiltonian Monte Carlo (HMC).

\subsubsection{Random Walk Metropolis (RWM)}
This method is the easiest to sample from, as it uses a simple random walk to propose the next value: if the current iterate is $x$, it proposes $y = x+ \epsilon$, where $\epsilon$ is multivariate normal distributed, $\epsilon \sim \mathtt{Normal}(0, \rho I)$, where $I$ is the identity matrix and $\rho$ is a scale parameter. Other covariance matrices can also be used instead of the identity but it must be the same for every $x$. The scale parameter is tuned to match the overall variance of the desired distribution. Too small a $\rho$ and successive samples and there will be too much serial correlation; too large a $\rho$ and acceptance probability might be near zero so the chain may get stuck. We tune $\rho$ by multiplying it up or down so that the average acceptance ratio since last tuning is between $0.4$ and $0.6$, which is the ball park value suggested by the literature.\footnote{See ~\citeasnoun{roberts-et-al:97}.} The number of steps we wait before tuning increases exponentially, so that after our burn-
in sample until 
our last iteration there is no tuning. 

This method performs well when the target distribution has not too many dimensions, and has approximately the same scale in each dimension. However, when there are many dimensions, it becomes exponentially harder to guess the right direction, and the method may take very long to converge; when there are dimensions that are at very different scales, then there may exist no $\rho$ that is good for all dimensions.

\subsubsection{Metropolis Within Gibbs (MWG)}
This is a simple extension of RWM that allows various sub-blocks of coordinates to have different scales. It is simply to sample each sub-block iteratively, conditional on the others, much like running several RWM within a Gibbs sampling framework. This also reduces the number of dimensions sampled at each step. The draw back is that more samples are needed. 

Precisely speaking, instead of sampling all dimensions of vector $x$ simultaneously, write it in terms of sub-vectors $x = \spvec{x_1;x_2;\vdots;x_k}$. Each sub-vector may represent several coordinates. Initialize $x^0$ arbitrarily and for $t \ge 1$, sample
\[\begin{array}{rcl}
 x_1^t & | & x_2^{t-1}, \cdots x_k^{t-1} \\
 x_2^t & | & x_1^t, x_3^{t-1}, \cdots x_k^{t-1} \\
  & \cdots \\
 x_k^t & | & x_1^t, \cdots x_{k-1}^{t} 
\end{array}\]

Each of the above is sampled using RWM, perhaps with different scale parameters for different sub-vectors. In each Gibbs iteration, for each of the variables, we only take one step of Metropolis-Hasting, which involves one proposal and possible acceptance. Because of detailed balance, embedding Metropolis-Hasting into Gibbs sampling in this way also works.

\subsubsection{Hamiltonian Monte Carlo (HMC)}
This method uses the gradient of the log likelihood function to inform the proposals, which can significantly improve the acceptance probabilities in high dimensions. The drawback is that each iteration is slower as several gradient calls is needed. The method is motivated by Hamiltonian dynamics in physics. It models the current iterate $x$ as a location vector, and treats the negative log likelihood function as an energy potential. In each step, it samples a random momentum vector and simulates the trajectory of the object by discretizing time and alternatively updating the momentum using the potential function and updating the position using the momentum. To make detailed balance work out, the first and last steps of simulation are half-steps. Precisely speaking, let the gradient of the log likelihood function be $G(x) = \nabla (\log(L(x)))$. Let $\epsilon$ and $\Delta$ be tuning parameters, representing the discretization in time and the number of steps to simulate respectively. The proposal is based on 
the pseudocode in this algorithm (this is taken from~\citeasnoun{neal:2011}):

\begin{algorithm}
\label{alg:hmc-step}
 \caption{Pseudocode for one step of HMC}
 \begin{algorithmic}
  \STATE Function HMC\_STEP($x$):
	\STATE Draw momentum $p_0 \sim \mathtt{Normal}(0,I)$.
	\STATE Initialize $y=x, p=p_0$.
	\STATE Update $p=p - \epsilon G(y)/2$.
	\FOR {$\Delta-1$ iterations}
	\STATE Update $y = y + \epsilon p$
	\STATE Update $p = p - \epsilon G(y)$.
	
	\ENDFOR
	\STATE Update $y=y+\epsilon p$.
	\STATE Update $p=p - \epsilon G(y)/2$.
	\RETURN $\begin{cases}
	         y & \text{with prob. } A(y|x)=\min(1, \frac{L(y)}{L(x)}\exp(\frac{\|p_0\|^2-\|p\|^2}{2}) \\
	         x & \text{otherwise}
	        \end{cases}$

 \end{algorithmic}
\end{algorithm}

Note that the chance of proposing $y$ given $x$ is simply the chance of drawing momentum $p_0$. Moreover, by the reversibility of the intermediate steps of discrete simulation, if we started at $y$ and drew a momentum of $-p$ (where $p$ is the final momentum vector in HMC\_STEP), then the proposal would be $x$. This implies that 
\[\frac{T(y|x)}{T(x|y)} = \frac{\exp(-\frac{1}{2}\|p_0\|^2)}{\exp(-\frac{1}{2}\|-p\|^2)},\]
which implies that
\[\frac{T(y|x)A(y|x)}{T(x|y)A(x|y)} = \frac{\exp(-\frac{1}{2}\|p_0\|^2)}{\exp(-\frac{1}{2}\|-p\|^2)}\frac{L(y)}{L(x)}\exp(\frac{\|p_0\|^2-\|p\|^2}{2}) = \frac{L(y)}{L(x)}.\]

So detailed balance holds and the following is a valid Metropolis-Hasting sampler: Initialize $x^0$ arbitrarily. For $t \ge 1$, set $x^t=\text{HMC\_STEP}(x^{t-1})$.

One can show that as the time discretization $\epsilon \to 0$, for any fixed total simulation time $\epsilon \Delta$, the acceptance probability goes to $1$. Hence, we would like $\epsilon$ to be small enough so the chain does not get stuck and $\epsilon \Delta$ large enough so that successive samples are not too serially correlated. In practice, we fix $\Delta=20$ and tune $\rho$ so that the empirical acceptance rate since last tuning is between 0.5 and 0.8. As before, we increase the interval between tuning times exponentially so that no tuning happens in the sample we keep (after burn-in and before the last iteration). Another detail is that to prevent cases in which $\epsilon \Delta$ is exactly what makes the proposal $y$ go back to original point $x$, instead of using the same $\epsilon$, we draw $\tilde{\epsilon}\sim \mathtt{Uniform}(0.85\epsilon, 1.15\epsilon)$ before each call to HMC\_STEP, and use $\tilde{\epsilon}$ as the step size throughout that call. Because this distribution is a-priori fixed, 
detailed balance is also 
preserved. All these are according to the best practices for applying HMC as outlined in~\citeasnoun{neal:2011}. 

\subsection{Our MCMC Sampler}
\label{app:mcmc-our}
Our MCMC procedure is based on the one in~\citeasnoun{train:03} but breaking up the estimation of the fixed coefficients into two steps, one step using Hamiltonian Monte Carlo (HMC) and the other Metropolis Within Gibbs (MWG). We use HMC to estimate the school fixed effects and MWG to estimate the other fixed coefficients. These techniques allow us to accommodate the large number of school fixed effects and the unequal scales across the other fixed coefficients. 

To sample from the full likelihood function $\Phi(\bs{\alpha},\bs{\beta},\mathbf{b},\mathbf{W})$ (Equation~\ref{eq:Phi}), we initialize $\bs{\alpha}^0, \bs{\beta}^0, \mathbf{b}^0, \mathbf{W_1}^0, \mathbf{W_2}^0, \mathbf{W_3}^0$ arbitrarily. For each $t \ge 1$, we do a few layers of Gibbs sampling. In some of the layers we embed a form of Metropolis-Hasting; but in each Gibbs iteration we only take one step of Metropolis-Hasting, much as it is in MWG. Furthermore, let $T$ be a parameter indicating how long we wait before tuning. We initialize $T$ to be $1$ and increase this parameter steadily, so that tuning becomes exponentially less frequent. For $t\ge 1$, each MCMC step is as follows:

\begin{enumerate}
 \item Draw $\bs{\gamma_i}^t | \bs{\alpha}^{t-1}, \bs{\beta}^{t-1}, \mathbf{b}^{t-1}, \mathbf{W}^{t-1}$. This is done using one iteration of RWM with likelihood function
 \[L(\mathbf{x}) = \phi_i (\bs{\alpha}^{t-1},\bs{\beta}^{t-1},\mathbf{x}) \exp(-\frac{1}{2}(\mathbf{W}^{t-1})^{-1} \| \mathbf{x} - \mathbf{b}^{t-1}\|^2)\]
 and starting value $\bs{\gamma_i}^{t-1}$. (See Equation~\ref{eq:phi} for definition of $\phi_i$.) We initialize $\rho=0.05$ and initially to tune for each $i$ every $\mathtt{Uniform}(1000T,1500T)$ steps.
 
 \item Draw $\mathbf{b}^t | \bs{\gamma_i}^t, \mathbf{W}^{t-1}$. This is sampling from $\mathtt{Normal}(\frac{1}{n}\sum_{i=1}^n \bs{\gamma_i}^t, \frac{1}{m} \mathbf{W}^{t-1})$.
 \item Draw $\mathbf{W}^t | \bs{\gamma_i}^t, \mathbf{b}^{t}$. This can be done as follows: For $l \in \{1,2,3\}$, let $\mathbf{C}_l^t$ be the covariance matrix of the $l$th block of $\gamma_i^t$ assuming mean as in the $l$th block of $\mathbf{b}^t$. (Recall that the random coefficients are organized into 3 blocks, with ell match being the first block, walk zone being the second, and distance, mcas, and \% white/asian being the third.) Let $k_l$ be the number of variables in the $l$th block. Draw $\mathbf{W}_l^t$ according to the Inverse Wishart Distribution with degree of freedom $\nu = k_l + n$ and scale matrix $\bs{\Psi} = k_l I_{l \times l} + n \mathbf{C}_l^t$. 
 \item Draw $\bs{\alpha}^t | \bs{\gamma_i}^t, \bs{\beta}^{t-1}$. This is done using one step of HMC with likelihood function
 \[L(\mathbf{x}) = \prod_{i=1}^n \phi_i (\mathbf{x}, \bs{\beta}^{t-1} |  \bs{\gamma_i}^t).\]
 We initialize $\epsilon = 0.015$, and $\Delta=20$. We tune every $1000T$ steps.
 \item Draw $\bs{\beta}^t | \bs{\gamma_i}^t, \bs{\alpha}^t$. This is done using one iteration of MWG with likelihood function
 \[L(\mathbf{x}) = \prod_{i=1}^n \phi_i (\bs{\alpha}^t, \mathbf{x} | \bs{\gamma_i}^t).\]
 We break the fixed coefficients $\bs{\beta}$ into 6 subvectors: 1) ``continuing;'' 2) ``sibling;'' 3) ``ell language match;'' 4) ``distance*black/hispanic'' and ``distance*income est.''; 5) ``mcas*black'' and ``mcas*income est.''; 6) ``\% white/asian*black/hispanic'' and ``\% white/asian*income est.'' We initialize the scales $\rho$ for each subvector to be .5, .5, .1, .1, .5, and .5 respectively. We tune every $\mathtt{Uniform}(100T,150T)$ steps.
\end{enumerate}

We run these steps 1,000,000 times, increasing the tuning interval parameter $T$ by a factor of $1.2$ every 5000 iterations. We throw out the first 500,000 iterations as burn-in. Note that in the interval we keep, no tuning happens. This ensures the correctness of the Markov chain in this period.

For robustness check, we re-ran this procedure 6 times, sometimes with different initial values, and we found near identical results each time.

\bibliographystyle{economet}
\bibliography{bps}

\end{document}